\documentclass[a4paper]{article}
\pdfoutput=1
\usepackage{jcappub}
\usepackage{url}
\usepackage{graphicx}
\usepackage{subcaption}
\usepackage{booktabs}
\usepackage{amsmath}
\usepackage{amssymb}
\usepackage{amstext}
\usepackage{amsfonts}
\usepackage{amsthm}
\usepackage{hyperref}
\usepackage{appendix}
\usepackage{physics}
\usepackage[utf8]{inputenc}
\usepackage{soul}
\usepackage{hhline}
\usepackage{comment}
\usepackage{color}
\usepackage{bm}
\usepackage[normalem]{ulem}

\newcommand{\darkph}{\ensuremath{{\rm A^{\prime}}}}
\newcommand{\res}{\ensuremath{{\rm res}}}
\newcommand{\vir}{\ensuremath{{\rm vir}}}
\newcommand{\NFW}{\ensuremath{{\rm NFW}}}
\newcommand{\obs}{\ensuremath{{\rm obs}}}
\newcommand{\sky}{\ensuremath{{\rm sky}}}
\newcommand{\rmmin}{\ensuremath{{\rm min}}}
\newcommand{\rmexp}{\ensuremath{{\rm exp}}}
\newcommand{\lin}{\ensuremath{{\rm lin}}}
\newcommand{\knee}{\ensuremath{{\rm knee}}}
\newcommand{\gas}{\ensuremath{{\rm gas}}}
\newcommand{\reio}{\ensuremath{{\rm reio}}}
\newcommand{\halo}{\ensuremath{{\rm -halo}}}
\newcommand{\MW}{\ensuremath{{\rm MW}}}
\newcommand{\Th}{\ensuremath{{\rm Th}}}
\newcommand{\Sc}{\ensuremath{{\rm Sc}}}
\newcommand{\dSc}{\ensuremath{{\rm dSc}}}

\usepackage{tabularx}
\newcolumntype{C}{>{\centering\arraybackslash}X}
\newcolumntype{R}{>{\raggedleft\arraybackslash}X}

\newcommand{\ie}{i.e.\ }
\newcommand{\eg}{e.g.\ }
\newcommand{\etal}{et~al.\ }

\newcommand{\perimeter}{Perimeter Institute for Theoretical Physics, 31 Caroline St N, Waterloo, ON N2L 2Y5, Canada}

\newcommand{\UW}{Department of Physics and Astronomy, University of Waterloo, Waterloo, ON N2L 3G1, Canada}

\newcommand{\york}{Department of Physics and Astronomy, York University, Toronto, ON M3J 1P3, Canada}

\def\eqref#1{(\ref{#1})}

\title{Patchy Screening of the CMB from Dark Photons}

\author[a,b]{Dalila~P\^irvu}

\author[a]{, Junwu~Huang}

\author[a,c]{and Matthew~C.~Johnson}

\affiliation[a]{\perimeter}
\affiliation[b]{\UW}
\affiliation[c]{\york}

\emailAdd{dpirvu@perimeterinstitute.ca}
\emailAdd{jhuang@perimeterinstitute.ca}
\emailAdd{mjohnson@perimeterinstitute.ca}

\abstract{
We study anisotropic (patchy) screening induced by the resonant conversion of cosmic microwave background (CMB) photons into dark-sector massive vector bosons (dark photons) as they cross non-linear large scale structure (LSS). Resonant conversion takes place through the kinetic mixing of the photon with the dark photon, one of the simplest low energy extensions to the Standard Model. In the early Universe, resonant conversion can occur when the photon plasma mass, obtained as the photon propagates through the ionized interstellar and intergalactic media, matches the dark photon mass. After the epoch of reionization, resonant conversion occurs mainly in the ionized gas that occupies virialized dark matter halos, for a range of dark photon masses between $10^{-13} {\rm \, eV} \lesssim m_\darkph \lesssim 10^{-11}$ eV. This leads to new CMB anisotropies that are correlated with LSS, which we refer to as patchy dark screening, in analogy with anisotropies from Thomson screening. Its unique frequency dependence allows it to be distinguished from the blackbody CMB. In this paper, we use a halo model approach to predict the imprint of dark screening on the CMB temperature and polarization anisotropies, as well as their correlation with LSS. We then examine the two- and three-point correlation functions of the dark-screened CMB, as well as correlation functions between CMB and LSS observables, to project the sensitivity of future measurements to the kinetic mixing parameter and dark photon mass. We demonstrate that an analysis with existing CMB data can improve upon current constraints on the kinetic mixing parameter by up to two orders of magnitude with the two-point correlation functions, while data from upcoming CMB experiments and LSS surveys can improve the reach by up to three orders of magnitude with two- and three-point correlation functions.}

\begin{document}
\maketitle

\section{Introduction}\label{sec:introduction}

Cosmic microwave background (CMB) and large scale structure (LSS) surveys have provided some of the most important evidence for physics beyond the Standard Model (BSM) of particle physics -- dark matter and dark energy, as well as early-Universe cosmological models such as inflation. These observations also provide powerful constraints on BSM physics, such as the number of relativistic degrees of freedom, neutrino masses, and the strength of interactions during an inflationary era. The next generation of CMB experiments such as the upcoming Simons Observatory~\cite{Ade:2018sbj}, as well as CMB-S4~\cite{abazajian_cmb-s4_2016} and CMB-HD~\cite{Sehgal:2019ewc}, and LSS surveys like DESI~\cite{DESI:2019jxc}, Euclid~\cite{Laureijs2011}, and LSST~\cite{0912.0201} promise to further expand our ability to detect and characterize BSM physics. These new surveys motivate the exploration of new observables that can exploit their full potential. In particular, a promising avenue of current and future effort is to use the CMB as a back-light through which to study the intervening LSS (see \eg\cite{Basu:2019rzm,Scott:2019fot}).

Within the Standard Model of cosmology and particle physics, CMB photons can interact with LSS via gravity (\eg weak lensing or the integrated Sachs-Wolfe effect) and electromagnetism (\eg Sunyaev Zel'dovich effects). These effects lead to additional temperature and polarization anisotropies in the CMB, the so-called secondary CMB, as well as new correlations with tracers of LSS such as galaxy surveys. The slew of associated new observables, and in particular cross-correlations between CMB and LSS, can be used to extract valuable information about the initial conditions and the formation and growth of structure in the early Universe. Several examples include: lensing reconstruction (see \eg\cite{Weak_lensing} for a review), kinetic Sunyaev Zel'dovich velocity reconstruction~\cite{Zhang:2015uta,Terrana2016,Deutsch:2017ybc,smith_ksz_2018,Cayuso:2021ljq}, moving-lens velocity reconstruction~\cite{Hotinli:2018yyc}, and patchy reionization optical depth reconstruction~\cite{dvorkin_reconstructing_2009}.

In this paper, we discuss the possibility of using the CMB and its cross-correlation with tracers of LSS to extract information about BSM physics that manifest in the low-redshift Universe. In particular, we will discuss how a new type of CMB secondary anisotropy and its correlation with LSS can be used to extend the reach in the parameter space of kinetically mixed dark photons~\cite{Okun:1982xi,Holdom:1985ag} by orders of magnitude.

The dark photon is a hypothetical vector boson that arises in various extensions of the Standard Model~\cite{Arvanitaki:2009fg,Goodsell:2009xc}. An ultra-light dark photon is an essential ingredient in dark matter models, either as a light bosonic dark matter candidate~\cite{Pospelov:2008jkf,Arias:2012az,Graham:2015rva}, or as a mediator to a sector of dark matter particles (see~\cite{Knapen:2017xzo} and reference within). Despite recent evidence of various collective effects which cast doubt on the validity of some of the models (vortex production during production of dark photon dark matter~\cite{East:2022rsi} and the two stream instability in the case of freeze-in dark matter  models~\cite{Lasenby:2020rlf}), it is still of great interest to probe the existence of ultra-light dark photons regardless of their cosmological abundance through superradiance~\cite{Baryakhtar:2017ngi,Siemonsen:2019ebd,East:2022ppo,East:2022rsi,Siemonsen:2022ivj}, cosmology~\cite{Mirizzi:2009iz,McDermott:2019lch,Caputo:2020rnx,caputo_dark_2020}, stellar objects~\cite{An:2013yua,Hardy:2016kme,Lasenby:2020goo} and laboratory searches~\cite{Chaudhuri:2014dla,Baryakhtar:2018doz,Chiles:2021gxk,Romanenko:2023irv,SENSEI:2020dpa}.

A dark photon and its coupling to the Standard Model can be described by the Lagrangian
\begin{align}
    \mathcal{L}= -\frac{1}{4}{F}_{\mu\nu}{F}^{\mu\nu}-\frac{1}{4}{F}^{\prime}_{\mu\nu}{F}^{\prime\mu\nu}
    -\frac{m_\darkph^2}{2}\darkph_\mu \darkph^\mu- \frac{\varepsilon}{2} {F}_{\mu\nu}{F}^{\prime\mu\nu}+ {\rm A}^{\mu} J_{\mu},
    \label{eq:darkphoton}
\end{align}
where ${\rm A}^\mu$ and $\darkph^\mu$ are the photon and dark photon fields respectively, with ${F}^{\mu\nu}$ and ${F}^{\prime\mu\nu}$ their field strengths, and $J^{\mu}$ is the Standard Model electromagnetic current. The dark photon has a mass $m_\darkph$ and couples to the Standard Model photon through a kinetic mixing parameter $\varepsilon$. This simple coupling leads to a plethora of observable consequences (see~\cite{Caputo:2021eaa} and references within). Most of these are based on the conversion, in particular resonant conversion, between the photon and the dark photon in a medium. In the lab, resonant conversion is facilitated with carefully prepared small scale experiments. In the early Universe, resonant conversion happens in different astrophysical and cosmological environments, as the dispersion relation of the photon (plasma frequency) is naturally scanned. When the plasma frequency of the photon $m_{\gamma}^2$ over its trajectory $\vec{x}$ matches the mass of the dark photon
\begin{equation}
    m_{\gamma}^2 (\vec{x}) = m_\darkph^2,
\end{equation}
CMB photons resonantly convert into dark photons.

In~\cite{Mirizzi:2009iz,McDermott:2019lch,Caputo:2020rnx,caputo_dark_2020,Garcia:2020qrp}, the resonant conversion between CMB photons and dark photons was studied both in the homogeneous early-time and inhomogeneous late-time limits. Cosmic expansion and the inhomogeneous distribution of ionized gas were identified as important scanners of the plasma frequency. In this paper, we examine the conversion from CMB photons into dark photons inside non-linear structure, after the epoch of reionization, where the amplitude of the density profile of ionized gas within dark matter halos provides the primary scanner.

Resonant conversion leads to a frequency-dependent disappearance of CMB photons that traces the distribution of matter in the Universe. Hence it can be treated as a frequency ($\omega$) and angle ($\hat{n}$) dependent optical depth $\tau(\varepsilon,\omega, \hat{n})$, after integrating along the line of sight. The optical depth from resonant conversion can be extracted or constrained from cosmological data, and we present five methods to search for dark photons, along with the projected sensitivity on $\varepsilon$:

\begin{itemize}
\item {\bf Spectral distortions of the CMB}: The spatially averaged $\ev{\tau(\varepsilon,\omega)}$ manifests as a distortion of the blackbody spectrum of the CMB, and as a result is constrained by COBE/FIRAS~\cite{COBEFixsen}. This effect, the late-time component of the effect studied in~\cite{Mirizzi:2009iz,McDermott:2019lch,Caputo:2020rnx,caputo_dark_2020}, scales as $\varepsilon^2$.

\item {\bf CMB temperature and polarization anisotropies}: The optical depth correlation function $\ev{\tau(\varepsilon,\omega, \hat{n})\,\tau(\varepsilon,\omega, \hat{n}^{\prime})}$ can be obtained from the measured CMB through the screening of the temperature and polarization anisotropies by resonant conversion. Data from CMB experiments can be used to extract the amplitude of dark screening, which scales like $\varepsilon^4$. We show how the large signal-to-noise ratio of CMB surveys, along with the characteristic frequency dependence of the screening signal, implies that this method outperforms the COBE/FIRAS constraint.

\item {\bf Correlating CMB anisotropies with templates from LSS}: As described in greater detail below, the morphology of the dark photon optical depth anisotropy depends on the distribution of ionized gas in halos. With assumptions about the galaxy-gas connection, a galaxy survey can be used to create a template for the dark photon optical depth field $\hat{\tau}(\omega, \hat{n})$. Cross-correlating the CMB measurement with this template $\ev{\tau(\varepsilon,\omega,\hat{n}) \, \hat{\tau}(\omega,\hat{n}^{\prime})}$ scales as $\varepsilon^2$, improving greatly on the CMB-only reach.

\item {\bf Correlation with Thomson screening}: The standard optical depth due to Thomson scattering by free electrons (we will denote by $\tau^\Th $) is also present as a source of screening in the measured CMB. Since this anisotropic signal traces the same distribution of ionized matter as the dark screening component, the two signals will be correlated yet distinguishable due to the latter being frequency dependent. The cross-correlation $\ev{\tau (\varepsilon,\omega, \hat{n}) \, \tau^\Th (\hat{n}^{\prime})}$ also scales as $\varepsilon^2$.

\item {\bf The CMB bispectrum and optical depth reconstruction}: Both conversion to dark photons and Thomson screening induce non-Gaussian statistics in the CMB anisotropies. The combined effect can be modeled via three-point correlation functions (bispectra) that also scale as $\varepsilon^2$. These bispectra hold additional information compared to two-point functions since there are more modes; the associated statistical anisotropy can additionally be used to reconstruct the dark screening optical depth, allowing for its study at the field level.
\end{itemize}

In this paper, we demonstrate the possibility of using the aforementioned methods to improve the reach on kinetically mixed dark photon in the mass range ($10^{-13} {\rm \, eV} \lesssim m_\darkph \lesssim 10^{-11}$ eV). The paper is organized as follows. We first review resonant photon to dark photon conversion and compute the properties of conversion inside individual halos in Section~\ref{sec:DPconversionprobability}, before summing over halos to obtain a frequency dependent dark screening optical depth in Section~\ref{sec:monopole}. In Section~\ref{sec:darkscreening}, we discuss the anisotropies of this dark screening optical depth,  correlation functions, and CMB observables. In Section~\ref{sec:crossc} we study the cross-correlation between this dark screening optical depth and the LSS of our Universe, and construct two-point cross-correlation functions between the CMB and LSS, as well as three-point cross-correlation functions of CMB observables. In Section~\ref{sec:forecasts}, we present a forecast of the sensitivity of existing and future CMB data-sets to the various correlation functions studied in this paper. The result of these forecasts are shown in Section~\ref{sec:conclusions}, along with a discussion of the prospect for constructing similar correlation functions in other new physics scenarios. In the appendix, we present details about the modeling of dark matter and gas halos (Appendix~\ref{appx:modeling}) and the computation of correlation functions of dark screening (Appendix~\ref{appx:2pf}), as well as a list of useful two-point correlation functions and quadratic estimators for the optical depth and other quantities (Appendix~\ref{appx:qes}).

\section{Photon to dark photon conversion}\label{sec:DPconversionprobability}

In this section, we discuss how photons resonantly convert into dark photons within non-linear structure in the context of the halo model of LSS, where matter is organized into virialized dark matter halos populated by gas (see \eg\cite{cooray_halo_2002, asgari2023halo} for a review). Resonant conversion can be modeled via the same formalism that describes neutrino oscillation in medium, that is, the Mikheyev-Smirnov-Wolfenstein (MSW) effect~\cite{Mikheyev:1985zog,Wolfenstein:1977ue}. In the following, we review the resonant conversion of photons into dark photons and present our prescription for modeling the conversion probability as a sum over halos.  

\subsection{Resonant conversion probability}

In an ionized medium, the Lagrangian in Eq.~\eqref{eq:darkphoton} leads to resonant conversion of photons to dark photons. This can be described by the Schr${\rm \ddot{o}}$dinger equation~\cite{Mirizzi:2009iz}:
\begin{equation}
    i\frac{\dd}{\dd t}\begin{pmatrix} {\gamma} \\ \darkph \end{pmatrix} =\frac{1}{4\omega(t)} \begin{pmatrix} {m_{\gamma}^2 (\vec{x}(t)) -m_\darkph^2} & 2\varepsilon m_\darkph^2 \\ 2\varepsilon m_\darkph^2 & {-m_{\gamma}^2 (\vec{x}(t)) +m_\darkph^2} \end{pmatrix}\begin{pmatrix} {\gamma} \\ \darkph \end{pmatrix},
\end{equation}
where $\gamma$ is an incident photon with frequency $\omega(t)$ that follows a trajectory $\vec{x}$ parameterized by time $t$. The photon acquires an effective mass $m_{\gamma}^2 (\vec{x}(t))$ (plasma frequency) as it crosses an ionized medium due to its interaction with the collective oscillations in the free electron density. Hence, to first order, the mass depends on the number density of electrons $n_{e} (\vec{x}(t))$ along its trajectory:
\begin{equation}\label{eq:mass2nden}
	m_{\gamma}^2 (\vec{x}(t)) \simeq 1.4 \times 10^{-21} {\rm \, eV}^2 \left( \frac{n_{e}(\vec{x}(t))}{{\rm cm}^{-3}} \right).
\end{equation}
Here we assume all baryonic matter is ionized and therefore ignore an additional negative contribution due to interactions with neutral atoms~\cite{Mirizzi:2009iz}.

In the small-$\varepsilon$ limit where conversion from dark photons back to photons can be safely neglected, the conversion probability is given by
\begin{equation}\label{eq:totalprobHEP}
	P_{\gamma \to \darkph} = \sum_{t_\res} \frac{\pi \varepsilon \, m_\darkph^2}{\omega (t_\res)} \times \varepsilon\left| \frac{\dd}{\dd t} \ln m_{\gamma}^2(\vec{x}(t) )\right|^{-1}_{t=t_\res},
\end{equation}
where $t_\res$ are the times when the resonance condition $m_{\gamma}^2(t_\res) = m_\darkph^2$ is met along the path $\vec{x}$. This expression for the total probability is a combination of the conversion rate $\Gamma_\res = \pi \varepsilon \, m_\darkph^2 / \omega (t_\res)$ and the resonance time scale $\Delta t_\res \simeq \varepsilon\left| \frac{\dd}{\dd t} \ln m_{\gamma}^2(\vec{x}(t)) \right|^{-1}_{t=t_\res}$. 

\subsection{Photon to dark photon conversion in non-linear structure}

In the homogeneous and weakly inhomogeneous early Universe ($z \gg 10$), the slowly diluting charged particle density caused by cosmic expansion provides a natural scanner of the dark photon mass, and ensures efficient conversion for a wide range of dark photon masses. At low redshift ($z \lesssim 10$), the scanner is mainly provided by the spatially varying electron density inside non-linear structure, \eg halos, which also ensures efficient conversion over a range of dark photon masses due to the large density contrast.

Expression~\eqref{eq:totalprobHEP} is the integrated probability to convert along the line of sight. Working in the halo model for LSS~\cite{cooray_halo_2002}, this expression becomes a sum over halos, where each term represents the probability that a photon converts within each. We re-write the probability per halo in terms of the mass and redshift: 
\begin{equation}
	P_i (\vec{x} | z_i, m_i) = \frac{\pi \varepsilon^2 m_\darkph^2}{\omega (z_i)} \left| \frac{\dd}{\dd t} \ln m_{\gamma}^2(\vec{x} | z_i, m_i ) \right|^{-1}_{t=t_\res}.
\end{equation}
This expression holds for any type of photon, but in what follows we will focus on the conversion of CMB photons along their path from the surface of last scattering to the Earth. In the remainder of this sub-section we explain how to simplify the term in the modulus to account for a photon's path across each halo. 

The effective photon mass $m_{\gamma}^2 (\vec{x})$ depends on the baryon number density as well as the ionization fraction. In a galactic halo, baryonic matter represents a fraction $\Omega_b / \Omega_c \sim 0.19$ of the total halo mass $m$. Since baryonic matter is predominantly protons by mass, and the Universe is electrically neutral, we approximate the number density of the electrons to be the same as the number density of baryons. Furthermore, we are interested in the period after reionization ($z \lesssim 6-10$) therefore we assume throughout that the ionization fraction is unity everywhere. Where relevant, we treat reionization as instantaneous at a redshift in the range $6<z<10$ to encapsulate uncertainties about the history of reionization. Finally, we neglect the impact of He reionization. Further details of our modeling of reionization and other assumptions can be found in Appendix~\ref{appx:modeling}.

For the density profile of baryons, we use the Battaglia \etal `AGN Feedback' gas density profiles introduced in~\cite{Battaglia_2016}, which are based on hydrodynamic cosmological simulations. We use a version of the profile where the fit parameters are based on simulations that include a sub-grid model for active galactic nuclei (AGN) feedback. The profile is given by an expression that parametrically resembles the standard Navarro-Frenk-White (NFW) density profile of dark matter in halos~\cite{1996ApJ462563N}:
\begin{equation}\label{eq:rhogas}
    \rho_\gas = \frac{\Omega_b}{\Omega_c} \rho_c(z) \rho_0(z,m)\left(\frac{x}{x_c}\right)^\gamma\left[1+\left(\frac{x}{x_c}\right)^{\alpha(z,m)}\right]^{-\frac{\beta(z,m)+\gamma}{\alpha(z,m)}}, \quad x \equiv \frac{r}{r_{200}(z,m)}.
\end{equation}
The quantity $\rho_c(z)$ is the critical density for a flat FRW Universe and $r_{200}$ is the radius where the gas density reaches $200\rho_c$. The exponents $\alpha, \beta, \gamma$ fix the slope in the regimes where $x\sim 1, x \gg 1$ and $x \ll 1$, respectively. There are two fixed quantities $\gamma=-0.2$ and the core scale $x_c = 0.5$ that control the central region in each halo. The remaining functions $\rho_0(z,m), \alpha(z,m)$ and $\beta(z,m)$ are fit with power laws:
\begin{equation}
    A = a \, \left(\frac{m_{200}}{10^{14} M_{\odot}}\right)^{b}(1+z)^{c}, \quad m_{200} = \frac{4\pi}{3} r^3_{200},
\end{equation}
where the best-fit parameters $\{a,b,c\}$ are in each case as follows: for $\rho_0$ $\{4000,0.29,-0.66\}$, for $\alpha$ $\{0.88, -0.03, 0.19\}$, and finally for $\beta$ $\{3.83, 0.04, -0.025 \}$. These values are taken from Table 2 of~\cite{Battaglia_2016}\footnote{Throughout the paper, unless $m_{200}$ is written explicitly (\eg when we define the gas profile in Eq.~\eqref{eq:rhogas} from~\cite{Battaglia_2016}), the symbol $m$ denotes the halo virial mass.}.

As discussed in detail in later sections, the assumptions about how gas inhabits dark matter halos has a significant effect on the signal and the resulting sensitivity. An extreme case to contrast with is to assume that baryons track the dark matter density everywhere. The parametric expression for the standard NFW profile is~\cite{1996ApJ462563N}:
\begin{equation}\label{eq:rhoNFW}
	\rho_\NFW = \frac{\rho_s}{\frac{r}{r_s} \left( 1 + \frac{r}{r_s} \right)^2}.
\end{equation}
Each halo has a physical scale radius and density that depend on its mass and redshift, \ie $r_s(z_i, m_i)$ and $\rho_s(z_i,m_i)$. The assumption that the baryons follow dark matter is expected to be reasonable in the outer regions of halos (\eg beyond the scale radius). However it is no longer valid in the inner regions where baryonic feedback processes are non-negligible. We present a discussion on the model uncertainty in our sensitivity due to varying assumptions about the electron profile in Appendix~\ref{appx:modeling}, using the NFW and AGN Feedback models as a proxy for the span of models.

The effective mass of a photon crossing a halo $i$ centered at redshift $z_i$ with mass $m_i$ depends on the number density of electrons $n_e$ along its path as defined in Eq.~\eqref{eq:mass2nden}. Assuming all baryonic mass inside halos is contained within protons, that are as numerous as electrons, we can write this in terms of the halo gas density profile:
\begin{equation}
\begin{aligned}
	m_{\gamma}^2 (\vec{x} | z_i, m_i ) 
			&=  \kappa  \, \left[\frac{\rho_\gas(\vec{x} | z_i, m_i )}{\rm M_\odot / Mpc^3}\right],
\end{aligned}
\end{equation}
where the term in brackets is dimensionless and we have defined $\kappa = 5.7\times10^{-38} {\rm \, eV^2}$.

\begin{figure}[ht!]
    \centering
    \includegraphics[width=0.6\textwidth]{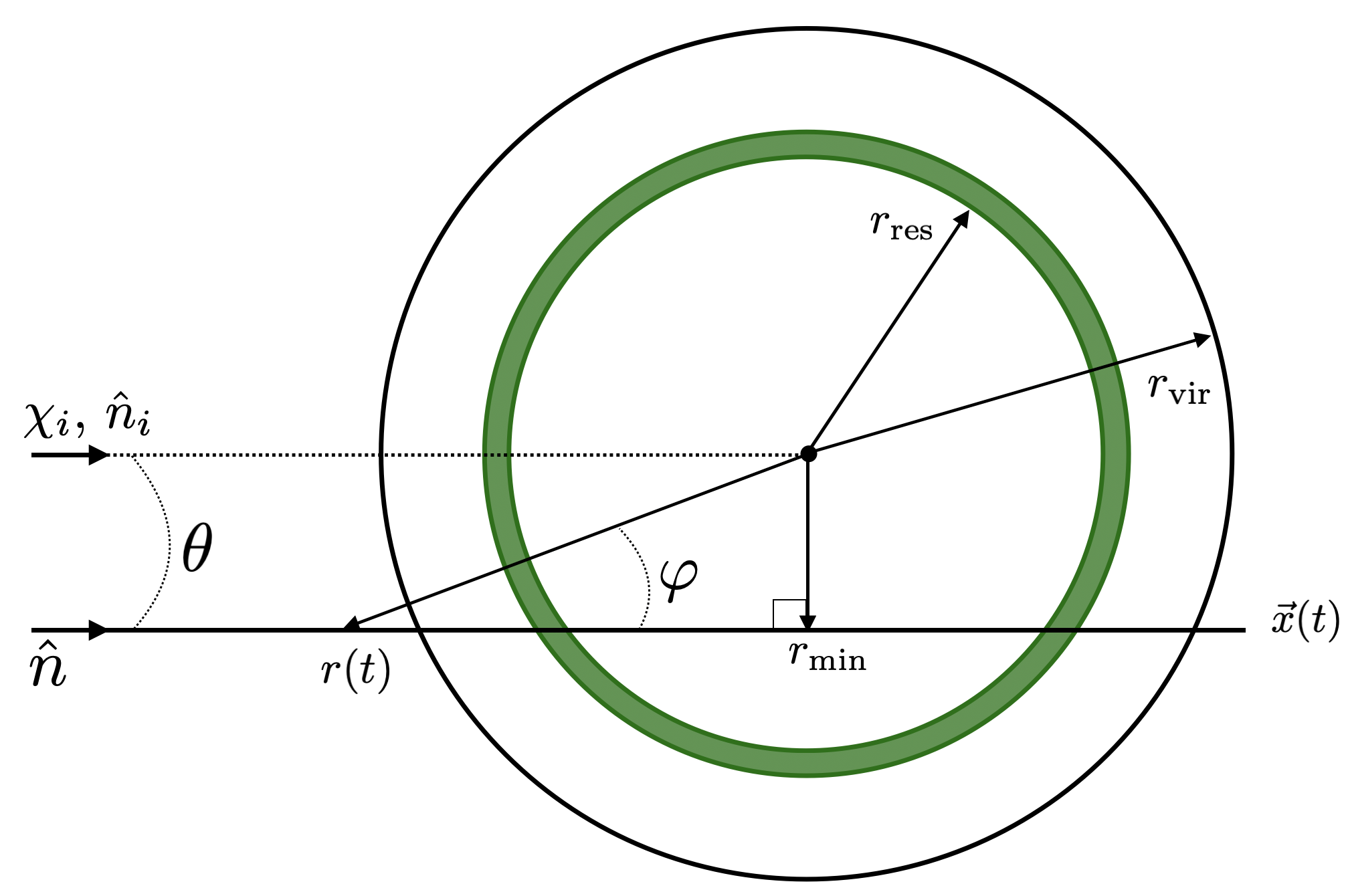}
    \caption{The trajectory of a photon $\vec{x}(t)$ through a dark matter halo centered at $\chi_i, \hat{n}_i$ on an observer's sky. In a coordinate system whose origin is at the halo center, the photon trajectory follows $r(\vec{x}(t))\equiv r(t)$. Resonant conversion to a dark photon occurs when the photon crosses through a spherical shell at $r_\res$ over a timescale $\Delta t_\res$; each trajectory has two crossings. We define the boundary of each halo by the virial radius $r_\vir$.}\label{fig:halodiagram}
\end{figure}

For resonant conversion that occurs at a radius $r_\res$, the resonance time scale can be broken into a radial and an angular part:
\begin{equation}\label{eq:chainrule}
	\Delta t_\res\equiv \varepsilon \left|\frac{\dd \ln m_{\gamma}^2}{\dd t}\right|^{-1}_{t_\res}  =  \frac{\varepsilon}{\kappa} m_\darkph^2 \left|\frac{\dd \rho (\vec{x})}{\dd t}\right|^{-1}_{t_\res} = \frac{\varepsilon}{\kappa} \frac{m_\darkph^2}{\left| \dd r(t_\res)/\dd t \right|} \left| \frac{\dd \rho}{\dd r} \right|^{-1}_{r_\res},
\end{equation}
where the velocity term $\left| \dd r(t)/\dd t \right|$ in Eq.~\eqref{eq:chainrule} depends on the precise photon trajectory through the halo and encapsulates the angular dependence of the probability. As shown in Fig.~\ref{fig:halodiagram}, the direction to the center of halo $i$ is $\hat{n}_i$ and the direction of the test photon is $\hat{n}$. We assume each halo is located at a single redshift $z_i$ throughout the photon's crossing time (\eg the halo size does not encompass cosmologically relevant distances), and take advantage of azimuthal symmetry around the $\hat{n}_i$ direction.

In the small angle approximation where $\hat{n} || \hat{n}_i$, the minimum comoving distance between the halo center and the photon's path is $(1+z_i) \, r_\rmmin =  \chi_i \, \theta$, where $\chi_i$ is the radial comoving distance to the halo center at redshift $z_i$ and $\theta $ is a small angle. The velocity $\left|{\dd r(t)/\dd t}\right| = \cos \varphi(t) $, where $\sin \varphi(t) = r_\rmmin/r(t)$.

Using this geometry, the conversion probability due to halo $i$ is separable into a radial part and an angular part. The latter one is a measure of how long the photon spends inside the resonance region $\sim \Delta t_\res$. We have:
\begin{equation}
	P^i_{\gamma \to \darkph} (\chi_i, m_i) = P(\chi_i, m_i) \, u(\hat{n} - \hat{n}_i  | \chi_i, m_i),
\end{equation}
where we define
\begin{equation}\label{eq:separable}
\begin{aligned}
	P(\chi_i, m_i) &= \frac{2 \pi \varepsilon^2 m_\darkph^4}{\kappa \, \omega (z_i)} \, \frac{\dd \rho_i}{\dd r}\Big|^{-1}_{r_\res} \, \Theta(r_\res - r_\vir), \\
	u(\hat{n} - \hat{n}_i  | \chi_i, m_i)  &= \left[ 1 - \frac{\left(\chi_i \theta / r_\res\right)^2}{(1+z_i)^2}  \right]^{-1/2}.
\end{aligned}
\end{equation}
The Heaviside step function $\Theta(r_\res - r_\vir)$ arises since we consider only photon to dark photon conversion happening inside the boundary of each halo, which we take here to be the virial radius $r_\vir$. The step function is normalized so that $\Theta(r_\res - r_\vir)=2$ for $r_\res < r_\vir$ and $\Theta(r_\res - r_\vir)=1$ for $r_\res = r_\vir$ to account for the fact that a photon crosses a resonance twice, going in and then out of a halo, except when it exactly grazes the edge of the virial radius. The effect of the sharp truncation at the boundary of the halo is significant for masses $m_{\gamma} \sim 10^{-13}$ eV that probe low densities. Note that the function $u(\hat{n} - \hat{n}_i  | \chi_i, m_i)$ has the apparent singularity when $\theta = (1+z_i) r_\res / \chi_i$, where the conversion probability blows up. However, the integral over the profile is finite and equal to
\begin{equation}
    \int d^2 \hat{n} \, u(\hat{n} - \hat{n}_i  | \chi_i, m_i) = 2 \pi (1+z_i)^2 r_\res^2 / \chi_i^2 \, . 
\end{equation}

We close this section by noting that within the halo model, we make the rather drastic assumptions that each halo (dark matter and gas) is spherically symmetric, identical at each mass and redshift, and has properties independent of their formation history and local environment. These assumptions will fail for individual halos. However, we expect that quantities dependent on the statistical properties of the full distribution of halos, such as power spectra, will be well-approximated (see \eg recent analyses such as~\cite{Kusiak_2022}). We are primarily concerned with such statistical quantities in the following.

\section{The photon to dark photon conversion monopole}\label{sec:monopole}

The total probability for a photon to convert over its trajectory is the sum of contributions from each individual halo. In this section we focus on the sky-averaged probability and explain why this effect induces a new type of optical depth. In the next section we introduce the two-point function of optical depth fluctuations owing to the halos' shapes on the sky.

The overall conversion probability is \begin{equation}\label{eq:totalprob}
\begin{aligned}
	\mathbf{P_{\gamma \to \darkph}}
	&= \ev{\sum_i  P^i_{\gamma \to \darkph} (\chi_i, m_i)}  
    = \int \dd m \int \dd^3\chi \, n(\chi, m) P(\chi, m) \, u(\hat{n} | \chi, m),
\end{aligned}
\end{equation}
where $\dd^3\chi = \chi^2\dd\chi \dd^2 \hat{n}$ and we have separated the resonant conversion probability $P_{\gamma \to \darkph}$ into its radial and angular components that depend on each halo's characteristics. To obtain the expression above we also identified the number density of halos of mass $m$ at redshift $\chi$ as 
\begin{equation}\label{eq:ndens}
	\ev{\sum_i \delta(m-m_i) \delta(\chi - \chi_i) \delta^2(\hat{n}-\hat{n}_i)}   \equiv n(\chi, m),
\end{equation}
where $\ev{\dots}$ denotes a sky-wide ensemble average and the delta functions are evaluated at each halo position and mass. The halo number density per volume per halo mass $n(\chi, m)$ is the halo mass function. We use the Tinker mass function throughout~\cite{tinker2008}.

Integrating the angular profile $u(\hat{n} | \chi, m)$ over the sky we implicitly weight the probability by the effective projected area of each halo. Evaluating the angular integral and simplifying we obtain:
\begin{equation}\label{eq:totalprob2}
    \mathbf{P_{\gamma \to \darkph}} = 4\pi \int_{0.01}^{z^\reio} \dd z \, \frac{\chi(z)^2}{H(z)} \int \dd m \, n(z, m) P(z, m) \, u_{00}(z, m),
\end{equation}
where $u_{00} = \int \dd^2 \hat{n} \, u(\hat{n}) / 4\pi$ is the monopole of $u(\theta)$, and we changed the integration variable to redshift $z$ from radial comoving distance $\chi$. Within our assumption of instantaneous reionization, we impose a sharp upper limit on the integral over redshift at $z^\reio$.

Photon to dark photon conversion manifests itself as a frequency dependent optical depth, encoding the removal of CMB photons along their path from recombination to CMB telescopes on Earth. The sky-averaged magnitude of this optical depth is the integrated probability in Eq.~\eqref{eq:totalprob2}:
\begin{equation}
    \bar{\tau} (\varepsilon, \omega) \equiv \mathbf{P_{\gamma \to \darkph}} \propto \varepsilon^{2} \omega^{-1},
\end{equation}
where from here onward the bar notation stands for the projected, \ie integrated over redshift, sky average. We introduce a new notation for the dimensionful average optical depth
\begin{equation}\label{eq:eta_definition}
    \bar{\eta} = \bar{\tau} \varepsilon^{-2} \omega \, [{\rm eV}],
\end{equation}
such that $\bar{\eta}$ depends only on the dark photon mass $m_\darkph^2$ and the cosmology. It is useful to make this distinction because both $\varepsilon$ and $\omega$ are parameters we vary later.

\begin{figure}[ht!]
    \centering
    \includegraphics[width=0.65\textwidth]{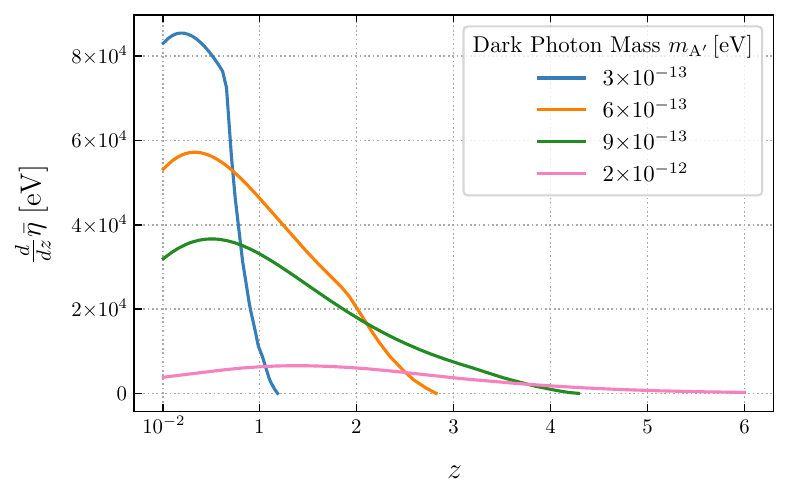}
    \caption{The average dimensionful optical depth defined in Eq.~\eqref{eq:eta_definition} as a function of redshift, for a range of dark photon masses. To obtain the full optical depth, one needs to integrate over redshift and multiply by the unknown mixing parameter divided by the frequency of the photon,  $\varepsilon^2/\omega$. Notice that the lighter masses produce the strongest signal at low redshift. Meanwhile, heavier dark photons probe redshifts all the way to reionization but require crossing large densities in order for the resonant conversion to take place. These are found within more massive halos whose number density is suppressed.}\label{fig:dtaudz}
\end{figure}

In Fig.~\ref{fig:dtaudz}, we show several examples of the differential optical depth along the line of sight for a range of dark photon masses. Notice in this plot that the light dark photons are only produced at low redshift. Light dark photons probe the outer-most regions of halos and the abrupt fall-off is due to the truncation of halos at the virial radius. Physically, this falloff would be broadened by the softer boundaries between halos and the intergalactic medium. In contrast, the heaviest dark photons probe regions near the core of halos where the gas density is highest. Since the gas profile is nearly flat near the core of halos, only the heaviest and rarest halos contribute to the optical depth of photons with $m_\darkph \geq 10^{-12}$ eV.

\begin{figure}[ht!]
    \centering
    \includegraphics[width=0.65\textwidth]{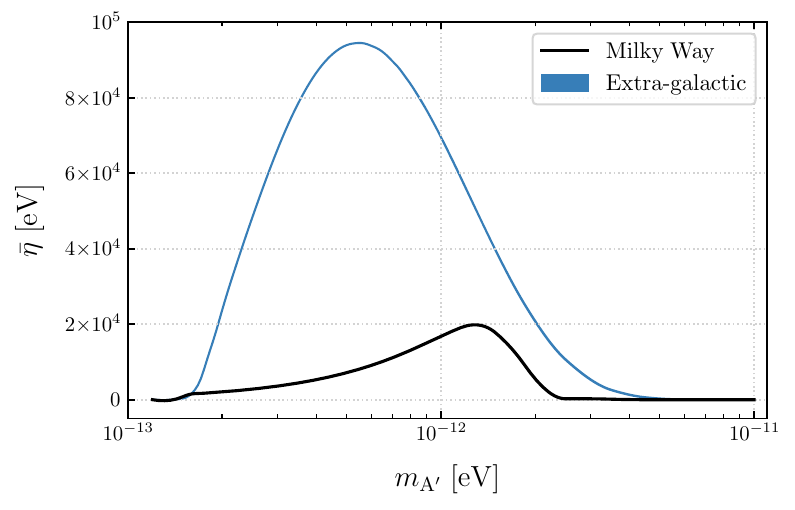}
    \caption{The dimensionful optical depth monopole as a function of dark photon mass $m_\darkph$. The extra-galactic contribution $\bar{\eta} \, (blue)$ is defined by Eq.~\eqref{eq:eta_definition}. Under our assumptions, resonant conversion occurs once reionization is completed and here we plot the case where $z^\reio = 6$ and where $z^\reio = 10$. The maximum change in the amplitude of $\bar{\eta}$ is of order $10^2 $ eV, less than the thickness of the blue line, and we conclude that the model is insensitive to the details of reionization over the range of masses we are probing. We therefore assume $z^\reio=6$ in the remainder of this work. At low mass, $\bar{\eta}\to 0$ due to the constraint that conversion only happens within the virial radius in each halo. The upper bound is set by the shape of the gas profile and details about the halo model. The contribution due to gas in the Milky Way $\eta^\MW \, (black)$ is defined in Eq.~\eqref{eq:tau_MW}.}\label{fig:monopole}
\end{figure}

In Fig.~\ref{fig:monopole} we plot the total (\eg integrated over the line of sight) dimensionful optical depth assuming both $z^\reio=6$ and $z^\reio=10$. The change in magnitude is less than the thickness of the blue curve, meaning that our model is insensitive to uncertainties related to when the end of reionization takes place. From now on, unless otherwise stated, we assume that reionization takes place instantaneously at $z^\reio=6$.  Notice that, unless $\varepsilon \ll 1$ photon to dark photon conversion is in the optically thick regime. For example, $\bar{\tau} = \bar{\eta} \, \varepsilon^{2}/\omega = 1$ for $\sim 100 $ GHz photons (near the peak of the CMB blackbody) at $\varepsilon \sim 10^{-4}$. This is a preliminary indicator that photon to dark photon conversion at low redshift can be a sensitive probe. The dark photon mass range over which there is a significant effect spans roughly one order of magnitude, peaking at $m_\darkph \sim 6 \times 10^{-13}$ eV.

Baryonic feedback can affect the accessible dark photon mass range and strength of the photon to dark photon conversion monopole, in principle contributing a source of modeling uncertainty in the expected signal. We explore this modeling uncertainty in more detail in Appendix~\ref{appx:modeling} by comparing the AGN gas model used here~\cite{Battaglia_2016} with the results obtained by assuming that baryons trace the dark matter NFW density profile~\cite{1996ApJ462563N}. We show that the two models give nearly identical results for low masses $m_\darkph$ where the resonant conversion condition is met near the outermost region of a halo. This can be understood as a consequence of the fact that the AGN gas profile and the NFW profile are only different near the core of halos where the NFW density increases without bound, while the gas density achieves a maximum; beyond the scale radius, the two density profiles are nearly the same. Also for this reason, the predictions for the two models differ most at large dark photon mass. Here, details about the gas population in each halo can have a significant effect on the projected signal and resulting sensitivity, and it is important to incorporate this modeling uncertainty in the interpretation of our results below. As we will demonstrate in more detail below, the monopole signal is more strongly influenced by this modeling uncertainty than correlation functions.

\subsection{Contribution from the Milky Way}\label{sec:MW}

The gas halo surrounding the Milky Way is also a source of resonant photon to dark photon conversion. In this section we model its contribution to the overall optical depth. The Earth's distance from the galactic center is well below the scale radius of the Milky Way halo, therefore in our model we assume that there is only an appreciable contribution to the optical depth monopole and not to optical depth anisotropies. We further assume that the Milky Way is an average spherical halo, with average AGN feedback and model it with the same gas profile of Eq.~\eqref{eq:rhogas} computed for $z^\MW=0$ and $m^\MW$ from~\cite{posti_mass_2019}. In reality, the details of the gas distribution can affect both the monopole and anisotropies in the optical depth. Nevertheless, we can estimate the relative magnitude of the galactic versus extra-galactic conversion under these assumptions.

The expression for the optical depth is simply the contribution from a single halo with appropriate properties:
\begin{equation}
\begin{aligned}\label{eq:tau_MW}
    \tau^\MW \equiv P_{\gamma \to \darkph}^\MW &= \frac{2 \pi \varepsilon^2 m_\darkph^4}{\kappa \, \omega} \, \frac{\dd \rho_\gas}{\dd r}\Big|^{-1}_{r_\res} \, \Theta(r_\res - r^\MW_\vir),
\end{aligned}
\end{equation}
where we discard the angular component by assuming the Earth is near the center of the halo, and $\omega$ is the frequency of the CMB photons today. Assuming a virial radius and virial mass for the Milky Way as found in~\cite{posti_mass_2019}, as well as the concentration-mass relation at redshift zero from~\cite{2014MNRAS.441.3359D}, we calculate the conversion probability for the relevant range of dark photon masses. The result is plotted in Fig.~\ref{fig:monopole}. Given the assumptions made in this paper, the extra-galactic optical depth dominates the contribution due to conversion in the Milky Way for all dark photon masses considered. 

Since the Milky Way does not host an AGN, the gas profile assumed here may be too diffuse. In the extreme scenario where gas traces dark matter, the magnitude of the Milky Way contribution increases monotonically for  higher dark photon masses. The increase is cutoff at the core when gas no longer traces dark matter. The high dark photon mass regime is where the extra-galactic contribution to the optical depth becomes small, and the Milky Way therefore introduces additional modeling uncertainties at the upper end of dark photon masses we consider. Note that such effects will only increase the reach in sensitivity to conversion, making the neglect of contributions from the Milky Way a conservative assumption.

\section{Patchy dark screening}\label{sec:darkscreening}

When photon to dark photon conversion occurs in non-linear structure, the associated optical depth for conversion is strongly anisotropic on the sky. These anisotropies in the optical depth serve as a screen of varying opacity through which the CMB must propagate on the way from decoupling to our telescopes here on Earth. In addition to the sky-averaged suppression in the intensity of CMB photons, new temperature and polarization anisotropies are introduced due to the different conversion probability of CMB photons to dark photons across different lines of sight.

The analogous effect in the standard cosmological model is the `screening' of CMB anisotropies due to the Thomson scattering of CMB photons by free electrons. For Thomson scattering, anisotropies in the optical depth couple to {\em anisotropies} in the CMB temperature and polarization. This is known as `patchy screening' of the CMB. Notably, there is no coupling of optical depth anisotropies to the CMB temperature monopole -- Thomson scattering doesn't change the energy of photons and for every photon scattered out of the line of sight, another is scattered into the line of sight. Therefore, patchy screening of the CMB is always a small effect, \ie second order in perturbations of the CMB and optical depth anisotropies. Although small, the detection of {\em patchy screening} during~\cite{dvorkin_reconstructing_2009} and after~\cite{Roy:2022muv} reionization is within the reach of future CMB experiments. 

Photon to dark photon conversion produces {\em patchy dark screening}, which has  two crucial differences to Thomson screening: conversion does not preserve the blackbody spectrum and conversion only removes photons from the line of sight. {\em Therefore, patchy dark screening couples the CMB monopole to fluctuations in the optical depth, and is a 1st order effect} (in anisotropies). Since the CMB monopole is $\sim 10^4$ times larger than the temperature anisotropies, patchy dark screening is far stronger at fixed optical depth than Thomson screening. Furthermore, the characteristic frequency-dependence of patchy dark screening can be used to separate it from the primary CMB and astrophysical foregrounds. In the remainder of this section and the next section, we derive various correlation functions that will be used to forecast the sensitivity of CMB experiments to the kinetic mixing parameter and dark photon mass, assuming the frequency dependent dark screening anisotropies can be {\it separated} from  the primary CMB anisotropies. A detailed discussion of how well this separation can be performed is presented in Section~\ref{sec:noise}.

\subsection{Anisotropic screening}

Before computing CMB correlators, we must first describe anisotropies in the photon to dark photon conversion optical depth, the anisotropies in the Thomson optical depth, and the cross-correlation between these two fields. The halo model for large-scale structure is a useful tool for these computations, since it is straightforward to populate dark matter halos with the electron density (for Thomson screening) and the conversion probability (for dark screening).

The standard optical depth to reionization is the integrated electron density along the line of sight:
\begin{equation}
    \tau^\Th  = \sigma_T \int \dd\chi \, a(\chi) n_e(\chi, \hat{n}),
\end{equation}
where $ \sigma_T $ is the Thomson cross-section and $a$ is the local scale factor. The inhomogeneous matter distribution introduces spatial fluctuations in the electron number density $n_{e}(\chi, \hat{n})$. This can be measured via the directional dependence they introduce on the optical depth field. The local perturbations in the electron density induce small fluctuations $\delta \tau^\Th (\hat{n})$ in the optical depth profile 
\begin{equation}
    \tau^\Th (\hat{n}) = \bar{\tau}^\Th  + \delta \tau^\Th (\hat{n}),
\end{equation}
where $\bar{\tau}^\Th $ is the standard cosmological parameter. Once again we work under the assumptions of the halo model where the integral above can be written as an average over halos whose number density depends on mass and redshift, where the electron number density fluctuations are related to the gas density profile. In the Limber approximation, the multipole expansion of the electron density from one halo centred on the north celestial pole (such that the azimuthal angular momentum $m=0$) is~\cite{HillPajer}:
\begin{equation}
    \rho^e_{\ell}(z, m) = \frac{a(z)}{\chi(z)^2} \int 4\pi r^2 \dd r \frac{\sin\left(\left(\ell + 1/2\right) r/\chi(z)\right)}{\left(\ell + 1/2\right) r/\chi(z)} \rho^\gas(r | z, m),
\end{equation}
where the $\rho^\gas$ is the Battaglia \etal AGN gas density profile~\cite{Battaglia_2016} and $\rho_e = n_e \, m_p$. The Limber approximation works best for $\ell \gg 1$, which is the regime we are in.

We now introduce the optical depth for photon to dark photon conversion. Consider small anisotropic perturbations to the average extra-galactic optical depth $\bar{\tau}$:
\begin{equation}
	\tau(\hat{n},\varepsilon,\omega) = \bar{\tau}(\varepsilon, \omega) \left( 1 + \delta\tau(\hat{n}) \right).
\end{equation}
Note that with this convention $\delta\tau(\hat{n})$ is independent of $\varepsilon$ and $\omega$. 
The next step is to compute the two-point angular power spectra for these fields. Projecting onto spherical harmonics, we define the power spectrum $C_{\ell}^{\delta\tau\delta\tau}$:
\begin{equation}
    \ev{\delta\tau_{\ell m} \, \delta\tau^{*}_{\ell^{\prime} m^{\prime}}} = \delta_{\ell \, \ell^{\prime}} \delta_{m \, m^{\prime}} C_{\ell}^{\delta\tau\delta\tau},
\end{equation}
In the context of the halo model, the power spectrum can be described by a sum of an intra-halo ($1$-halo) and an inter-halo ($2$-halo) contributions:
\begin{equation}\label{eq:Celltautau}
	C_\ell^{\delta\tau\delta\tau} = C_{\ell}^{\tau\tau} / \bar{\tau}^2 = C_\ell^{1\halo} + C_\ell^{2\halo} , \quad \ell\geq 1.
\end{equation}
The quantity $C_{\ell}^{\tau \tau}$ is the angular power spectrum of the optical depth due to photon to dark photon conversion (which depends on $\varepsilon$ and $\omega$). The terms on the right hand side of Eq.~\eqref{eq:Celltautau} are computed in detail in Appendix~\ref{appx:2pf}. The final expressions are: 
\begin{equation}
\begin{gathered}\label{eq:halomodelmasterspectra}
	\bar{\tau}^2 C_\ell^{1\halo} = \frac{4\pi}{2l+1} \int \dd z \, \frac{\chi(z)^2}{H(z)} \int \dd m \, n(z,m) \left[ P(z, m) \, u_{\ell 0}(z, m) \right]^2, \\
	\bar{\tau}^2 C_\ell^{2\halo} = \frac{4\pi}{2\ell+1} \left[ \prod_{i=1,2}\int \dd z_i \, \frac{\chi(z_i)^2}{H(z_i)} \int \dd m_i \, n(z_i, m_i) b(z_i, m_i) P(z_i, m_i) u_{\ell 0}(z_i, m_i) \right] C_\ell^\lin(z_1,z_2), \\
	C_\ell^\lin(z_1,z_2) = \frac{2}{\pi} \int \dd k \, k^2 j_\ell(k \chi_1)\, j_\ell(k \chi_2)\, \sqrt{P^\lin(k, \chi_1)P^\lin(k, \chi_2)}.
\end{gathered}
\end{equation}
The quantity $j_\ell(k \chi)$ is the spherical Bessel function and $b(z, m)$ is the linear halo bias. $P^\lin(k, \chi)$ is the linear matter power spectrum calculated using CAMB~\cite{Lewis_2000}. All quantities are computed for fixed cosmology. Throughout this work we used the following set of parameters: dark matter density $\Omega_c h^2 = 0.12$, baryon density $\Omega_b h^2 = 0.022$, Hubble constant $H_0 = 67.3 {\rm \, km \, s^{-1} \, Mpc^{-1}}$, scalar spectral index $n_s = 0.96$, scalar amplitude $A_s = 2.2\times10^{-9}$, and optical depth to reionization $\bar{\tau}^\Th =0.06$.

\begin{figure}[ht!]
    \includegraphics[width=1\textwidth]{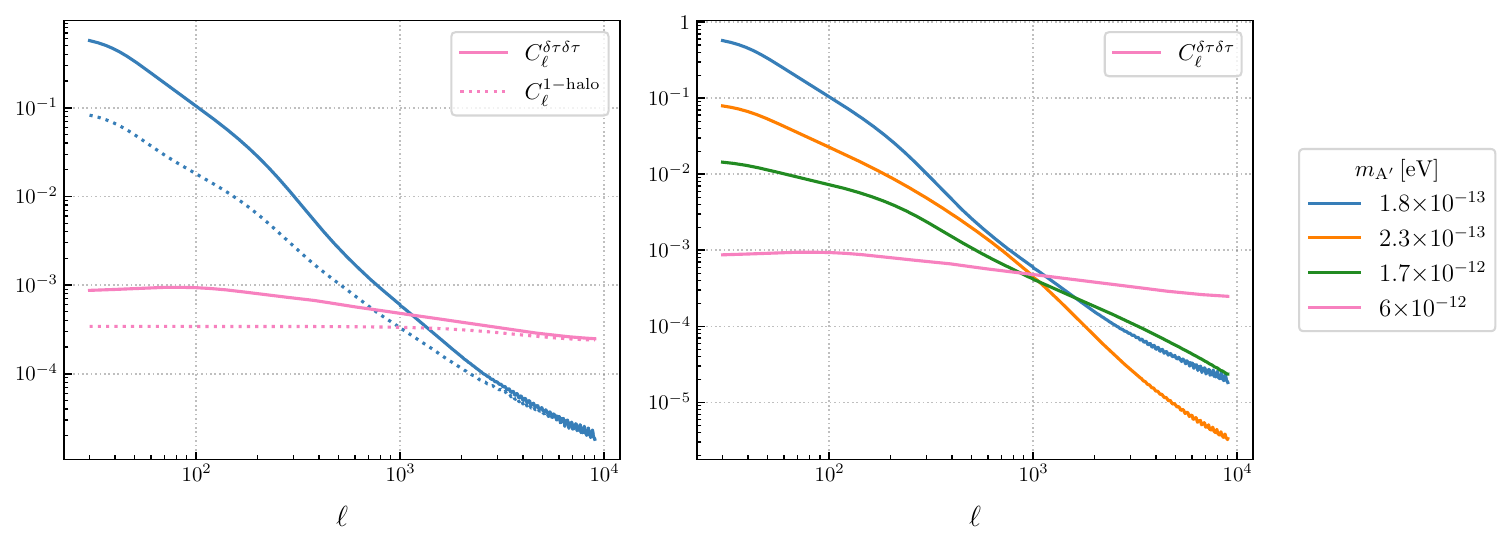}
    \caption{The angular power spectrum of dark screening fluctuations, $C_{\ell}^{\delta\tau\delta\tau}$, which is independent of $\varepsilon$ and $\omega$, for a range of dark photon masses that span the parameter space we are probing. In the left panel, we compare the $1$-halo term with the sum in Eq.~\eqref{eq:Celltautau} to show that most power on large scales is due to the $2$-halo term, while on small scales the $1$-halo term dominates. At large mass, when the transition from photon to dark photon occurs nearer to the halo core where gas densities are largest, the two terms become comparable and the spectrum is scale invariant. This feature also affects the signal amplitude hierarchy: although at large $m_\darkph > 10^{-12}$ eV the monopole is subdominant (\eg Fig.~\ref{fig:monopole}), the additional structure on small scales causes $C_{\ell}^{\tau\tau}$ in this regime to be equivalent in terms of constraining power for $\varepsilon$ to spectra for $m_\darkph < 10^{-12}$ eV. This will be relevant in Section~\ref{sec:forecasts} where we present our forecasts.}\label{fig:Cells}
\end{figure}

In Fig.~\ref{fig:Cells} we plot the optical depth power spectrum of $\delta\tau$ fluctuations for a range of dark photon masses, assuming $z^\reio=6$. In the left panel, we show the relative importance of the $1$-halo term compared to the total. The $1$-halo term dominates at high $\ell$, while the $2$-halo term dictates the shape and amplitude on large scales $\ell \lesssim 1000$. The $1$-halo term is larger at the upper end of the dark photon mass range we consider. This is consistent with the fact that conversion happens near the halo core in this regime. In the right panel we show the total power spectrum $C_{\ell}^{\delta\tau\delta\tau}$ for $4$ choices of $m_\darkph$ that span the parameter space we probe for. Any changes can be attributed to the radius of the resonance surface at that mass given the gas density profile $\rho^\gas$ as well as details about the halo model, for \eg the population of halos $n(z,m)$, which also dictates the bias function $b(z,m)$. The total magnitude of the power spectrum, $C_{\ell}^{\tau\tau}$, depends strongly on the dark photon mass through $\bar{\eta}$. Hence, it will be maximized by masses near the peak in the monopole $\bar{\tau}(m_\darkph)$, which was depicted in Fig.~\ref{fig:monopole}.

Repeating the computation presented in Appendix~\ref{appx:2pf} for the dark screening case, we find the $1$-halo and $2$-halo contributions to the Thomson screening auto-power spectrum. The full expression is
\begin{equation}
\begin{aligned}
    C_{\ell}^{\tau^\Th  \tau^\Th} &= \frac{\sigma_T^2}{m_p^2} \left( \int \frac{\dd z}{H(z)} \frac{\chi(z)^2}{(1+z)^2} \int \dd m \, n(z, m) \, \rho^e_{\ell 0}(z, m)^2 \textcolor{white}{\prod_{i=1,2}} \right. \\
    &+ \left. \left[ \prod_{i=1,2}\int \frac{\dd z_i}{H(z_i)}\frac{\chi(z_i)^2}{(1+z_i)} \int \dd m_i \, n(z_i, m_i) b(z_i, m_i) \rho^e_{\ell 0}(z_i, m_i) \right] C_\ell^\lin(z_1,z_2)\right).
\end{aligned}
\end{equation}
The quantities have the same meaning as in expressions~\eqref{eq:halomodelmasterspectra} above.
 
\begin{figure}[ht!]
    \centering
   \includegraphics[width=0.6\textwidth]{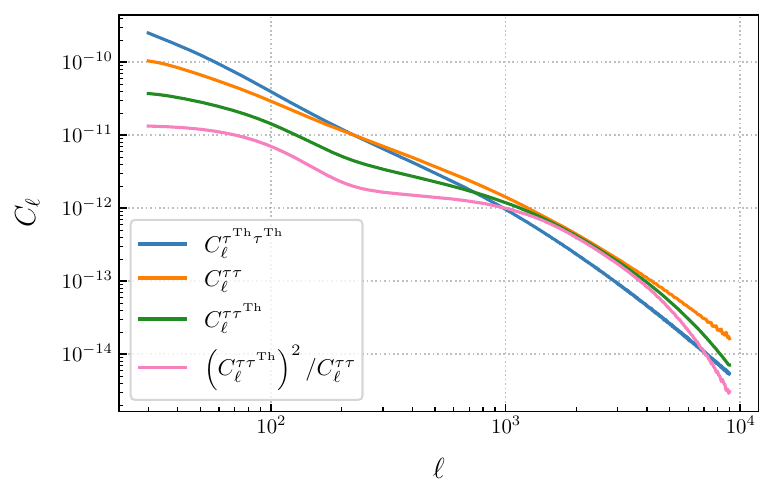}
   \caption{We show the angular power spectrum of the two-point function of Thomson screening (\textit{blue}), dark screening (\textit{orange}), as well as their cross-correlation (\textit{green}). We chose the value of $\varepsilon = 10^{-6}$ and fixed $m_\darkph = 6\times 10^{-13}$ eV and $\omega=30 $ GHz. We also present the $\varepsilon$-independent combination $\left(C_{\ell}^{\tau \tau^\Th}\right)^2/C_{\ell}^{\tau \tau}$ (\textit{pink}), which will be useful for the sensitivity forecast in Section~\ref{sec:forecasts}.}\label{fig:screening_corrs}
\end{figure}

Anisotropies in the Thomson screening and dark screening optical depth fields trace similar matter density profiles over the sky and across the line of sight. It is therefore expected that the fluctuations in either field are correlated. The two-point function $\ev{\tau \tau^\Th} \sim \varepsilon^2$ has the following angular power spectrum:
\begin{gather}\label{eq:crossspectra}
    \begin{aligned}
        \bar{\tau} C_{\ell}^{\delta\tau \tau^\Th} &= \frac{\sigma_T}{m_p} \sqrt{\frac{4\pi}{2\ell+1}} \left( \int \frac{\dd z}{H(z)} \frac{\chi(z)^2}{(1+z)} \int \dd m \, n(z, m) \, \mathcal{F}(z,z,m,m) \textcolor{white}{\prod_{i=1,2}} \right. \\
        &+ \left. \left[ \prod_{i=1,2} \int \frac{\dd z_i}{H(z_i)}\frac{\chi(z_i)^2}{(1+z_i)} \int \dd m_i \, n(z_i, m_i) b(z_i, m_i)  \right] \mathcal{F}(z_1,z_2,m_1,m_2) \, C_\ell^\lin(z_1,z_2) \right),
    \end{aligned} \\
    \mathcal{F}(z_1,z_2,m_1,m_2) = P(z_1, m_1) \, u_{\ell 0}(z_1, m_1) \, \rho^e_{\ell 0}(z_2, m_2),
\end{gather}
where all quantities are real. The Thomson auto-power spectrum is shown in Fig.~\ref{fig:screening_corrs}, alongside examples for the power spectrum of the dark screening optical depth and their cross-correlation for fixed $m_\darkph$ and $\varepsilon$ ( $C_{\ell}^{\tau^\Th  \tau^\Th}$ is independent of $\varepsilon$). The power spectra involving dark screening scale differently with $\varepsilon$, so that for $\varepsilon$ around $7\times 10^{-7}$, $C_{\ell}^{\tau \tau^\Th} > C_{\ell}^{\tau \tau}$. The magnitude of $C_{\ell}^{\tau \tau}$ varies over the range of dark photon masses we study such that it peaks around $m_\darkph \simeq 10^{-12}$ eV and falls abruptly for mass values towards both ends of the interval.

\subsection{Patchy dark screening of the CMB} \label{sec:CMBscreeningAngPS}

We now describe in detail how the optical depth anisotropies of the previous section manifest as anisotropic spectral distortions in the CMB temperature and polarization. The largest effect arises from patchy dark screening of the CMB temperature monopole. In addition to this, there are new spectral anisotropies from the screening of temperature and polarization anisotropies. An additional novel signature of patchy dark screening is the production of B-modes (curl) from pure E-mode (divergence) polarization anisotropies, as occurs in any scenario with screening~\cite{dvorkin_b_2009}. This leads to statistically anisotropic correlations between the temperature, E modes, and B modes dependent on the specific realization of the anisotropic optical depth and CMB anisotropies in our Universe. 

In analogy with patchy screening from Thomson scattering, the photon to dark photon optical depth fluctuations suppress the CMB temperature fluctuations and polarization Stokes parameters. The combined effect of dark and Thomson screening on the observed temperature and polarization are 
\begin{equation}\label{eq:generalobs}
\begin{aligned}
	T^\obs(\hat{n},\omega) &= \bar{T} + T^\Sc (\hat{n}) + T^\dSc (\hat{n},\omega), \\
	(Q \pm i U)^\obs(\hat{n},\omega) &= (Q \pm i U)^\Sc (\hat{n}) + (Q \pm i U)^\dSc (\hat{n},\omega),
\end{aligned}
\end{equation}
where 
\begin{equation}\label{eq:generalscr}
\begin{aligned}
	T^\Sc (\hat{n}) &\simeq \left[1 - \tau^\Th (\hat{n})\right] T(\hat{n}), \\   
    T^\dSc (\hat{n},\omega) &\simeq- \tau(\hat{n},\omega) \left[\bar{T} + T(\hat{n})\right],  \\
 (Q \pm i U)^\Sc (\hat{n}) &\simeq \left[1 - \tau^\Th (\hat{n})\right] (Q \pm i U)(\hat{n}), \\ 
 (Q \pm i U)^\dSc (\hat{n},\omega) &\simeq  - \tau(\hat{n},\omega) (Q \pm i U)(\hat{n}). 
\end{aligned}
\end{equation}
The label \textit{obs} stands for observed anisotropies, \textit{Sc} for screening from Thomson scattering, and \textit{dSc} for dark screening. We have explicitly isolated the dependence on the CMB blackbody temperature monopole $\bar{T} = 2.725 K$, where in our notation the sky average of $T(\hat{n})$ is zero while the sky average of $T^{(\obs)}$ is not. As our first approximation, we assume that $T(\hat{n})$ and $(Q \pm i U)(\hat{n})$ are the lensed CMB temperature and polarization anisotropies. This neglects the lensing of the screened CMB along the line of sight, which is a small higher order effect. We neglect extra-galactic foregrounds, such as the cosmic infrared background, point sources, and Sunyaev Zel'dovich effects. We assume that no significant kinetic mixing happened between recombination and the end of reionization. Finally, we work in the limit where $\tau(\hat{n},\omega), \tau^\Th (\hat{n}) \ll 1$.

The total spectral distortion to the blackbody spectrum due to photons' conversion into dark photons is given by
\begin{equation}\label{eq:spectral_distortion}
    B(\hat{n}, \omega, T) = B^0(\omega, T) \left( 1 - P_{\gamma \to \darkph}(\hat{n},\omega) \right),
\end{equation}
where $P_{\gamma \to \darkph}$ is the overall conversion probability and $B^0(\omega, T)$ represents the intensity of the theoretical Planckian spectrum. In natural units this is:
\begin{equation}\label{eq:a_blackbody}
    B^0(\omega, T) = \frac{\omega^3}{2\pi^2} \left(e^{\frac{\omega}{T}} -1 \right)^{-1}.
\end{equation}

The dependence on frequency $\omega$ in Eq.~\eqref{eq:generalscr} is relative to the blackbody spectrum of the CMB in units of temperature, not intensity (see, \eg also~\cite{PhysRevD.91.085015}). To convert, we take the leading order expansion in temperature fluctuations $\delta T(\hat n)$ of Eq.~\eqref{eq:spectral_distortion}:
\begin{equation}
    B( \omega, \hat{n}) = B^0(\omega, \bar{T}) + \delta B( \omega, \hat{n})  \simeq B^0(\omega, \bar T) + \frac{\partial B^0(\omega, T)}{\partial T}\bigg{|}_{T = \bar T} \delta T(\hat{n})  -  B^0(\omega, \bar T ) P_{\gamma \to \darkph} (\hat{n},\omega).
\end{equation}
Therefore, in CMB temperature units, the temperature fluctuation function of frequency is
\begin{equation}
    \frac{1}{\frac{\partial B(\omega, T)}{\partial T}\bigg{|}_{T = \bar T}}\delta B = \delta T - \frac{1}{\frac{\partial B(\omega, T)}{\partial T}\bigg{|}_{T = \bar T}}B(\omega, \bar T ) P_{\gamma \to \darkph} 
    =  \delta T - \frac{1-e^{-x}}{x} \bar T P_{\gamma \to \darkph},
\end{equation}
where $x=\omega/\bar{T}$ in natural units. A factor of
\begin{equation}\label{eq:zeta_freq}
    \zeta(\omega) = \frac{1-e^{-x}}{x},
\end{equation}
appears when converting the frequency dependence in the absorption optical depth to the frequency dependence in the standard temperature units, and the full frequency dependence of the dark screening signal in units of temperature is $\zeta(\omega) / \omega$.

Next, we decompose the lensed CMB temperature and optical depths into spherical harmonics and the Stokes parameters into spin-2 spherical harmonics:
\begin{equation}\label{eq:harmonicdefinition}
\begin{aligned}
	T(\hat{n}) &= \sum_{\ell m} T_{\ell m} 	Y_{\ell m}(\hat{n}), \\ 
    \tau(\hat{n}) &= \sum_{\ell m} \tau_{\ell m} 	Y_{\ell m}(\hat{n}), \\ 
	(Q \pm i U)(\hat{n}) &= \sum_{\ell m} \left[E_{\ell m} \pm i B_{\ell m} \right] {}_{\pm 2} Y_{\ell m}(\hat{n}),
\end{aligned}
\end{equation}
where $E_{\ell m}$ and $B_{\ell m}$ are the moments of the E- and B-mode polarization anisotropies. Note that the latter is induced only by lensing in the absence of primordial gravitational waves and is therefore substantially smaller than E-mode polarization. We neglect the dark screening of lensing B-modes below.

Given multi-frequency observations of the CMB, it is possible to separate the blackbody and dark screening components of the temperature and polarization anisotropies. Assuming perfect separation of the blackbody and patchy dark screening components (we discuss the scenario where this separation is imperfect in Section~\ref{sec:noise}), we construct correlation functions between Thomson-screened and dark-screened temperature and polarization anisotropies, that is, $T^\Sc $ and $T^\dSc $, respectively. Note that there will be both statistically isotropic and statistically anisotropic components of the correlation functions.

\subsection{Dark screening auto-correlation functions}\label{sec:twoptcorrs}

As a warm up, in the Standard Model ($\varepsilon = 0 $ and hence $T^\dSc =0$), the anisotropies in optical depth $\tau^\Th $ lead to temperature anisotropies, which are captured in the auto-correlation function for the Thomson-screened blackbody anisotropies:
\begin{eqnarray}\label{eq:TScTSc}
    \ev{T^\Sc_{\ell m} T^\Sc_{\ell^{\prime} m^{\prime}}} &=& (-1)^m C_{\ell}^{T^\Sc T^\Sc} \delta_{\ell \ell^{\prime}} \delta_{m-m^{\prime}} \nonumber \\ 
    &-& \sum_{\ell_1 m_1} (-1)^{m_1}                                  \begin{pmatrix}
									 \ell & \ell^{\prime} & \ell_1 \\
								          m & m^{\prime} &  -m_1
								\end{pmatrix}
        \sqrt{2\ell_1+1} W_{\ell \ell^{\prime} \ell_1}^{0 0 0} \left[ C_{\ell}^{TT} + C_{\ell^{\prime}}^{TT} \right]  \tau^\Th_{\ell_1 m_1},
\end{eqnarray}
where we have defined
\begin{equation}
W_{\ell \ell^{\prime} \ell_1}^{m m^{\prime} m_1} = \sqrt{\frac{(2 \ell + 1) (2 \ell^{\prime} +1)}{4 \pi}}  
								\begin{pmatrix}
									\ell & \ell^{\prime} & \ell_1 \\
								         m & m^{\prime} & -m_1 
								\end{pmatrix}.
\end{equation}
The statistically isotropic component of the correlator is~\cite{dvorkin_b_2009}
\begin{equation}\label{eq:isotsctsc}
    C_{\ell}^{T^\Sc T^\Sc} \equiv C_\ell^{TT} + \sum_{\ell^{\prime} \ell^{\prime\prime}} C_{\ell^{\prime\prime}}^{\tau^\Th  \tau^\Th} C_{\ell^{\prime}}^{TT} (W_{\ell \ell^{\prime} \ell^{\prime\prime}}^{000} )^2 + N_{\ell}^{T^\Sc T^\Sc}.
\end{equation}
$C_\ell^{TT}$ is the lensed primary CMB temperature power spectrum, $C_{\ell}^{\tau^\Th  \tau^\Th}$ is the Thomson optical depth power spectrum, and $N_{\ell}^{T^\Sc T^\Sc}$ encompasses all other contributions to the blackbody CMB such as instrumental noise, foregrounds, etc. The statistically anisotropic component of the correlator is induced by the particular realization of patchy Thomson screening in our Universe. We have explicitly kept this term, which would vanish for a full ensemble average over all fields. As we describe in detail below, this statistical anisotropy can be used to reconstruct the anisotropies of optical depth $\tau^\Th $, and induces various three-point correlation functions.

Similarly, anisotropies in the dark screening optical depth $\tau(\hat{n},\omega)$ lead to anisotropies in $T^\dSc (\omega)$, which is captured by the auto-correlation function of the dark screening component of the observed CMB temperature:
\begin{equation}
\begin{aligned}\label{eq:TdScTdSc}
    \ev{T^\dSc_{\ell m} (\omega) T^\dSc_{\ell^{\prime} m^{\prime}} (\omega^{\prime})} &= (-1)^m C_{\ell}^{T^\dSc T^\dSc} (\omega,\omega^{\prime}) \delta_{\ell \ell^{\prime}} \delta_{m -m^{\prime}} \\ 
    &+ \sum_{\ell_1 m_1} (-1)^{m_1}                                  \begin{pmatrix}
									 \ell & \ell^{\prime} & \ell_1 \\
								        m & m^{\prime} &  -m_1 
								\end{pmatrix}
        \sqrt{2\ell_1+1} W_{\ell \ell^{\prime} \ell_1}^{0 0 0} \left[ C_{\ell}^{\tau\tau} (\omega,\omega^{\prime}) + C_{\ell^{\prime}}^{\tau\tau} (\omega,\omega^{\prime}) \right]  \bar{T} T_{\ell_1 m_1},
\end{aligned}
\end{equation}
where $C_{\ell}^{\tau\tau} (\omega,\omega^{\prime})$ is the dark screening optical depth, for which we have retained the frequency dependence with respect to the blackbody, namely a factor of $\left( \zeta(\omega) / \omega \right)^2$. The statistically isotropic contribution to the correlator is
\begin{equation}
\begin{aligned}\label{eq:autoCl}
    C_{\ell}^{T^\dSc T^\dSc} (\omega,\omega^{\prime}) &\equiv   \bar{T}^2 \, C_{\ell}^{\tau\tau} (\omega,\omega^{\prime}) + \sum_{\ell^{\prime} \ell^{\prime\prime}} C_{\ell^{\prime\prime}}^{\tau\tau} (\omega,\omega^{\prime})  C_{\ell^{\prime}}^{TT} (W_{\ell \ell^{\prime} \ell^{\prime\prime}}^{000} )^2  + N_{\ell}^{T^\dSc T^\dSc} (\omega,\omega^{\prime}) \\
    &\simeq \bar{T}^2 \, C_{\ell}^{\tau\tau} (\omega,\omega^{\prime}) + N_{\ell}^{T^\dSc T^\dSc} (\omega,\omega^{\prime}).
\end{aligned}
\end{equation}
The contribution from instrumental noise and foregrounds is $N_{\ell}^{T^\dSc T^\dSc} (\omega,\omega^{\prime})$. Because the CMB anisotropies are so small in comparison to the monopole, the first term completely dominates the statistically isotropic contribution to the correlator. 
As a result, measuring $C_{\ell}^{T^\dSc T^\dSc}$ can be extremely sensitive to photon to dark photon conversion. However, such an auto-correlation function is proportional to the small kinetic mixing parameter $\varepsilon^4$, which limits the reach of measurements of the dark-screened CMB power spectrum. We forecast the reach of such an analysis in Section~\ref{sec:forecastauto}.

Similar dark-screened auto-correlations can be computed for the $TE$, $EE$ and $BB$ CMB spectra. These are listed in Appendix~\ref{appx:qes}. In determining the sensitivity of CMB experiments to  $\varepsilon$ in Section~\ref{sec:forecasts}, we use all isotropic components of the two-point auto-correlators. However, the temperature auto-correlation provides the best sensitivity because of the coupling of $\tau$ to $\bar{T}$.

The statistically anisotropic component of the dark screening auto-correlation is proportional to the un-screened temperature anisotropies, and we explore in the next section how this can be used to search for photon to dark photon conversion. We can also construct cross-correlations functions in the form of $\ev{T^\dSc_{\ell m} (\omega) T^\Sc_{\ell^{\prime} m^{\prime}}}$, the discussion of which we also leave to the next section.

\section{Cross-correlating dark screening}\label{sec:crossc}

In this section, we study the cross-correlation between observables that contain the dark screening optical depth $\tau(\hat{n},\omega)$ and those that do not. The essential qualitative understanding that motivates the construction of these correlation functions is the following: {\it Dark screening occurs in halos, and is therefore correlated with observables (within the Standard Model) that are sensitive to either the halos' locations or their electron density distributions.} As we showed, both Thomson screening and photon to dark photon conversion depend on the electron density profile. Therefore, there is a non-zero cross-correlation between the Thomson-screened and dark-screened temperature and polarization anisotropies. In the following, we discuss two ways of combining these maps: we consider first cross-correlation between the Thomson-screened and dark-screened CMB, and then their correlation with LSS tracers. We compute the relevant two- and three- point correlation functions for each method and identify the ones that are most sensitive to dark screening.

\subsection{Two-point cross-correlation}

Based on the discussions in the last section, an obvious candidate for the correlation function we want is:
\begin{equation}
\begin{aligned}\label{eq:TdScTSc}
    \ev{T^\dSc_{\ell m} (\omega) T^\Sc_{\ell^{\prime} m^{\prime}}} &= (-1)^m C_{\ell}^{T^\dSc T^\Sc} (\omega) \delta_{\ell \ell^{\prime}} \delta_{m-m^{\prime}} \\ 
    &- \sum_{\ell_1 m_1} (-1)^{m_1} \begin{pmatrix}
									 \ell & \ell^{\prime} & \ell_1  \\
								        m & m^{\prime} &  -m_1 
								\end{pmatrix}
        \sqrt{2\ell_1+1} W_{\ell \ell^{\prime} \ell_1}^{0 0 0}  C_{\ell^{\prime}}^{TT} \tau_{\ell_1 m_1} (\omega) \\
    &+ \bar{T} \sum_{\ell_1 m_1} (-1)^{m_1} \begin{pmatrix}
									\ell  & \ell^{\prime} & \ell_1 \\
								    m  & m^{\prime} & -m_1
								\end{pmatrix}
        \sqrt{2\ell_1+1} W_{\ell \ell^{\prime} \ell_1}^{0 0 0} C_\ell^{\tau \tau^\Th} (\omega) T_{\ell_1 m_1},
\end{aligned}
\end{equation}
where $C_\ell^{\tau \tau^\Th} (\omega)$ was defined in Eq.~\eqref{eq:crossspectra} and 
\begin{eqnarray}\label{eq:crosstdsctsc}
    C_{\ell}^{T^\dSc T^\Sc} (\omega) &\equiv&    \sum_{\ell^{\prime} \ell^{\prime\prime}} C_{\ell^{\prime\prime}}^{\tau\tau^\Th} (\omega)  C_{\ell^{\prime}}^{TT} (W_{\ell \ell^{\prime} \ell^{\prime\prime}}^{000} )^2.
\end{eqnarray}
Note that for this correlator there are statistically anisotropic contributions proportional to both the dark screening optical depth and the un-screened temperature anisotropies. The cross-correlation between the Thomson-screened and dark-screened temperature scales as $\bar{\tau} \sim \varepsilon^2$. This is more favorable than the $\varepsilon^4$ scaling found in Eq.~\eqref{eq:TdScTdSc}. However, unlike for Eq.~\eqref{eq:TdScTdSc}, only the second statistically anisotropic term depends on the temperature monopole $\bar{T}$ (Thomson screening doesn't couple to the temperature monopole). The consequence  is that this cross-correlation is less competitive than the dark-screened auto-correlation at fixed noise.

Before moving on, let's turn to polarization. $E$ and $B$ modes are defined with the relation in Eq.~\eqref{eq:harmonicdefinition}. Combined with the assumption that the Thomson-screened temperature anisotropies can be separated from the dark-screened anisotropies through the frequency dependence, we can construct a variety of two-point correlation functions -- 12 in total! These can be found in Appendix~\ref{appx:qes}. Unlike for temperature auto- and cross-correlations, some correlators involving polarization vanish in the absence of screening, and only receive statistically anisotropic contributions. These include
\begin{equation}\label{eq:TBanisotropic}
    \ev{T^\dSc_{\ell_1 m_1} i B^\Sc_{\ell_2 m_2}} = \sum_{\ell m} E_{\ell m} (-1)^m \begin{pmatrix}
				\ell_1 & \ell_2 & \ell \\
				m_1 & m_2 & -m 
	\end{pmatrix} \sqrt{2\ell+1} o_{\ell_1 \ell_2 \ell} W^{2 2 0}_{\ell_2 \ell \ell_1} \bar{T} C_{\ell_1}^{\tau \tau^\Th} \\
\end{equation}
and 
\begin{equation}\label{eq:EBanisotropic}
    \ev{E^\Sc_{\ell_1 m_1} i B^\dSc_{\ell_2 m_2}} = - \sum_{\ell m} \tau_{\ell m} (-1)^m \begin{pmatrix}
				\ell_1 & \ell_2 & \ell \\
				m_1 & m_2 & -m 
	\end{pmatrix} \sqrt{2\ell+1} o_{\ell_1 \ell_2 \ell} W^{2 2 0}_{\ell_2 \ell_1 \ell} C_{\ell_1}^{EE},
\end{equation}
where $C_{\ell}^{EE}$ is the lensed primary CMB E-mode power spectrum and 
\begin{equation}
o_{\ell \ell^{\prime} \ell^{\prime\prime}} \equiv \frac{1}{2} \left[ 1 - (-1)^{\ell+\ell^{\prime}+\ell^{\prime\prime}} \right].
\end{equation}
These correlations are also potential sensitive probes of patchy dark screening, as demonstrated in greater detail below.

\subsection{Correlating patchy dark screening with LSS}\label{sec:lssxds}

Expanding our focus beyond CMB observables, photon to dark photon conversion happens inside halos, and therefore the patchy dark screening signal is correlated with the LSS. It is most natural to look for correlations between patchy dark screening and tracers of LSS, such as galaxy redshift surveys. Given a tracer and various model assumptions, one can build a template for patchy dark screening, which we will denote by $\hat{\tau}(\varepsilon_0)$. In this section we build the intuition of how to use such a template to improve our sensitivity to $\varepsilon$. As one example, the template can be built in the following way:
\begin{equation}\label{eq:templateequation}
    \hat{\tau}_{\ell m}^g (\varepsilon_0, \omega) = \left[ (C^{\tau\tau}_{\ell}  (\varepsilon_0, \omega) \mathbf{C}^{gg}_\ell)^{-1} \cdot \mathbf{C}^{g \tau}_\ell (\varepsilon_0, \omega)  \right]  \cdot \mathbf{g}_{\ell m},
\end{equation}
where $\mathbf{g}_{\ell m}$ are the moments of the galaxy overdensity field and the vector notation denotes the redshift information; $C_\ell^{\tau\tau}$ is the model photon to dark photon optical depth power spectrum, $\mathbf{C}^{gg}_\ell$ is the redshift $\times$ redshift galaxy overdensity covariance matrix, and $\mathbf{C}^{g \tau}_\ell$ is a vector of the dark photon optical depth $\times$ galaxy overdensity cross-spectra at each redshift. We have explicitly indicated that the model power spectra involving the patchy dark screening optical depth depends on the fiducial choice $\varepsilon_0$ for the kinetic mixing parameter. Note that the template defined in Eq.~\eqref{eq:templateequation} can be improved by going beyond this simple linear filter, for example using machine learning techniques as in Ref~\cite{Kvasiuk:2023nje}.

The largest contribution to the cross-correlation of the template with the patchy dark screening component of the CMB is statistically isotropic and given by
\begin{equation}\label{eq:TdSctaug}
    \ev{T_{\ell m}^\dSc \hat{\tau}^g_{\ell^{\prime} m^{\prime}} (\varepsilon_0, \omega)}_{\rm isotropic} = - \bar{T} C_{\ell}^{\tau \hat{\tau}^g} \left(\varepsilon,\varepsilon_0,\omega \right) \, \delta_{\ell\ell^{\prime}}\delta_{m m^{\prime}},
\end{equation}
where $C_{\ell}^{\tau \hat{\tau}^g}$ is the cross-power spectrum between the template and real photon to dark photon optical depth. Importantly, $C_{\ell}^{\tau \hat{\tau}^g} \left(\varepsilon,\varepsilon_0,\omega \right) \propto \varepsilon^2$, and therefore this quantity scales more favorably with $\varepsilon$ in the small-$\varepsilon$ limit than the monopole contribution to the temperature auto-spectrum (which scales $\propto \varepsilon^4$). In addition to the more favorable scaling with the mixing parameter, cross-correlation with a template can be beneficial for mitigating systematic effects and galactic or extra-galactic foregrounds in the observed CMB. We forecast the reach of such analysis in Section~\ref{sec:CMBwithtemplate}.

\subsection{Reconstruction}\label{sec:reconstruction}

Coming back to the CMB, the discussion in the previous section suggests searching for correlation functions that would allow us to reconstruct the map $ \tau^\Th (\hat{n}) $ from CMB observables. $ \tau^\Th (\hat{n}) $, depending on the same electron density distribution in the Universe as $ \tau(\hat{n},\omega)$, would be correlated with $T^\dSc $, just like $\hat{\tau}^g$. The statistically anisotropic components of the two-point correlation functions in Eq.~\eqref{eq:TScTSc} can be used to construct quadratic estimators for patchy Thomson screening optical depth $ \tau^\Th (\hat{n}) $. Similarly, patchy dark screening, as well as the un-screened primary CMB temperature and polarization anisotropies can be reconstructed \eg from equations~\eqref{eq:TdScTdSc},~\eqref{eq:TdScTSc},~\eqref{eq:EBanisotropic}. An exhaustive list of quadratic estimators is presented in Appendix~\ref{appx:qes}. Similar quadratic estimators are used to measure weak lensing of the CMB (see \eg\cite{Weak_lensing} for a review), and are employed in a wide variety of other contexts in CMB science, in particular for the reconstruction of patchy Thomson screening during reionization~\cite{dvorkin_reconstructing_2009} and kinetic Sunyaev Zel'dovich velocity reconstruction~\cite{Zhang:2015uta,Terrana2016,Deutsch:2017ybc,smith_ksz_2018,Cayuso:2021ljq}. 

Starting again with a Standard Model example, the reionization optical depth can be found from Eq.~\eqref{eq:TScTSc} as:
\begin{align}
    \hat{\tau}^\Th_{LM} =& - N_L^{\tau^\Th ; T^\Sc T^\Sc} \sum_{\ell m} \sum_{\ell^{\prime} m^{\prime}} (-1)^M 
    \begin{pmatrix}
				\ell & \ell^{\prime} & L \\
				m & m^{\prime} & -M 
	\end{pmatrix} \sqrt{2L+1} \, G_{\ell \ell^{\prime} L}^{\tau^\Th ; T^\Sc T^\Sc} T^\Sc_{\ell m} T^\Sc_{\ell^{\prime} m^{\prime}},
\end{align}
where 
\begin{align}
    N_{L}^{\tau^\Th ; T^\Sc T^\Sc} =& \left[ \sum_{\ell \ell^{\prime}}  \frac{\left( W^{0 0 0}_{\ell \ell^{\prime} L} \left[C_{\ell}^{TT} + C_{\ell^{\prime}}^{TT} \right] \right)^2}{2 \, C^{T^\Sc T^\Sc}_{\ell} C^{T^\Sc T^\Sc}_{\ell^{\prime}}} \right]^{-1}, & G_{\ell \ell^{\prime} L}^{\tau^\Th ; T^\Sc T^\Sc} =& \frac{W^{0 0 0}_{\ell \ell^{\prime} L} \left[C_{\ell}^{TT} + C_{\ell^{\prime}}^{TT} \right]}{2 \, C^{T^\Sc T^\Sc}_{\ell} C^{T^\Sc T^\Sc}_{\ell^{\prime}}}.
\end{align}
The weights and prefactor are chosen such that the estimator is unbiased if the input power spectra provide an accurate model, \ie
\begin{eqnarray}
    \ev{\hat{\tau}^\Th_{LM}} = \tau^\Th_{LM},
\end{eqnarray}
as well as has minimum variance when all fields in the problem are Gaussian:
\begin{eqnarray}
    \ev{\hat{\tau}^\Th_{LM} \hat{\tau}^\Th_{L^{\prime}M^{\prime}}} = (C_{L}^{\tau^\Th  \tau^\Th} + N_L^{\hat{\tau}^\Th ; T^\Sc T^\dSc} )\delta_{LL^{\prime}}\delta_{MM^{\prime}}.
\end{eqnarray}
The prefactor $N_L^{\hat{\tau}^\Th ; T^\Sc T^\dSc}$ is the noise on the reconstruction. 

Similarly, one can reconstruct $\hat{\tau}^\Th $ from measurements of polarization, notably from $ E^\Sc B^\Sc $. The relevant quadratic estimators are presented in Appendix~\ref{appx:qes}. Estimators for the Thomson optical depth found there are equivalent those presented in Ref.~\cite{dvorkin_reconstructing_2009}. The two-point function that would result from cross-correlating the dark-screened CMB temperature with the Thomson optical depth map, \ie $\ev{\hat{\tau}^\Th  T^\dSc}$, is another means to construct the correlation function in Eq.~\eqref{eq:TdSctaug}.

More quadratic estimators can be built from the dark screening anisotropies $T^\dSc $. The leading term that contains the dark screening optical depth is:  
\begin{equation}\label{eq:ScdSctau}
    \hat{\tau}^{qe \, *}_{LM} = - N_L^{\hat{\tau}; T^\dSc T^\Sc} \sum_{\ell m} \sum_{\ell^{\prime} m^{\prime}} (-1)^M 
    \begin{pmatrix}
				\ell & \ell^{\prime} & L \\
				m & m^{\prime} & -M 
	\end{pmatrix} \sqrt{2L+1} \,G_{\ell \ell^{\prime} L}^{\hat{\tau}; T^\dSc T^\Sc} T^\dSc_{\ell m} T^\Sc_{\ell^{\prime} m^{\prime}},
\end{equation}
where
\begin{align}
    N_L^{\hat{\tau}^{qe}; T^\Sc T^\dSc} &=  \left(\sum_{\ell \ell^{\prime}} \frac{(W_{L \ell \ell^{\prime}}^{0 0 0})^2  (C_{\ell^{\prime}}^{TT})^2}{C_{\ell}^{T^\Sc T^\Sc}   C_{\ell^{\prime}}^{T^\dSc T^\dSc}} \right)^{-1}, \\
    G_{\ell \ell^{\prime} L}^{\hat{\tau}^{qe}; T^\Sc T^\dSc} &=   \frac{C_{\ell^{\prime}}^{T^\Sc T^\Sc} C_{\ell}^{T^\dSc T^\dSc} W_{L \ell \ell^{\prime}}^{0 0 0}  C_{\ell^{\prime}}^{TT} - (-1)^{\ell+\ell^{\prime}+L}C_{\ell}^{T^\Sc T^\dSc} C_{\ell^{\prime}}^{T^\Sc T^\dSc}W_{L \ell^{\prime} \ell}^{0 0 0}  C_{\ell}^{TT}}{C_{\ell}^{T^\Sc T^\Sc}C_{\ell^{\prime}}^{T^\Sc T^\Sc} C_{\ell}^{T^\dSc T^\dSc} C_{\ell^{\prime}}^{T^\dSc T^\dSc} - (C_{\ell}^{T^\Sc T^\dSc})^2 (C_{\ell^{\prime}}^{T^\Sc T^\dSc})^2} \nonumber \\
    &\simeq \frac{W_{L \ell \ell^{\prime}}^{0 0 0}  C_{\ell^{\prime}}^{TT}}{C_{\ell}^{T^\Sc T^\Sc}   C_{\ell^{\prime}}^{T^\dSc T^\dSc}}.
\end{align}
We used the fact that the product of the screened auto-spectra is larger than the product of the screened cross-spectra in the second line.

It is important to note that in constructing the weights $G_{\ell \ell^{\prime} L}^{\hat{\tau}; T^\Sc T^\dSc}$ and the prefactor $N_L^{\hat{\tau}; T^\Sc T^\dSc}$ that the input power spectra come from a {\em theoretical} model for the un-screened CMB temperature power spectrum $C_\ell^{TT}$ as well as contributions from noise and foregrounds to both $C_{\ell}^{T^\Sc T^\Sc}$ and $C_{\ell^{\prime}}^{T^\dSc T^\dSc}$. The model for $C_{\ell}^{T^\Sc T^\Sc}$ and $C_{\ell^{\prime}}^{T^\dSc T^\dSc}$ can be checked against the measured power spectra of these maps. However, since we cannot directly measure $C_\ell^{TT}$, there is inevitably some residual model uncertainty. This manifests as a bias on the reconstruction: 
\begin{eqnarray}
    \ev{\hat{\tau}^{qe}_{LM}} = b_L^{\tau} \, \tau_{LM}, \quad \quad b_L^\tau = N_L^{\hat{\tau}; T^\Sc T^\dSc} \sum_{\ell \ell^{\prime}} \frac{(W_{L \ell \ell^{\prime}}^{0 0 0})^2  (C_{\ell^{\prime}}^{TT})_{\rm model} (C_{\ell^{\prime}}^{TT})_{\rm actual}}{C_{\ell}^{T^\Sc T^\Sc}   C_{\ell^{\prime}}^{T^\dSc T^\dSc}}.
\end{eqnarray}
When the model and actual power spectra are identical, the bias factor is unity. Importantly, it is possible to measure this bias by comparing \eg $\hat{\tau}^{qe}_{LM}$ to $\hat{\tau}^{g}_{LM}$. In principle, the bias can be measured without cosmic variance since the comparison is done at the level of the modes and not the power spectra; this is an example of `sample variance cancellation' (\eg\cite{Seljak_2009}). This procedure is elucidated in greater detail in Section~\ref{sec:forecasts}. 

As a third example, a quadratic estimator for the un-screened temperature anisotropies is
\begin{align}\label{eq:TQE}
    \hat{T}_{LM} =& N_L^{\hat{T}; T^\dSc T^\Sc} \sum_{\ell m} \sum_{\ell^{\prime} m^{\prime}} (-1)^M \begin{pmatrix}
				\ell & \ell^{\prime} & L \\
				m & m^{\prime} & -M 
	\end{pmatrix} \sqrt{2L+1} \,G_{\ell \ell^{\prime} L}^{\hat{T}; T^\dSc T^\Sc} T^\dSc_{\ell m} T^\Sc_{\ell^{\prime} m^{\prime}},
\end{align}
where
\begin{align}
    N_{L}^{\hat{T}; T^\dSc T^\Sc} =& \left[ \sum_{\ell \ell^{\prime}} \frac{\bar{T}^2 \left| W^{0 0 0}_{\ell \ell^{\prime} L} C_{\ell}^{\tau\tau^\Th} \right|^2}{C^{T^\dSc T^\dSc}_{\ell} C^{T^\Sc T^\Sc}_{\ell^{\prime}}} \right]^{-1}, & G_{\ell \ell^{\prime} L}^{\hat{T}; T^\dSc T^\Sc} \simeq& \frac{W^{0 0 0}_{\ell \ell^{\prime} L} \bar{T} C_{\ell}^{\tau\tau^\Th}}{C^{T^\dSc T^\dSc}_{\ell} C^{T^\Sc T^\Sc}_{\ell^{\prime}}}.
\end{align}
The weights and reconstruction noise in this case depend on a theoretical model for $C_{\ell}^{\tau\tau^\Th}$, an object we have no prior knowledge of and hope to search for. This implies that the reconstruction of the un-screened temperature anisotropies will be significantly biased. However, as described above, one can measure the bias by comparing to a template for $T_{LM}$, which at least on large angular scales, can be provided by $T^\Sc $. That is, a correlation function $\ev{\hat{T} T^\Sc}$, can also be used to search for anisotropic dark screening.

\subsection{Three-point correlation functions}\label{sec:bispectra}

The discussion in the previous section suggests that one should correlate maps reconstructed from CMB observables, such as $\hat{\tau}^\Th $, $\hat{\tau}$ and $\hat{T}$, with CMB observables (\eg $T^\Sc $ or $T^\dSc $) or itself. In terms of CMB observables, these correlation functions will be three-point or four-point correlation functions. In other words, the statistically anisotropic contributions to the two-point correlation functions in the previous section imply that there are many non-vanishing three-point correlation functions even in the case where the temperature, polarization, and optical depth fields are Gaussian.

For example, the correlation functions 
\begin{align}
    \ev{\hat{\tau}^\Th T^\dSc} \sim \ev{\ev{T^\Sc T^\Sc} T^\dSc}   \quad \text{and} \quad \ev{\hat{T}T^\Sc} \sim \ev{\ev{T^\dSc T^\Sc} T^\Sc}
\end{align}
both come from the three-point correlation function $\ev{T^\dSc T^\Sc T^\Sc}$. Therefore, rather than working with the more intuitive two-point correlation functions involving reconstructed maps, we can forecast the sensitivity directly using three-point functions; we present this forecast in Section~\ref{sec:bispectrumforecast}.

Three-point functions are described by the angle-averaged bispectrum, defined as
\begin{eqnarray}
B_{\ell \ell^{\prime} \ell^{\prime\prime}}^{XYZ} = \sum_{m m^{\prime} m^{\prime\prime}} 
	\begin{pmatrix}
			\ell & \ell^{\prime} & \ell^{\prime\prime} \\
			m & m^{\prime} & -m^{\prime\prime} 
	\end{pmatrix}
	\ev{X_{\ell m} Y_{\ell^{\prime} m^{\prime}} Z_{\ell^{\prime\prime} m^{\prime\prime}}}.
\end{eqnarray}
There are many bispectra to consider between temperature, polarization, and templates for the optical depth. Here we focus on the largest bispectra that scale like $\varepsilon^2$, since these will be most sensitive to photon to dark photon conversion. 

The most relevant bispectra involving only CMB temperature and polarization are (see Eq.~\eqref{eq:TdScTSc} and Eq.~\eqref{eq:TBanisotropic})
\begin{equation}\label{eq:bispectrumTdScTScTSc}
B_{\ell \ell^{\prime} \ell^{\prime\prime}}^{T^\dSc T^\Sc T^\Sc} = \bar{T} \sqrt{2\ell^{\prime\prime} + 1} W_{\ell \ell^{\prime} \ell^{\prime\prime}}^{0 0 0} \left( C_{\ell^{\prime}}^{TT} + C_{\ell^{\prime\prime}}^{TT}   \right) C_\ell^{\tau \tau^\Th} (\omega) 
\end{equation}
and
\begin{equation}\label{eq:TEB_bispectrum}
B_{\ell \ell^{\prime} \ell^{\prime\prime}}^{T^\dSc E^\Sc  B^\Sc} = \bar{T} \sqrt{2\ell^{\prime\prime} + 1} W_{\ell \ell^{\prime} \ell^{\prime\prime}}^{0 2 2} o_{\ell \ell^{\prime} \ell^{\prime\prime}} \left( C_{\ell^{\prime}}^{EE} + C_{\ell^{\prime\prime}}^{EE}  \right) C_\ell^{\tau \tau^\Th} (\omega).
\end{equation}
Note that both are proportional to the CMB monopole. The most relevant bispectra that scale like $\varepsilon^2$ and involve both the CMB and an optical depth template are 
\begin{equation}
B_{\ell \ell^{\prime} \ell^{\prime\prime}}^{E^\Sc B^\dSc \hat{\tau}} = \sqrt{2\ell^{\prime\prime} + 1} W_{\ell \ell^{\prime} \ell^{\prime\prime}}^{2 2 0} o_{\ell \ell^{\prime} \ell^{\prime\prime}} \left( C_{\ell}^{EE} + C_{\ell^{\prime}}^{EE} \right) C_{\ell^{\prime\prime}}^{\tau \hat{\tau}} (\omega) 
\end{equation}
and
\begin{equation}
B_{\ell \ell^{\prime} \ell^{\prime\prime}}^{T^\Sc E^\dSc \hat{\tau}} = B_{\ell \ell^{\prime} \ell^{\prime\prime}}^{T^\dSc E^\Sc \hat{\tau}}  = \sqrt{2\ell^{\prime\prime} + 1} W_{\ell \ell^{\prime} \ell^{\prime\prime}}^{0 0 0} e_{\ell \ell^{\prime} \ell^{\prime\prime}} \left( C_{\ell}^{TE} + C_{\ell^{\prime}}^{TE} \right) C_{\ell^{\prime\prime}}^{\tau \hat{\tau}} (\omega).
\end{equation}

These bispectra, similar to the cross-correlation with LSS in Eq.~\eqref{eq:TdSctaug}, are proportional to the CMB monopole $\bar{T}$, while at the same time, scales as $\varepsilon^2$. This, as we will demonstrate in more detail in Section~\ref{sec:forecasts}, makes these bispectra almost as sensitive as the two-point functions $\ev{T^\dSc T^\dSc}$ (scaling as $\varepsilon^4$), and more sensitive than $\ev{T^\dSc T^\Sc}$ (proportional to $ C_{\ell}^{TT}$).

\section{Forecast}\label{sec:forecasts}

In this section we forecast the projected sensitivity of several CMB experiments to the mixing parameter $\varepsilon$ over a range of fixed values for the dark photon mass $m_\darkph$. We consider each of the five techniques mentioned in the introduction, and identify the most promising observables for each experimental configuration. These relevant correlation functions are summarized in Table~\ref{table:correlation}. We first compute the constraint on $\varepsilon$ from the monopole  spectral distortion using COBE/FIRAS. We then describe our forecast assumptions, and compute the reach of existing and future CMB anisotropy experiments.  

\begin{table}[ht!]
    \begin{center}
        \begin{tabular}{|c|c|c|c|c|} 
            \hhline{~----}
           \multicolumn{1}{c|}{} & $\varepsilon$ & $C_\ell^{TT}$ & $\tau^\Th $ & $C_{\ell}^{EE}$\\
           \hhline{-====}
          $\ev{T^\dSc} $ & 2 & 0 & 0 & - \\
          $\ev{T^\Sc T^\Sc} $  & 0 & 1 & 2 & -\\
          $\ev{T^\dSc T^\dSc} $ & 4 & 0 & 0 & - \\
          $\ev{T^\Sc T^\dSc} $  & 2 & 1 & 1 & -\\ 
          $\ev{T^\dSc \hat{\tau}^{g}} $  & 2 & 0 & 0 & -\\
          $\ev{T^\Sc T^\Sc T^\dSc} $  & 2 & 1 & 1& - \\
          $\ev{T^\dSc E^\Sc B^\Sc} $  & 2 & - & 1& 1 \\
          \hline
        \end{tabular}
        \caption{The scaling of various correlation functions with the small parameters: kinetic mixing $\varepsilon$, primary CMB power $C_\ell^{TT}$, primary $E$-mode polarization power $C_{\ell}^{EE}$ and optical depth of Thomson screening $\tau^\Th $. The correlation functions $\ev{T^\Sc T^\Sc}$ and $\ev{T^\Sc T^\dSc}$ are also shown for comparison.}\label{table:correlation}
    \end{center}
\end{table}

\subsection{FIRAS bounds}\label{sec:firas}

First we look at the CMB monopole constraint given by the COBE satellite. This method has been used in the past to forecast the constraint on $\varepsilon$ from different models for the distribution of ionized gas since recombination~\cite{caputo_dark_2020, Mirizzi:2009iz,Garcia:2020qrp}.

The CMB monopole was measured by the FIRAS instrument on the COBE satellite and was discovered to have a near perfect blackbody spectrum~\cite{COBEFixsen}. The data consists of $43$ measurements of the sky-averaged temperature over frequencies in the range $\omega = 68.05 - 639.46 $ GHz~\footnote{Data was taken from  \url{https://lambda.gsfc.nasa.gov/data/cobe/firas/monopole_spec/firas_monopole_spec_v1.txt}}. This gives a best fit blackbody temperature of $\bar{T}=2.725\pm 0.002$ K, with residuals of order roughly $10^{-4}$ below the peak intensity and $1\sigma$ uncertainties of the same magnitude. This remarkable precision already gives a constraint on the amplitude of the conversion probability for CMB photons of order the uncertainty $P_{\gamma \to \darkph} \lesssim 10^{-4}$. In this section we improve this bound by considering the full available frequency spectrum. To constrain our model we use the method described in~\cite{Mirizzi:2009iz}.

\begin{figure}[ht!]
    \centering
    \includegraphics[width=0.5\textwidth]{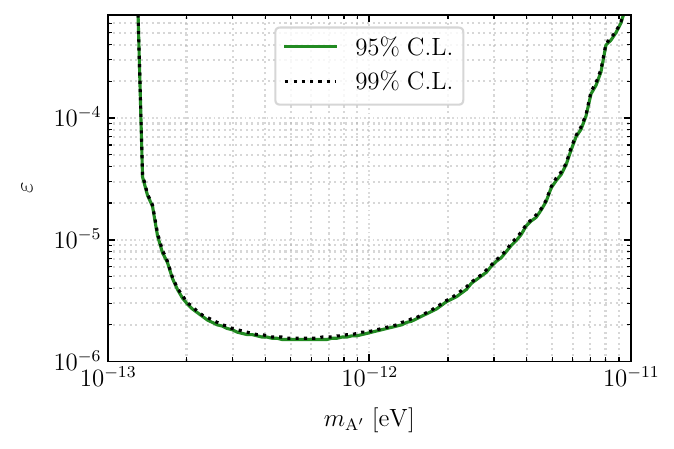}
    \caption{Exclusion contours on coupling constant $\varepsilon$ function of dark photon mass $m_\darkph$ from COBE/FIRAS~\cite{COBEFixsen} data. We see that the uncertainty in the measurement of $\bar{T}$ gives a constraint on the magnitude of the dark-screened temperature monopole $\ev{T^\dSc}$. Here we used the total dark screening optical depth due to both galactic and extra-galactic contributions $\bar{\tau} + \tau^\MW$. This constraint provides an upper bound on $\varepsilon$ of at most $10^{-6}$.}\label{fig:FIRASexclusion}
\end{figure}

Assuming an isotropic conversion of photons into dark photons, the CMB blackbody spectrum is distorted according to
\begin{equation}
    B(\omega, T, \varepsilon, m_\darkph) = B^0(\omega, T) \left( 1 - P_{\gamma \to \darkph}(\bar{\tau}_{\rm tot}, m_\darkph) \right),
\end{equation}
where $B^0(\omega, T)$ is the theoretical blackbody spectrum defined in Eq.~\eqref{eq:a_blackbody} and $\bar{\tau}_{\rm tot}$, the total dark screening monopole including the galactic component, implicitly depends on $\omega$ and $\varepsilon$. The reduced $\chi^2$ estimator is an average over all available frequency channels of the difference between the measured data and the expected signal $B^\rmexp$ in each frequency bin:
\begin{equation}
    \chi^2 = \frac{1}{N-1} \sum_i^N \frac{1}{\sigma^\rmexp_i}\left(B^\rmexp_i - B_i(T, \varepsilon, m_\darkph)\right).
\end{equation}
This estimator is minimized by some value of $T$ at each point in the plane spanned by $\varepsilon$ and $m_\darkph$. In Fig.~\ref{fig:FIRASexclusion} we show the $95\%$ and $99\%$ confidence limit contours in this parameter space, for the $\chi^2$ of a distribution with $42$ degrees of freedom. The exclusion regions are similar in both cases, indicating that the regression method is robust, \ie the $\chi^2$ changes rapidly with temperature around the minimum. 

Over the accessible range of dark photon masses, the constraint on $\varepsilon$ reaches up to $\simeq 10^{-6}$, roughly two orders of magnitude better than the naive limit set by the error bar on the blackbody temperature. In Appendix~\ref{appx:modeling} we explore the model uncertainty implicit in our constraint, which arises primarily from our lack of knowledge of the gas profile. Comparing the constraints obtained in our fiducial model for gas profiles with those obtained for a model in which gas perfectly tracks dark matter (e.g. a model without any baryonic feedback), we conclude that the constraints at low dark photon mass are robust. At high dark photon mass, the constraint is strengthened in a model where baryons perfectly track dark matter. We can extrapolate that if our fiducial gas model has too little baryonic feedback, the constraints could further weaken at high dark photon mass. Incorporating information from measurements of the Sunyaev Zel'dovich effect (\eg\cite{ACTgasthermo}), the dispersion measure of fast radio bursts (\eg\cite{Prochaska_2019,Madhavacheril:2019buy})), 21 cm intensity mapping (\eg\cite{Bernal_2022}), and other tracers of baryons will be helpful in mitigating this modeling uncertainty and will be explored in future work.

\subsection{Forecast assumptions for CMB anisotropy experiments}\label{sec:noise}

In Section~\ref{sec:CMBscreeningAngPS}, we assumed that the frequency-dependent anisotropies due to resonant conversion ($T^\dSc $, $E^\dSc $, $B^\dSc $) could be perfectly separated from the blackbody Thomson-screened anisotropies ($T^\Sc $, $E^\Sc $, $B^\Sc $). Here, we explore the degree to which this separation can be made in the presence of instrumental noise and measurements in only a small number of frequency channels. We estimate the residual noise on the dark-screened and Thomson-screened maps achievable with existing and future CMB experiments, which is used in the following forecasts based on two- and three-point correlation functions. 

We consider three different CMB experiments: the combination of the Low Frequency Instrument (LFI)~\cite{planck2015LFI} and High Frequency Instrument (HFI)~\cite{planck2015HFI} on the Planck satellite, CMB Stage-4~\cite{abazajian_cmb-s4_2016} and CMB-HD~\cite{Sehgal:2019ewc, brinckmann_snowmass2021_2021}. In the context of our analysis, a CMB experiment is characterized by the sensitivity and resolution at a set of measured frequencies.Throughout, we assume Gaussian beam and white uncorrelated noise for all instruments, as well as full-sky coverage and no foregrounds.

Before proceeding, it is important to comment on the potential impact of CMB foregrounds. There are a variety of galactic foregrounds (see \eg\cite{dickinson2016cmb} for an overview) that fall with frequency, with a similar power law in the power spectra to the patchy dark screening signal (falling as $\omega^{-2} - \omega^{-3}$) including: synchrotron, free-free, and spinning dust. These are strongest in the galactic plane, and their influence can be mitigated by masking the most contaminated regions of the sky, or by incorporating information about the morphology of the signal. Nevertheless, these foregrounds can in principle add significant extra power at low-frequencies and on large angular scales. Extra-galactic radio point sources, which are dominated by synchrotron emission, are also a potentially important foreground to consider. Resolved point sources can again be dealt with by masking, however the unresolved point sources can add power at low frequencies and on small angular scales. This signal is also correlated with other tracers of LSS, limiting the power of cross-correlations to mitigate foregrounds. The extent to which galactic and extra-galactic foregrounds degrade the forecast we present below requires a detailed analysis, which we defer to future work.

The instrumental noise considered throughout our analysis is modeled as:
\begin{equation}\label{eq:noises}
\begin{gathered}
	N^{TT}_{\ell} = \Delta_T^2 \exp \left[\ell(\ell+1) \, \frac{\theta_{\rm FWHM}^2}{8 \log 2}\right] \left[ 1 + \left(\ell/\ell_\knee\right)^{\alpha_\knee} \right], \\
    N^{EE}_{\ell} = N^{BB}_{\ell} = \sqrt{2} N^{TT}_{\ell}, \quad N^{TE}_{\ell} = 0.
\end{gathered}
\end{equation}
Here, $\Delta_T \left[ \rm \mu K \, rad\right]$ represents the sensitivity in temperature, while $\theta_{\rm FWHM} \left[ \rm rad\right]$ is the full width at half maximum of our assumed Gaussian beam, which characterizes the resolution of the instrument. The sensitivity and resolution vary with frequency. Furthermore, ground-based measurements are subject to atmospheric contamination on large angular scales. To account for this effect, in the analysis for both CMB-S$4$ and CMB-HD we include the additional `red noise' term in the second bracket with $\alpha_\knee=-3$ and $\ell_\knee=100$ in all frequency bins. This contribution diverges at low-$\ell$, and becomes increasingly irrelevant for $\ell > \ell_\knee$. The values we choose to represent each experiment are shown in Table \ref{tab:noisemodel}. Note, throughout this section, we denote by $\nu$ the photon frequency in GHz such that $\nu \equiv \omega / 2\pi $.

\begin{table}[ht!]
    \begin{subtable}{1\textwidth}

    \begin{center}
        \begin{tabular}{|l|c|c|c|c|c|c|c|c|c|} 
            \hline
          Frequency $\nu$ (GHz) & $30$ & $44$ & $70$ & $100$ & $143$ & $217$ & $353$ & $545$ & $857$\\
          \hline\hline
          $\Delta_T \, (\rm \mu K \, arcmin)$ & $195.1$ & $226.1$ & $199.1$ & $77.4$ & $33.$ & $46.8$ & $153.6$ & $818.2$ & $40090.7$ \\
          $\theta_{\rm FWHM}$ (arcmin) & $32.41$ & $27.1$ & $13.32$ & $9.69$ & $7.3$ & $5.02$ & $4.94$ & $4.83$ & $4.64$\\\hline
        \end{tabular}
        \caption{Parameters used for the Planck forecast. Sensitivities and temperatures taken from~\cite{planck2015HFI} for  frequencies $100 $ GHz and higher (HFI instrument), and~\cite{planck2015LFI} for the lowest three frequencies (LFI instrument).}\label{subtab:planck}
      \end{center} 
    \end{subtable}

    \bigskip
    
    \begin{subtable}{1\textwidth}
    \begin{center}
        \begin{tabular}{|l|c|c|c|c|c|c|c|} 
            \hline
          Frequency $\nu$ (GHz) & $20$ & $27$ & $39$ & $93$ & $145$ & $225$ & $278$ \\
          \hline\hline
          $\Delta_T \, (\rm \mu K \, arcmin)$ & $10.41$ & $5.14$ & $3.28$ & $0.50$ & $0.46$ & $1.45$ & $3.43$ \\
          $\theta_{\rm FWHM}$ (arcmin) & $11.0$ & $8.4$ & $5.8$ & $2.5$ & $1.6$ & $1.1$ & $1.0$ \\\hline
        \end{tabular}
        \caption{Sensitivity and resolution for CMB-S4 V3R0 configuration.}\label{subtab:s4}
    \end{center}
    \end{subtable}

    \bigskip

    \begin{subtable}{1\textwidth}
    \begin{center}
        \begin{tabular}{|l|c|c|c|c|c|c|c|} 
            \hline
          Frequency $\nu$ (GHz) & $30$ & $40$ & $90$ & $150$ & $220$ & $280$ & $350$ \\
          \hline\hline
          $\Delta_T \, (\rm \mu K \, arcmin)$ & $6.5$ & $3.4$ & $0.7$ & $0.8$ & $2.0$ & $2.7$ & $100.0$ \\
          $\theta_{\rm FWHM}$ (arcmin) & $1.25$ & $0.94$ & $0.42$ & $0.25$ & $0.17$ & $0.13$ & $0.11$ \\\hline
        \end{tabular}
        \caption{CMB-HD forecast parameters, taken from Table 2 of~\cite{brinckmann_snowmass2021_2021}.}\label{subtab:hd}
    \end{center}

    \end{subtable}
    \caption{Noise parameters for Fisher forecasts. We model the noise covariance as in Eq.~\eqref{eq:noises} where in the case of ground-based CMB-S$4$ and CMB-HD we include the red noise term with parameters $\alpha_\knee=-3$ and $\ell_\knee=100$ in all frequency bins.}\label{tab:noisemodel}
\end{table}

Later, when we compute the Fisher information, we will need to estimate the noise covariance for the dark screening two-point functions of Section~\ref{sec:twoptcorrs}. These power spectra have an intrinsic inverse frequency squared dependence $C_{\ell}^{X^\dSc X^\dSc} \propto \varepsilon^4 / \omega^2$. It is possible to disentangle this signal from the measured CMB by cross-correlating measurements across multiple frequency channels. We do this by applying a harmonic Internal Linear Combination (ILC) algorithm~\cite{ILCpaper} which we now describe.

Recall that the expected \textit{isotropic} measured CMB signal consists of the primary CMB which is screened by the inhomogeneous field $\tau$ plus instrumental noise. For example, in the temperature case we have from equations~\eqref{eq:generalobs},~\eqref{eq:generalscr},~\eqref{eq:autoCl}:
\begin{equation}
\begin{aligned}\label{eq:Scparam}
    C_{\ell}^{TT\, \obs}(\omega_i,\omega_j) &= C_{\ell}^{TT} - C_{\ell}^{T^\dSc T^\dSc} (\omega_i,\omega_j) + N_{\ell}^{TT}(\omega_i) \delta_{\omega_i \omega_j} \\
    &\simeq C_{\ell}^{TT} - \bar{T}^2 C_{\ell}^{\tau\tau} (\omega_i,\omega_j)+ N_{\ell}^{TT}(\omega_i) \delta_{\omega_i \omega_j},
\end{aligned}
\end{equation}

\begin{figure}[ht!]
    \centering
   \includegraphics[width=1\textwidth]{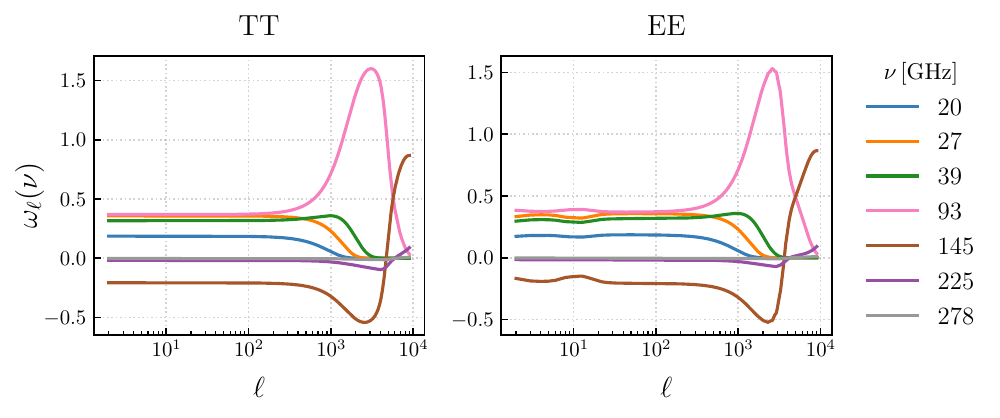}
    \caption{Illustration of the weight functions $\mathbf{w}_\ell$ as defined in Eq.~\eqref{eq:weightsss} for the frequencies of CMB-S$4$ with noise parameters defined in Table \ref{tab:noisemodel}. Notice that the dominant frequency is the $\omega = 93 $ GHz channel, but this changes at higher $\ell$. This is because the ILC favors the $N_{\ell}(\omega)$ with the lowest magnitude at a given $\ell$.}\label{fig:weights}
\end{figure}

The instrumental noise term is frequency dependent as displayed in Table \ref{tab:noisemodel}, where the $i$ and $j$ labels above denote each available channel. Let us re-write the dark screening two-point function term in the frequency explicit form $C_{\ell}^{\tau\tau} (\omega_i=1,\omega_j=1)/ \omega_i\omega_j$. To build the covariance matrix for all temperature measurements we simply choose a frequency for the signal, $\omega_0$, and measure all entries in reference to it. Overall we find:
\begin{equation}\label{eq:covmatILC}
    \mathbf{C}_{\ell} = \mathbf{\Omega}^{-1} \mathbf{C}^{TT}_{\ell} - \mathbf{e}\mathbf{e}^{\dagger} \bar{T}^2 \frac{\zeta(\omega_0)^2}{\omega_0^{2}} C_\ell^{\tau\tau}(\mathbf{\omega}=1) + \text{diag} \left( \mathbf{\Omega}^{-1} \mathbf{N}^{TT}_{\ell}(\mathbf{\omega}) \right).
\end{equation}
The frequency-dependent matrix has entries $\Omega^{-1}_{ij} = \left(\zeta(\omega_0)^2/\omega_0^2 \right)\left(\omega_i \omega_j / \zeta(\omega_i)\zeta(\omega_j)\right)$ and $\mathbf{e} = \left(1,1, \dots, 1\right)$.
The ILC method consists of weighting each matrix element appropriately in order to minimize a frequency-independent residual. This is constructed like
\begin{equation}\label{eq:ILCnoise}
    \tilde{N}^{T^\dSc T^\dSc}_{\ell} = \mathbf{w}_{\ell}^{\dagger} \cdot \left( \mathbf{\Omega}^{-1} C^{TT}_{\ell} + \mathbf{\Omega}^{-1} \mathbf{N}^{TT}_{\ell} \right) \cdot \mathbf{w}_{\ell},
\end{equation}
where the weights $\mathbf{w}_\ell$ satisfy
\begin{equation}\label{eq:weightsss}
    \mathbf{w}_\ell = \frac{\left( \mathbf{C}_{\ell} \right)^{-1} \mathbf{e}}{\mathbf{e}^{\dagger} \left( \mathbf{C}_{\ell} \right)^{-1} \mathbf{e}}.
\end{equation}
In our analysis, we use a fixed baseline frequency $\omega_0 = 30 $ GHz. For the $C_{\ell}^{XX}$ terms of the primary CMB we compute the lensed temperature and polarization power spectra using CAMB~\cite{Lewis_2000}, using the cosmological parameters listed in Section~\ref{sec:darkscreening}. Fig.~\ref{fig:weights} shows the weights computed for CMB-S$4$ specifications. On large scales, the $93$ and $145 $ GHz maps, which are the lowest noise, are used to subtract off the blackbody CMB, with the other channels weighted inversely with frequency.

A similar computation can be used to find the weights and residual noise for polarization. The ILC-cleaned spectrum of the temperature map includes the frequency-dependent term, computed at the baseline frequency $\omega_0$, plus the noise residual:
\begin{equation}
    \tilde{C}^{T^\dSc T^\dSc}_{\ell} = \frac{\zeta(\omega_0)^2}{\omega_0^{2}} C_\ell^{T^\dSc T^\dSc}(\omega=1) + \tilde{N}^{T^\dSc T^\dSc}_{\ell}.
\end{equation}
Henceforth, the $\tilde{.}$ notation will refer to an ILC-cleaned map, or its associated residual defined as in Eq.~\eqref{eq:ILCnoise}. Note the factor of $\zeta(\omega)$ defined in Eq.~\eqref{eq:zeta_freq}. The CMB peaks in intensity around $160 $ GHz. The factor of $\zeta(\omega)$ approaches unity for frequencies below the CMB peak, and vanishes for frequencies $\omega \gg 160 $ GHz. Since a $\omega^{-1}$ scaling weighs small frequencies more strongly, the additional $\zeta(\omega)$ factor does not affect the analysis significantly.

\begin{figure}[ht!]
    \flushleft
    \includegraphics[width=1.\textwidth]{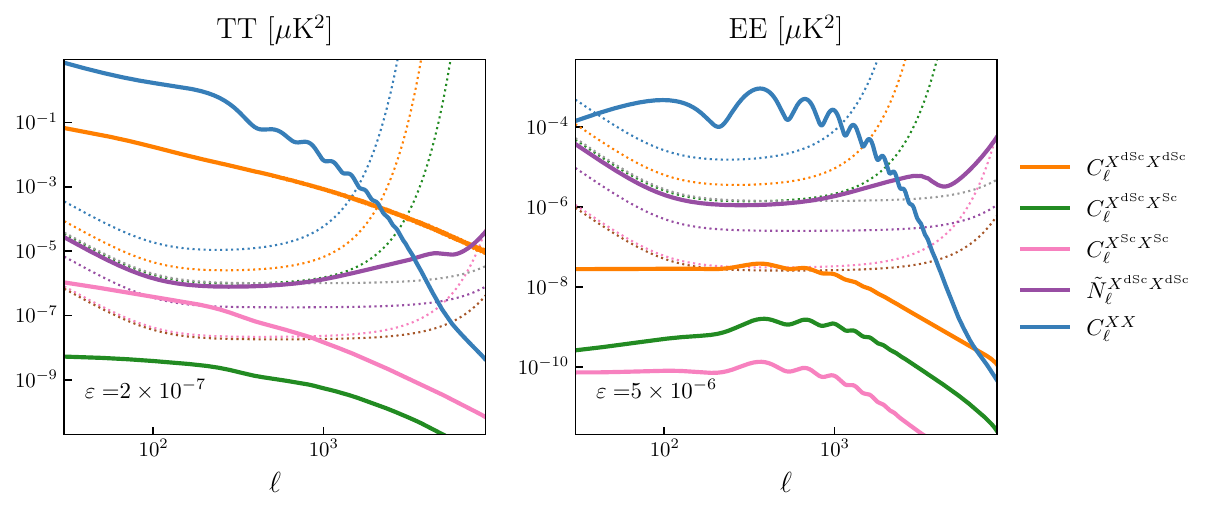}
    \caption{Scale comparison. We fixed $m_\darkph=10^{-12}$ eV and $\varepsilon$ as shown in each panel. We show the isotropic component to the $XX = TT, EE$ power spectra due to dark screening $C_{\ell}^{X^\dSc X^\dSc} \propto \varepsilon^4$ (\textit{orange}), Thomson screening $C_{\ell}^{X^\Sc X^\Sc}$ (\textit{pink}), and their cross-correlation $C_{\ell}^{X^\Sc X^\dSc} \propto \varepsilon^2$ (\textit{green}). Notice $C_{\ell}^{T^\dSc T^\dSc}$ dominates due to the additional factor $\propto \bar{T}^2$. Due to the different scaling with $\varepsilon$, the cross-spectrum $C_{\ell}^{T^\Sc T^\dSc}$ in \textit{green} is comparable in magnitude to the \textit{orange} curve around a value $\varepsilon \sim 8 \times 10^{-11}$. In the case of $EE$, the switch happens around $\varepsilon \sim 7 \times 10^{-7}$. The lensed primary CMB spectra (\textit{blue}) and the ILC leftover noise (\textit{purple}) computed at the baseline $\omega_0=30 $ GHz are shown for comparison. The dotted lines represent the instrumental noise $N_{\ell}(\omega)$ for each channel in CMB-S$4$, as defined in Table \ref{tab:noisemodel}. The colors of the dotted lines are the same as their corresponding weights' in Fig.~\ref{fig:weights}.}\label{fig:scalecomp}
\end{figure}

Not only does $\tilde{N}^{T^\dSc T^\dSc}_{\ell}$ do a good job of removing the primary CMB signal, it also gives a lower noise amplitude at high $\ell$ compared to the noisiest frequency channels. This can be seen by comparing the solid purple line in Fig.~\ref{fig:scalecomp} depicting $\tilde{N}^{T^\dSc T^\dSc}_{\ell}$ with the dotted lines that show $N^{TT}_{\ell}(\omega)$ for CMB-S$4$. In the limit of infinitely many frequency measurements, one could in principle fully isolate the frequency-dependent signal from the blackbody component.

Finally, to compute the ILC for the strictly blackbody signal of the Thomson-screened CMB of Eq.~\eqref{eq:TScTSc}, we set $\Omega^{-1}_{ij} = 1$. The residual $\tilde{N}^{T^\Sc T^\Sc}_{\ell}$ in this case is a linear combination of the instrumental noise spectra, once again weighted heavily by the middle frequencies with lowest magnitude, but this time all weights are positive. Again, if we had access to infinitely many frequency channels, the ILC would perfectly recover the Thomson-screened CMB.

\subsection{CMB auto-correlation}\label{sec:forecastauto}

The first method we look at considers the cross-correlation of all dark-screened CMB auto-power spectra. The covariance matrix takes the following form:

\begin{equation}
\mathbf{C}_\ell = \begin{pmatrix}
    \tilde{C}^{T^\dSc T^\dSc}_{\ell} & \tilde{C}^{T^\dSc E^\dSc}_{\ell} & 0 \\
    \tilde{C}^{T^\dSc E^\dSc}_{\ell} & \tilde{C}^{E^\dSc E^\dSc}_{\ell} & 0 \\
    0 & 0 & \tilde{C}^{B^\dSc B^\dSc}_{\ell}
\end{pmatrix},
\end{equation}
where each noise term contained in $\tilde{C}_{\ell}$ is the ILC residual for the particular $T$, $E$ or $B$ measurement. The Fisher matrix is defined as:
\begin{equation}
F_{ij} = f_\sky \sum_\ell \frac{2 \ell + 1}{2} {\rm Tr} \left[ (\mathbf{C}_\ell)^{-1} \cdot \partial_i \mathbf{C}_\ell \cdot (\mathbf{C}_\ell)^{-1} \cdot \partial_j \mathbf{C}_\ell  \right].
\end{equation}
where $(\mathbf{C}_\ell)^{-1}$ is the matrix inverse of $\mathbf{C}_\ell$ evaluated at a fiducial value for parameters $i$ and $j$; $\partial_i \mathbf{C}_\ell$ is the derivative of the covariance matrix with respect to parameter $i$ evaluated at the fiducial parameters $i$ and $j$. The observed sky-fraction is assumed to be $f_\sky = 0.7$ for Planck and $f_\sky =0.5$ for CMB-S4 and CMB-HD. We are interested in expanding the reach for the mixing parameter $\varepsilon$ at a variety of values for $m_\darkph$. Since in this case we only have access to dark screening power spectra, we are constraining $\varepsilon^4$. The Fisher `matrix' has a single entry given by $F_{\varepsilon^4 \varepsilon^4}$. To find the exclusion region, we compute the the 1-sigma constraint on $\varepsilon^4$, given a fiducial value $\varepsilon^4 = 0$: $\sigma_{\varepsilon^4} = 1/\sqrt{F_{\varepsilon^4 \varepsilon^4}}$. The Fisher matrix simplifies in this case to:
\begin{equation}
    F_{\varepsilon^4 \varepsilon^4} |_{\varepsilon^4 = 0} = f_\sky \sum_\ell \frac{2 \ell + 1}{2} \sum_{X = T,E,B} \left( \frac{C_{\ell}^{X^\dSc X^\dSc}(\varepsilon^4=1)}{\tilde{N}_{\ell}^{X^\dSc X^\dSc}} \right)^2,
\end{equation}
and the 1-sigma sensitivity is roughly given by $\sigma_{\varepsilon^4} \sim 1/(S/N)_{\varepsilon^4=0}$.

To obtain the variance on $\varepsilon$, we assume a Gaussian posterior for the probability distribution of the positive real-valued $\varepsilon^4$. The sensitivity on $\varepsilon$ is then well approximated by:
\begin{equation}\label{eq:sigmaCMBAUTO}
    \sigma_{\varepsilon} \simeq 0.7 \, \sigma_{\varepsilon^4}^{1/4} = 0.7 \left[ f_\sky \sum_\ell \frac{2\ell+1}{2} \left( \frac{\bar{T}^2 \, C_\ell^{\tau\tau}(\varepsilon^4=1)}{\tilde{N}^{T^\dSc T^\dSc}_\ell} \right)^2 \right]^{-1/8}.
\end{equation}
Notice from the expression above that the value of $\sigma_{\varepsilon}$ improves as $\sim \ell^{-1/4}$ with the number of modes with significant S/N. This feature is general, regardless of the observable we use in the forecast. In terms of signal, most of the sensitivity of the CMB auto-correlation is due to the $C_{\ell}^{T^\dSc T^\dSc} \sim \bar{T}^2$ term. Finally, the shape of the sensitivity as a function of mass $m_\darkph$ will trace the optical depth monopole as $\bar{\eta}^{-1/2}$. This sensitivity is shown in Fig.~\ref{fig:exclusionplot} for Planck and CMB S$4$. As explained above, the sensitivity is bounded at low dark photon mass from imposing a hard cutoff at the virial radius in each halo, while at high mass it is given by the fact that the halo mass function $n(z,m)$ falls to zero. The boundaries of the contours are also sensitive to our assumptions about the lower bound on the mass of halos described by the AGN gas profiles. We comment on the impact of this modeling uncertainty in Appendix~\ref{appx:modeling}. Overall, the sensitivity is superior to the FIRAS bound due to the dependence on the CMB monopole $\bar{T}$.

\subsection{CMB cross-correlation with a conversion template}\label{sec:CMBwithtemplate}

Cross-correlating the measured CMB with other probes increases the forecasted sensitivity to $\varepsilon$. To investigate the degree of improvement that could eventually be possible, we assume in this section that a perfect template for patchy screening occurring at $z<2$ is available. Such a template could be created from a massive galaxy survey and a detailed model for the relation between galaxy density and ionized gas density, as described in Section~\ref{sec:lssxds}. The choice of $z<2$ is motivated by the redshift range that will be covered by near-term surveys. When referring to the template, we will write $\hat{\tau}$ and its amplitude will depend on a fiducial choice of coupling strength denoted by $\varepsilon^2_0$. In the presence of a dark photon, the measured dark-screened CMB depends on an unknown `true' $\tau \propto \varepsilon^2$.

One can write the first order in $\delta\tau$ contributions to the two-point function between the template $\hat{\tau}$ and the dark-screened temperature $T^\dSc $ from Eq.~\eqref{eq:generalscr} as:
\begin{equation}
	C^{\hat{\tau}T^\dSc}_\ell = \varepsilon^{2}\varepsilon_0^{2} \, \bar{T} \, C^{\hat{\tau}\tau}_\ell(\varepsilon=1,\varepsilon_0=1).
\end{equation}
Notice that another advantage of this method is that it allows us to be sensitive to $\varepsilon^2$ directly, which is the same power of the coupling that appears in the dark screening optical depth monopole (in Section~\ref{sec:firas}). There is no statistically isotropic component of the cross-correlation between polarization and the template, and so we do not discuss them here. Note that for the template power spectrum, anisotropies are calculated up to $z=2$ but the monopole $\hat{\bar{\tau}}$ corresponds to the full contribution up to reionization, \ie $\bar{\tau} = \hat{\bar{\tau}}$. Breaking up the two-point function into an integral over redshift, one can see that $C^{\hat{\tau}\tau}_\ell$ simplifies to $C^{\hat{\tau}\hat{\tau}}_\ell$ for a perfect template. For an imperfect template, this result would include a (scale-dependent) correlation coefficient describing the imperfect overlap of the template with the actual dark screening optical depth.

The covariance matrix assuming a perfect template is the following:
\begin{equation}
    \mathbf{C}_\ell = \begin{pmatrix}
       C^{\hat{\tau} \hat{\tau}}_\ell  & C^{\hat{\tau} T^\dSc}_\ell  \\
       C^{\hat{\tau} T^\dSc}_\ell  & \tilde{C}^{T^\dSc T^\dSc}_\ell
    \end{pmatrix}.
\end{equation}
With a fiducial $\varepsilon \to 0$, the Fisher estimator in this case simplifies to:
\begin{equation}
	\varepsilon_0^4 F_{(\varepsilon/\varepsilon_0)^2 (\varepsilon/\varepsilon_0)^2} = f_\sky \sum_\ell \left(2 \ell + 1\right) \frac{\bar{T}^2 C^{\hat{\tau}\hat{\tau}}_\ell(\varepsilon^2=1)}{\tilde{N}^{T^\dSc T^\dSc}_\ell},
\end{equation}
and the uncertainty on the coupling constant is
\begin{equation}\label{eq:sigmaCMBXTEMP}
    \sigma_{\varepsilon} \approx 0.76 \, \sigma_{\varepsilon^2}^{1/2} = 0.76 \left[ f_\sky \sum_\ell \left(2\ell+1\right) \frac{\bar{T}^2 C_\ell^{\hat{\tau}\hat{\tau}}(\varepsilon^2=1)}{\tilde{N}^{T^\dSc T^\dSc}_\ell} \right]^{-1/4}.
\end{equation}
This sensitivity is plotted in Fig.~\ref{fig:exclusionplot} for the various CMB experiments we consider. Notice here that the sensitivity contour corresponding to this estimator is improved by around one order of magnitude compared to the dark-screened CMB-only result. In short, this is due to the $\varepsilon^2$ scaling instead of $\varepsilon^4$ that brings about a more favorable scaling $\sim \ell^{-1/2}$ with the number of modes that are measured with appreciable S/N.

\subsection{Correlations with Thomson screening, the bispectrum and reconstruction}\label{sec:bispectrumforecast}

Recall from Section~\ref{sec:twoptcorrs} that there are statistically anisotropic contributions to the two-point function between the CMB temperature and polarization anisotropies. These statistically anisotropic contributions form the basis of the reconstruction of the un-screened CMB temperature and polarization anisotropies as well as the Thomson and photon to dark photon optical depth introduced in Section~\ref{sec:reconstruction} and enumerated in Appendix~\ref{appx:qes}. These statistical anisotropies also imply the existence of the three-point functions introduced in Section~\ref{sec:bispectra}. In this section, we explore how this non-Gaussian information can be used to search for dark photons.

We first consider the $\ev{T^\dSc T^\Sc T^\Sc}$ bispectrum, which is proportional to $C_\ell^{\tau \tau^\Th} (\omega)$. Factoring out the dependence on $\varepsilon$:
\begin{equation}
B_{\ell \ell^{\prime} \ell^{\prime\prime}}^{T^\dSc T^\Sc T^\Sc} = \varepsilon^2 \bar{T} \sqrt{2\ell^{\prime\prime} + 1} W_{\ell \ell^{\prime} \ell^{\prime\prime}}^{0 0 0} \left( C_{\ell^{\prime}}^{TT} + C_{\ell^{\prime\prime}}^{TT}   \right) C_\ell^{\tau \tau^\Th} (\varepsilon=1),
\end{equation}
we see that forecasting the limits on $\varepsilon^2$ is a straightforward exercise in estimating the amplitude of this bispectrum, a problem whose optimal solution is already known from studies of primordial non-Gaussianity (for an overview, see \eg\cite{Liguori_2010}). The simplest bispectrum estimator is 
\begin{equation}
\hat{\varepsilon}^2 = \sigma_{\varepsilon^2}^2 \sum_{\ell m} \sum_{\ell^{\prime} m^{\prime}} \sum_{\ell^{\prime\prime} m^{\prime\prime}} \frac{B_{\ell \ell^{\prime} \ell^{\prime\prime}}^{T^\dSc T^\Sc T^\Sc} (\varepsilon^2 = 1)}{\tilde{C}^{T^\dSc T^\dSc}_{\ell} \tilde{C}^{T^\Sc T^\Sc}_{\ell^{\prime}} \tilde{C}^{T^\Sc T^\Sc}_{\ell^{\prime\prime}}} 	\begin{pmatrix}
			\ell & \ell^{\prime} & \ell^{\prime\prime} \\
			m & m^{\prime} & -m^{\prime\prime} 
	\end{pmatrix}
 T^\dSc_{\ell m} T^\Sc_{\ell^{\prime} m^{\prime}} T^\Sc_{\ell^{\prime\prime} m^{\prime\prime}},
\end{equation}
where the resulting constraint is related to the estimator variance by
\begin{equation}\label{eq:TTT_variance}
    \sigma_{\varepsilon} = 0.76 \sqrt{\sigma_{\varepsilon^2}} = 0.76 \left[ f_\sky \sum_{\ell \ell^{\prime} \ell^{\prime\prime}} \frac{1}{2} \frac{\left( B_{\ell \ell^{\prime} \ell^{\prime\prime}}^{T^\dSc T^\Sc T^\Sc} (\varepsilon^2 = 1)  \right)^2}{\tilde{N}^{T^\dSc T^\dSc}_{\ell} \tilde{N}^{T^\Sc T^\Sc}_{\ell^{\prime}} \tilde{N}^{T^\Sc T^\Sc}_{\ell^{\prime\prime}}} \right]^{-1/4}.
\end{equation}
Note that just as for the quadratic estimators discussed above, the weights in the bispectrum estimator are constructed from models for $C_{\ell}^{TT}$ and $C_\ell^{\tau \tau^\Th} (\omega,\varepsilon=1)$. The factor of 1/2 comes from having two indistinguishable $T^\Sc $ fields.

An alternative starting point would be to first use $T^\dSc $ and $T^\Sc $ to reconstruct $T$ (see Eq.~\eqref{eq:TQE}) and then correlate this with $T^\Sc $. If we assume that the model for $C_\ell^{\tau \tau^\Th}$ is known up to the value of $\varepsilon^2$ then the reconstruction of $T$ will have a bias of $\varepsilon^2$ such that:
\begin{equation}
    \ev{\hat{T}_{LM} \hat{T}_{LM}} \simeq \varepsilon^4 C_{L}^{TT} + N_{L}^{T; T^\dSc T^\Sc} (\varepsilon=1), \quad \quad \ev{\hat{T}_{LM} T^\Sc_{LM}} \simeq \varepsilon^2 C_{L}^{TT},
\end{equation}
where we have neglected small contributions that appear at higher order in $\varepsilon$, higher order in $\tau^\Th $ or lower order in $\bar{T}$. Computing the Fisher matrix we have:
\begin{equation}
\begin{aligned}
    F_{\varepsilon^2 \varepsilon^2} &= \sum_L f_\sky (2L+1) \frac{(C_{L}^{TT})^2}{N_{L}^{T; T^\dSc T^\Sc} (\varepsilon=1) \tilde{C}^{T^\Sc T^\Sc}_{L}} \\
    &= \sum_{\ell \ell^{\prime} L} \bar{T}^2 f_\sky (2L+1) (W_{\ell \ell^{\prime} L}^{0 0 0})^2 \frac{C_\ell^{\tau \tau^\Th} (\varepsilon=1)^2 (C_L^{TT})^2}{\tilde{C}^{T^\dSc T^\dSc}_{\ell} \tilde{C}^{T^\Sc T^\Sc}_{\ell^{\prime}} \tilde{C}^{T^\Sc T^\Sc}_{L}},
\end{aligned}
\end{equation}
It is informative to compare this result to the variance on the bispectrum estimator in Eq.~\eqref{eq:TTT_variance}. Since $\ell^2 C_\ell^{TT}$ falls with $\ell$ (at sufficiently high $\ell$), the two results agree in the limit where the dominant contributions to the sum in Eq.~\eqref{eq:TTT_variance} come from squeezed configurations with $\ell \gg 1$ -- the triangle rule then implies that either the term proportional to $C_{\ell^{\prime}}^{TT}$ or $C_{\ell^{\prime\prime}}^{TT}$ dominates the bispectrum. Said differently, because the reconstruction of $T$ improves by measuring many small-scale modes, we mainly capture information about squeezed configurations of the bispectrum where $T^\Sc $ is evaluated at low-$\ell$ while the other power of $T^\Sc $ and $T^\dSc $ are evaluated at high-$\ell$. We expect to be in this regime, since screening occurs mainly on small scales (\eg it is associated with halos). Note that a completely analogous situation arises in kinetic Sunyaev Zel'dovich velocity reconstruction, as described in detail in Ref.~\cite{smith_ksz_2018}. 

Including polarization, the best sensitivity on $\varepsilon$ can be obtained from the $\ev{T^\dSc E^\Sc B^\Sc}$ bispectrum, where the estimator variance is
\begin{equation}\label{eq:TEB_variance}
    \sigma_{\varepsilon} = 0.76\left[f_\sky \sum_{\ell \ell^{\prime} \ell^{\prime\prime}} \frac{\left( B_{\ell \ell^{\prime} \ell^{\prime\prime}}^{T^\dSc E^\Sc B^\Sc} (\varepsilon^2 = 1)  \right)^2}{\tilde{N}^{T^\dSc T^\dSc}_{\ell} \tilde{N}^{E^\Sc E^\Sc}_{\ell^{\prime}} \tilde{N}^{B^\Sc B^\Sc}_{\ell^{\prime\prime}}} \right]^{-1/4},
\end{equation}
where $B_{\ell \ell^{\prime} \ell^{\prime\prime}}^{T^\dSc E^\Sc B^\Sc}$ was defined in Eq.~\eqref{eq:TEB_bispectrum}. This bispectrum can yield a competitive sensitivity compared to the temperature-only bispectrum above since  $C^{B^\Sc B^\Sc}_{\ell^{\prime\prime}} \ll C^{EE}_{\ell^{\prime\prime}}$ in the signal-dominated regime.

The sensitivities to $\varepsilon$ using the $\ev{T^\dSc T^\Sc T^\Sc}$ and $\ev{T^\dSc E^\Sc B^\Sc}$ bispectra are shown in Fig.~\ref{fig:exclusionplot}. The sensitivity of the bispectrum estimator for the experimental configurations studied here is slightly weaker than the result from the CMB auto-correlation. This is due to the smallness of $\tau^\Th $, which, parametrically, suppresses the sensitivity compared to CMB auto-correlation as well as LSS cross-correlation by an $\varepsilon$-independent factor of $(C_\ell^{\tau \tau^\Th})^2/C_\ell^{\tau \tau} \sim 10^{-11} - 10^{-12}$, depending on $m_\darkph$ (see Fig.~\ref{fig:screening_corrs}). However, this estimator also scales as $\varepsilon^2$, and brings about the most favorable scaling $\sim \ell^{-3/4}$ with the number of modes measured at significant S/N. Notable is also the improvement of $\ev{T^\dSc E^\Sc B^\Sc}$ over $\ev{T^\dSc T^\Sc T^\Sc}$ between the analyses for Planck and HD-like noise, due to predicted noise drop especially in polarization measurements.

To explore the improved sensitivity on $\varepsilon$ from small-scale modes we evaluate he bispectrum estimators~\eqref{eq:TTT_variance} and~\eqref{eq:TEB_variance} for Planck and CMB-HD. For Planck we capture most of the SNR by summing from $\ell =2$ up to $\ell=3000$, while for CMB-HD there  is significant SNR up to $\ell=6000$. From Fig.~\ref{fig:Cells}, at large dark photon mass, there is a lot of structure at high $\ell$ in the dark screening optical depth coming from the $1$-halo term. Due to the favorable scaling with $\ell$ in the bispectrum constraints, the CMB-HD sensitivity is greatly enhanced by the additional support on small scales of the cross-power spectrum, despite the decreasing monopole $\bar{\tau}$ and the smallness of $\bar{\tau}^\Th $. Note that if we extended the range of halo masses in our halo model below the conservative $10^{11} M_{\odot}$, the sensitivity on $\varepsilon$ would increase. However, this would also bring about larger uncertainties in what the appropriate gas profile to be used on those small scales might be.

\section{Conclusion and remarks}\label{sec:conclusions}

\begin{figure}[ht!]
    \centering
    \includegraphics[width=0.85\textwidth]{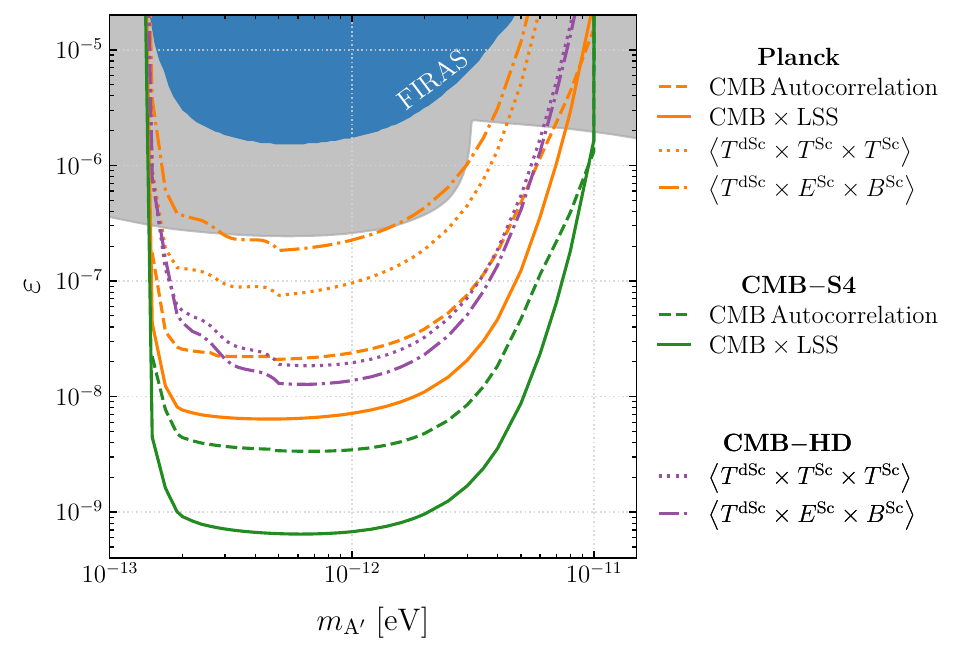}
    \caption{The projected sensitivity of several estimators for Planck, as well as future surveys CMB-S4 and CMB-HD. The gray shaded region is excluded from analysis~\cite{caputo_dark_2020, Mirizzi:2009iz} with data from COBE/FIRAS~\cite{COBEFixsen}. The blue shaded region shows the equivalent constraint using our model, as explained in Section~\ref{sec:firas}. The solid lines show the projected sensitivity of the CMB auto-correlation functions in Section~\ref{sec:twoptcorrs} and~\ref{sec:forecastauto} with uncertainty given by Eq.~\eqref{eq:sigmaCMBAUTO}. The dashed contours were computed using Eq.~\eqref{eq:sigmaCMBXTEMP} and show the projected sensitivity from the cross-correlation between CMB and LSS in Section~\ref{sec:lssxds} and~\ref{sec:CMBwithtemplate}. The dotted and dot-dashed contours given by Eq.~\eqref{eq:TTT_variance} and Eq.~\eqref{eq:TEB_variance} respectively show the projected sensitivities of the bispectra presented in Section~\ref{sec:bispectra} and~\ref{sec:bispectrumforecast}. The projected sensitivity from CMB-S4 and CMB-HD are similar for the CMB auto-correlation and cross-correlation between CMB and LSS, whereas bispectra sensitivity with CMB-HD is superior to CMB-S4. We used $f_\sky = 0.7$ for Planck and $f_\sky =0.5$ for CMB-S4 and CMB-HD.}\label{fig:exclusionplot} 
\end{figure}

In this paper, we have studied the resonant conversion of CMB photons to a hypothetical dark photon inside LSS at low redshift. Conversion leads to a frequency dependent patchy `dark' screening of the CMB. Observationally, dark screening manifests as an anisotropic and frequency dependent optical depth $\tau(\omega, \hat{n})$, which can in principle be extracted from CMB data. The sensitivity to $\varepsilon$ for a variety of two- and three-point correlation functions are shown in Fig.~\ref{fig:exclusionplot}, which we summarize in the following (see also Table~\ref{table:correlation}).

The patchy dark screening optical depth $\tau(\omega,\hat{n})$ can be measured using CMB data alone. The global signal, \ie the $\overline{\tau}(\omega)$ monopole, leads to a spectral distortion which is constrained by COBE/FIRAS~\cite{COBEFixsen} and can potentially be measured with future experiments targeting spectral distortions. The constraint from conversion at low-redshift in non-linear structure that we obtain using existing FIRAS data~\cite{COBEFixsen} is consistent with previous limits obtained from conversion over a wider range of redshift and a different treatment of inhomogeneities.

Extending previous analyses, we have demonstrated that CMB and LSS correlation functions are in principle a far more powerful probe of photon to dark photon conversion. The two-point function of dark screening ($\ev{\tau(\omega,\hat{n})\tau(\omega,\hat{n}^{\prime})}$) can be extracted from existing and future CMB data by measuring $\ev{T^\dSc (\omega,\hat{n})T^\dSc (\omega,\hat{n}^{\prime})}$. The patchy dark screening map $T^\dSc (\omega,\hat{n})$ can be separated from the blackbody CMB by taking advantage of multi-frequency observations of the CMB.  As shown in Fig.~\ref{fig:exclusionplot}, such a correlation function can be more sensitive than the existing FIRAS constraints by up to $2$ orders of magnitude despite scaling with the small kinetic mixing parameter as $\varepsilon^4$.

A better reach on the mixing parameter can be obtained by cross-correlating $\tau(\omega,\hat{n})$ with other observables that are sensitive to the underlying distribution of electron density in the Universe. One such correlation function is  $\ev{T^\dSc (\omega,\hat{n}) \hat{\tau}(\hat{n}^{\prime})}$, where $\hat{\tau}(\hat{n}^{\prime})$ is a template for the dark screening optical depth. This correlation function takes advantage of the fact that for a dark photon with mass $\lesssim 10^{-12}$ eV, conversion mostly happens at late times, creating a strong correlation between tracers of LSS and patchy dark screening. As shown in Fig.~\ref{fig:exclusionplot}, the reach obtained from this correlator can be an order of magnitude better than the CMB-only result.

Cross-correlation functions can also be constructed from the dark-screened and Thomson-screened CMB alone. Qualitatively, these correlation functions can be understood as a correlation between the dark screening $\tau(\omega,\hat{n})$ and the Thomson screening optical depth $\tau^\Th $ from halos. Three-point correlation functions offer the best sensitivity to this correlation. In particular, the $\ev{T^\dSc T^\Sc T^\Sc}$ and $\ev{T^\dSc E^\Sc B^\Sc}$ bispectra (both shown in Fig.~\ref{fig:exclusionplot}) offer the best sensitivity, and are comparable to the reach anticipated with CMB auto-correlation functions. Compared to the cross-correlation with a template, the reduction in sensitivity is mainly a result of the smallness of the Thomson screening optical depth $\tau^\Th $, which suggests that these three-point correlation functions could be better probes of photon to dark photon conversion in the weakly inhomogeneous Universe around recombination, when $\tau^\Th $ was much larger. Compared to the CMB auto-correlation functions, the sensitivity of these three-point correlation function scales much more favorably with $\ell$, and hence improves with increased  sensitivity and high resolution -- the regime targeted by future surveys. The similarity in the reach of the CMB auto-correlation functions and these three-point correlation functions for future CMB survey is a numerical coincidence, and the relative strength of these two methods in a real data analysis likely depend on systematics and foregrounds, an investigation that we postpone to future work. Finally, we note that in the event of a detection, a combination of the two-point and three-point functions can be used to break degeneracies between the dark photon mass, kinetic mixing parameter and electron density profile, which is essential for extracting detailed information about the dark screening optical depth and how it correlates with the distribution of ionized gas.
The methodologies we developed in this paper can be used to search for conversions of photon to dark photon in various other environment in the early-Universe, the details of which we will work out in a few follow up studies.

The study presented in this paper is a novel example of using cross-correlations between an observable in the Standard Model (SM) of cosmology ($\Lambda$CDM) and signals of a model beyond the Standard Model (BSM) of particle physics. Experimentally, the measurement of these correlators is enabled by the rapid improvement of cosmological experiments. Theoretically, these $\ev{{\rm SM} \times {\rm BSM}}$ correlators allow us to use the ultra-high precision cosmological data on the anisotropic Universe to study BSM signals at the same order of the small BSM parameter as the monopole signal $\ev{{\rm BSM}}$. 
We expect similar $\ev{{\rm SM} \times {\rm BSM}}$ correlators will allow us to better probe other BSM signals with the rapidly improving cosmological CMB and LSS datasets, and take advantage of the synergy between the upcoming CMB experiments like the Simons Observatory~\cite{Ade:2018sbj}, as well as future experiments CMB-S4~\cite{abazajian_cmb-s4_2016} and CMB-HD~\cite{Sehgal:2019ewc}, with upcoming LSS surveys like DESI~\cite{DESI:2019jxc}, Euclid~\cite{Laureijs2011}, and LSST~\cite{0912.0201}. Constructing new observables of this kind can allow us to better search for new interactions between the Standard Model and dark sectors, including dark matter annihilation, decay, and mixing between the visible and dark sector particles.

\acknowledgments

We thank Liang Dai, Neal Dalal, Hongwan Liu and Kendrick Smith for helpful conversations, and Gustavo Marques-Tavares and Cristina Mondino for comments on the draft. MCJ and JH are supported by the Natural Sciences and Engineering Research Council of Canada through a Discovery Grant. This research was supported in part by Perimeter Institute for Theoretical Physics. Research at Perimeter Institute is supported by the Government of Canada through the Department of Innovation, Science and Economic Development Canada and by the Province of Ontario through the Ministry of Research, Innovation and Science. 

\appendix

\section{Modeling dark screening in a dark matter halo}\label{appx:modeling}

\subsection{Dark matter halo models}

The halo model of large scale structure assumes that all matter in the Universe is stored in virialized halos whose physical properties are fully described by the mass contained within their boundary (defined e.g. by the virial radius). Galaxies occupy dark matter halos and in doing so they act as tracers for the underlying dark matter distribution. The halo model is a semi-analytical framework used for understanding the non-linear structure of the matter distribution. There are two principal quantities needed to make predictions: the halo mass-function and the halo density profile. The former describes the halo number density as a function of mass and redshift. The latter describes how mass is distributed within each halo. Unlike the mass-function, it is not universal, meaning that it depends on cosmology and astrophysics. To make matters more simple, these expressions are assumed to be a function of a few variables such as mass, redshift and halo radius, and the parameters that enter these expressions are generally obtained from a mix of analytic predictions, simulations and even data. Other quantities that need to be specified in a halo model are the halo bias function, which to first order is fully determined by the mass function, as well as a concentration-mass relation that gives a characteristic scale radius for the halo density profile. Useful reviews on halo models are \eg\cite{cooray_halo_2002, asgari2023halo}.

To perform halo model computations throughout this paper we assume the mass-function of~\cite{tinker2008} that fixes the bias function~\cite{Tinker_2010}, and the concentration-mass relation from~\cite{duffycon}, which fixes the free parameters in the halo density profile. We also work under the assumption that the halo boundary is its virial radius, so that the halo mass is defined in a sphere of radius $r_\vir$. Our halo-model computations are done using a modified version of the code \textit{hmvec}\footnote{\url{https://github.com/simonsobs/hmvec}}. A detailed description of the assumptions that enter this code can be found in the Appendix B of~\cite{smith_ksz_2018}.

For our modeling, we use $50$ redshift bins of equal comoving radial width in the range $ z= \left[0.01, z^{\rm max} \right] $, where the reionization redshift $z^{\rm max} \in \{2,6,10 \}$. The first case was used with the purpose of obtaining a template angular power spectrum for the distribution of galaxies, as measured by a futuristic LSS survey. The latter two provide a conservative range for when reionization was completed. Additionally, the cosmology (to be precise, the linear matter power spectrum) is defined for $10^4$ comoving wavenumber bins $k$ logarithmically spaced in the $10^{-4}-10^{3} {\rm \, Mpc^{-1}}$ range. Finally, we considered $100$ halo mass bins logarithmically spaced in the $10^{11} - 10^{17} {\rm \, M_{\odot}}$ interval. The lower bound is a conservative mass limit of halos with feedback processes significant enough to disrupt the gas profile. For the upper bound, the number density of halos in the halo model is exponentially suppressed with mass, and using the halo mass function in~\cite{tinker2008}, one can expect less than one halo with a mass $> 3\times 10^{16} M_{\odot}$ in a volume the size of the Hubble sphere. In other words, we consider all halos with mass $\geq 10^{11} M_{\odot}$.

\subsection{Charged particle density profiles}

In our calculations we considered an idealized scenario where reionization takes place instantaneously: the ionization fraction goes from zero to unity at exactly $z^\reio$. We compared the case where $z^\reio=6$ and $z^\reio=10$ and found no significant difference \eg see Fig.~\ref{fig:monopole}. However, if we relax the assumption about the lower mass bound of halos where conversion can happen, this no longer holds true. In fact, we found that there is an order $1-10\%$ difference in the magnitude of the optical depth monopole $\bar{\tau}$ for masses above $m_\darkph > 10^{-12}$ eV when we consider halos with mass down to $m = 10^{9} M_{\odot}$. This would not affect the contours in Fig.~\ref{fig:exclusionplot} significantly, but represents an example of a source of error in the modeling.

\begin{figure}[ht!]
    \centering
    \includegraphics[width=0.6\textwidth]{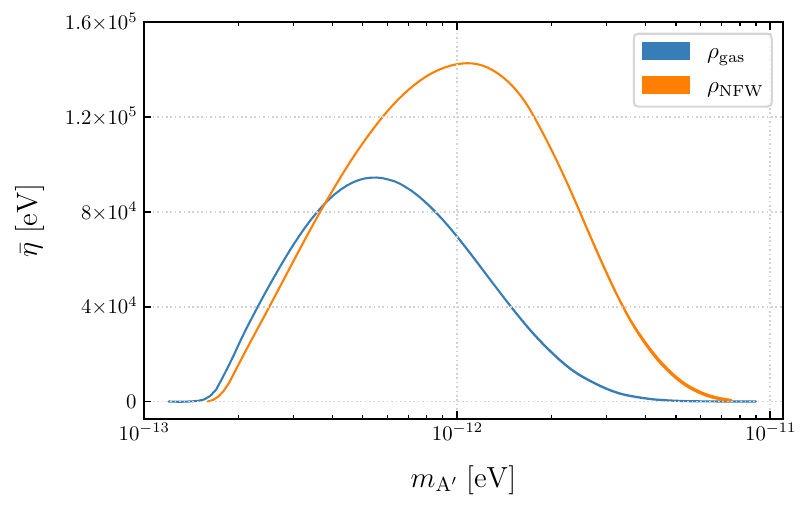}
    \caption{Comparison between the average dimensionful optical depth for the NFW and gas profiles from~\cite{1996ApJ462563N} and~\cite{Battaglia_2016}, respectively. Here two cases are plotted: the top edge in each line corresponds to taking $z^\reio=10$, while the bottom edge is computed for $z^\reio=6$. In either case, assumptions about the end of reionization are of little importance. As described in the main text, the shape and magnitude of the monopole $\bar{\eta}$ plays an important role in determining the shape and reach in sensitivity of $\varepsilon$, for any experiment and forecast method. Therefore, the difference in magnitude between the two curves depicted here illustrates the importance of modeling the charged electron density in halos.}\label{fig:compare_NFW_gas_monopoles}
\end{figure}

\begin{figure}[ht!]
    \centering
    \includegraphics[width=0.6\textwidth]{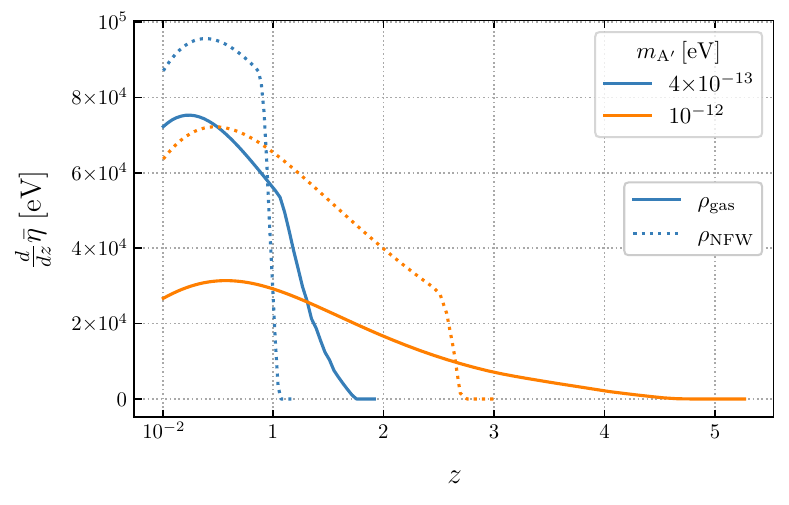}
    \caption{Comparison between the differential average optical depth per redshift bin for two different choices of density profile. Notice that the range of relevant redshift bins is sensitive to this choice, as well as the magnitude of the signal. For example, if we have an LSS template up to $z=2$ as assumed in the main text, then we expect the cross-correlation with CMB dark screening to be less strong at $m_\darkph \gtrsim 10^{-12}$ eV if gas traces dark matter in the absence of AGN feedback processes.}\label{fig:compare_NFW_gas_dtaudz}
\end{figure}

A potentially more important assumption has to do with the precise choice of density profile for the charged electrons on small scales. We explore this in more detail in the present section. As an example, the density of dark matter is consistently greater towards the core of halos (in fact, it is unbounded at $r \to 0$) than the gas density. Repeating all calculations under the assumption that electrons follow the NFW profile~\cite{1996ApJ462563N} in Eq.~\ref{eq:rhoNFW} instead of expression~\eqref{eq:rhogas} for gas from~\cite{Battaglia_2016} that we have been using in the main text, we obtain the results shown in Fig.~\ref{fig:compare_NFW_gas_monopoles} for the sky-averaged dark screening optical depth monopole, and in Fig.~\ref{fig:compare_NFW_gas_dtaudz} for the differential (dimensionful) monopole. Notice that the greatest disagreement between the curves lies in the upper half of the dark photon masses we considered. This is related to the differences in the two density profiles on small scales which we elaborate on further.

\begin{figure}[ht!]
    \centering
    \includegraphics[width=1.\textwidth]{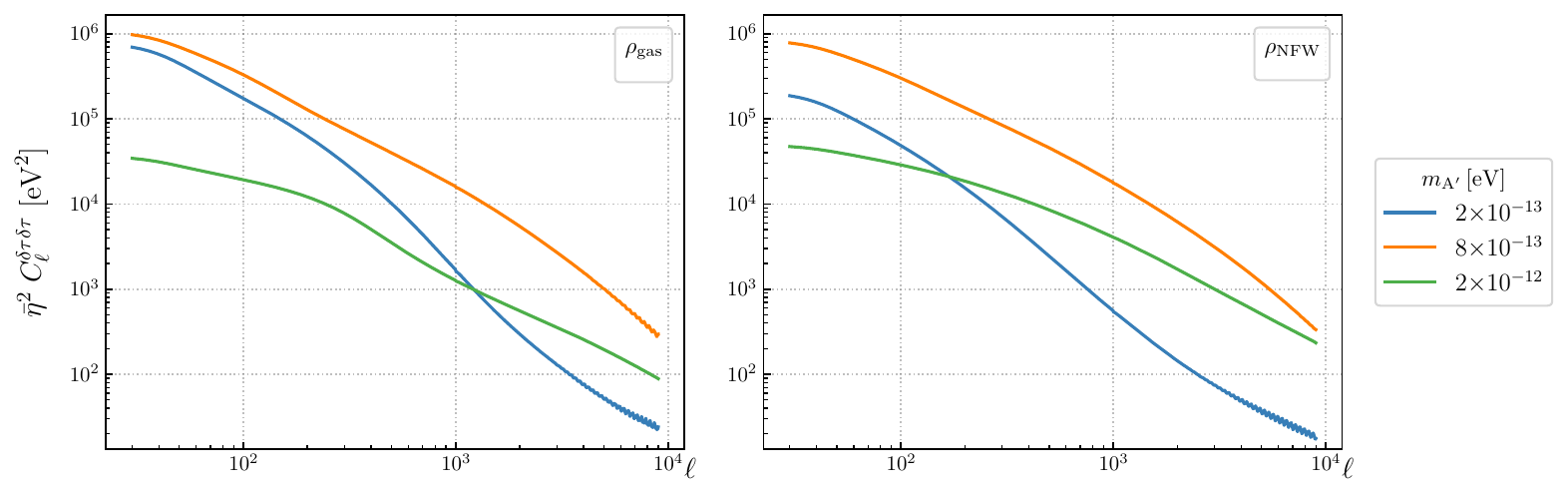}
    \caption{The left panel shows the dark screening power spectra for three choices of dark photon mass, given a conversion model that assumes electrons in halos follow $\rho_\gas$, as in the main text. On the right we show the equivalent spectra where $\rho_\NFW$ is assumed. The $\rho_\gas$ profile is flat near the core and drops slower with radius than $\rho_\NFW$, hence at low $m_\darkph$ we expect more power on larger scales on the left side. Meanwhile, $\rho_\NFW$ favours larger densities near the halo cores, so at large mass $m_\darkph$ we expect more power on smaller scales. All spectra are multiplied by their respective monopole $\bar{\eta}_\gas$ and $\bar{\eta}_\NFW$ depicted in Fig.~\ref{fig:compare_NFW_gas_monopoles}.}\label{fig:compare_NFW_gas_celltautau}
\end{figure}

Since the NFW profile is unbounded at $r\to 0$, we imposed that no resonant conversion happens below the scale radius $r_s$ of any halo, effectively adding a factor of $\Theta(r_s-r_\res)$ in the expression for the radial probability in~\eqref{eq:separable}. The NFW profile is monotonically decreasing with radius, hence this approximation excludes halos where the resonance condition is met near their core. As is evident in Fig.~\ref{fig:compare_NFW_gas_monopoles}, the constraint is most relevant for the production of heavy dark photons, which require the largest overdensities. The same quantitative difference between density profiles shows up in the shape of the angular two-point functions, which is depicted in Fig.~\ref{fig:compare_NFW_gas_celltautau}. Here, the relevant quantity is the ratio between the $1$-halo and $2$-halo terms. The distribution of power on small angular scales is influenced by the shape of the density profiles. However, the overall magnitude is given by the corresponding monopole at every mass.

For consistency reasons, we also assume that the NFW model breaks down at the scale radius of the Milky Way, which leads to a hard upper boundary on the range of dark photon masses that can be considered in this case, given by $m_\darkph \propto \sqrt{\rho^\NFW(r_s^\MW)} \approx 2.86\times 10^{-12}$ eV. As mentioned in the main text, for the Milky Way we assuming a virial radius and virial mass from~\cite{posti_mass_2019}, and the concentration-mass relation at $z=0$ from~\cite{2014MNRAS.441.3359D} which can be used to compute the scale radius via $r_\vir^\MW = c^\MW(m^\MW) r_s^\MW$.

\begin{figure}[ht!]
    \centering
    \includegraphics[width=0.7\textwidth]{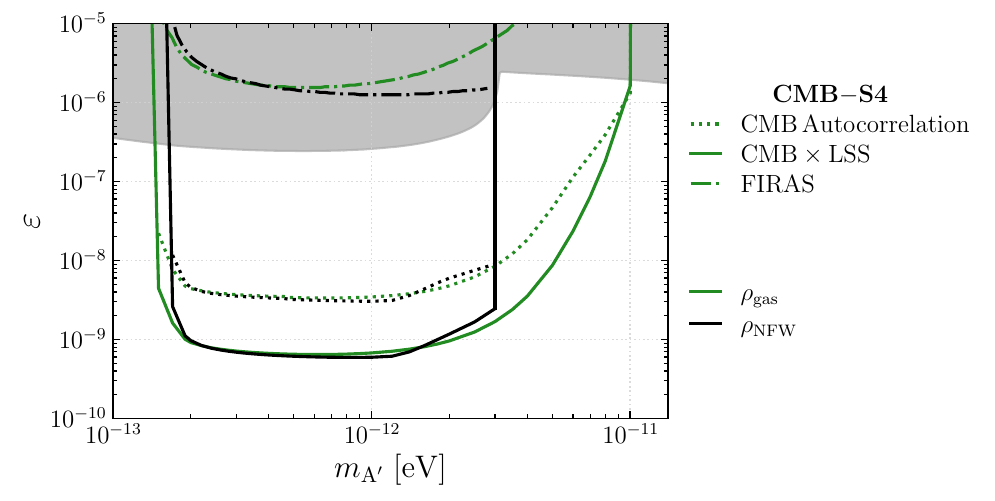}
    \caption{Comparison between sensitivity given by two different profiles for the distribution of electrons inside halos, assuming CMB-S$4$-like noise. The green contours are the same as those shown in Fig.~\ref{fig:exclusionplot} for the `AGN feedback' gas profile~\cite{Battaglia_2016}, while the black contours show the equivalent constraints starting from the NFW density profile~\cite{1996ApJ462563N}. As previously, the FIRAS constraint is computed taking into account the total average optical depth, including the Milky Way contribution, which for NFW is monotonically increasing with mass and dominates over the extra-galactic component at the upper mass end. The solid black boundary at $m_\darkph \approx 3\times 10^{-12}$ eV is imposed by assuming the NFW model breaks down where $m^2_\darkph$ meets the resonance condition on scales less than the Milky Way scale radius $r_s^\MW$. For the contours that take into account the CMB anisotropies, we point out that the match between the contours shows that our model assumptions are robust on scales between the virial radius and the scale radius. However, the precise modeling of the gas profile around the core regions of halos will ultimately dictate the sensitivity on $\varepsilon$ at larger dark photon masses.}\label{fig:exclusionplot_gasvsNFW}
    \end{figure}

Overall, it appears that the effect of changing the density profile modeling on the overall sensitivity on $\varepsilon$ is minimal, as shown in Fig.~\ref{fig:exclusionplot_gasvsNFW}. This suggests that our projection is relatively insensitive to the exact electron density model for masses below the mass limit set by the Milky Way. This is a indication that the assumptions going into our forecast analysis are reasonable and robust, at least on scales above the scale radii of halos. However, the precise modeling of the gas profile around the core regions of halos will ultimately dictate the sensitivity on $\varepsilon$ at larger dark photon masses. On the other hand, for lower dark photon masses, an improved sensitivity can be reached with improving knowledge of the electron density profile in the region right outside the virial radius of a halo~\cite{ACTgasthermo}. Improving the reach by more than half an order of magnitude in mass towards lower masses is likely with new data from ACT DR6 and DESI.

To conclude, in recent years there has been significant progress in our understanding of the electron density distributions and fluctuations in halos, from new experimental measurements~\cite{ACTgasthermo} and numerical simulation~\cite{Garaldi:2021gfo}. Prospective studies with cosmological data and CHIME/FRB~\cite{Madhavacheril:2019buy} will further shrink these uncertainties, and potentially allow us to have more robust projections on the sensitivity at higher dark photon masses.

\section{Correlation functions of dark screening}\label{appx:2pf}

In this appendix, we derive in detail the angular power spectrum of the two-point auto-correlation of optical depth fluctuations. We use the notation conventions from the main text. First, we recall that the optical depth can be written as a sum of a homogeneous and an anisotropic part:
\begin{equation}
    \tau(\chi, \hat{n}) = \bar{\tau}(\varepsilon, \omega) \left[1 + \delta\tau(\chi, \hat{n}) \right].
\end{equation}
To fix the notation, we define the volume elements $\dd^3\bm{k} = k^2 \, \dd k \, \dd^2 \hat{n}_k$ and $\dd^3\bm{\chi} = \chi^2 \dd\chi \, \dd^2 \hat{n}$, where the solid angle $\dd^2\hat{n} = \sin\theta \dd\theta \dd\phi$. The two-point function of the overall optical depth is defined in configuration space as follows:
\begin{equation}
    \ev{\tau^*(\bm{\chi_1})\tau(\bm{\chi_2})} = \bar{\tau}(\varepsilon, \omega)^2 \left(1 + \xi^{\delta\tau\delta\tau}(\bm{\chi_1}, \bm{\chi_2}) \right),
\end{equation}
where we introduced the notation $\xi^{\delta\tau\delta\tau} = \ev{\delta\tau^*(\bm{\chi_1})\delta\tau(\bm{\chi_2})}$ to represent the two-point auto-correlation of anisotropies. The physical interpretation is the following: given that a photon following trajectory $\hat{n}_1$ on the observer's sky undergoes resonant conversion in a halo with mass $m_1$ at redshift $z_1$, what is the chance that another photon traveling along $\hat{n}_2$ converts as well? In the halo model, this probability is a sum of two terms:
\begin{equation}
    \xi^{\delta\tau\delta\tau} = \xi^{1\halo}(\chi_1, \hat{n}_1, \chi_1, \hat{n}_2) + \xi^{2\halo}(\chi_1, \hat{n}_1, \chi_2, \hat{n}_2).
\end{equation}
The first describes the case where the two photon conversions happen within the same halo centered at $\chi(z_1)$, and the second term represents the contribution from two different halos at $\chi(z_1)$ and $\chi(z_2)$. To derive the expressions for the sky-averaged power spectra, we will need to perform a sum over all halos in the halo model, project onto spherical harmonics, then integrate over the Hubble volume. In the derivations below we sometimes leave the mass and redshift dependence implicit to simplify notation. We also use redshift $z$ and comoving distance $\chi(z)$ interchangeably to denote the parametric dependence on time of various quantities.

\subsection{One-halo term}\label{appx:2pf1halo}

We start by computing the $1$-halo term. Labeling all halos within each redshift/comoving distance bin by $i$, we get:
\begin{equation}\label{eq:2pf1haloconfig}
\begin{aligned}
	\bar{\tau}^2 \xi^{1\halo} &= \ev{\sum_i P^{i \, *}_{\gamma \to \darkph} (\chi, \hat{n}_1, m_i) P^i_{\gamma \to \darkph} (\chi, \hat{n}_2, m_i)} \\
		&= \ev{\sum_i \int \dd m \, \delta(m-m_i) \int \dd^2 \hat{n} \, \delta^2 (\hat{n}-\hat{n}_i) \left| P(\chi, m)\right|^2 \, u^*(\hat{n}_1 - \hat{n}) \, u(\hat{n}_2 - \hat{n})}  \\
		&= \int \dd m \, n(\chi,m) \left|P(\chi, m)\right|^2 \int \dd^2 \hat{n} \, u^*(\hat{n}_1 - \hat{n}) \, u(\hat{n}_2 - \hat{n}).
\end{aligned}
\end{equation}
In the last line, we used the expression for the average halo number density from Eq.~\eqref{eq:ndens}. Only the second integral is now angle-dependent, so we project it onto spherical harmonics:
\begin{equation}
\begin{aligned}\label{eq:1hbeast}
    \int \dd^2 \hat{n} \, &u^{*}(\hat{n}_1 - \hat{n} | \chi, m) \, u(\hat{n}_2 - \hat{n} | \chi, m) = \\
    &= \int \dd^2 \hat{n} \sum_{\ell^{\prime} m^{\prime}} u^{*}_{\ell^{\prime} m^{\prime}}(\chi, m) Y^{*}_{\ell^{\prime} m^{\prime}}(\hat{n}_1 - \hat{n} ) \sum_{\ell^{\prime\prime} m^{\prime\prime}} u_{\ell^{\prime\prime} m^{\prime\prime}}(\chi, m) Y_{\ell^{\prime\prime} m^{\prime\prime}}(\hat{n}_2 - \hat{n} ) \\
    &= \int \dd^2 \hat{n} \sum_{\ell^{\prime} \ell^{\prime\prime}} u^{*}_{\ell^{\prime} 0} \, Y^{*}_{\ell^{\prime} 0}(\hat{n} - \hat{n}_1) \, u_{\ell^{\prime\prime} 0} \, Y_{\ell^{\prime\prime} 0}(\hat{n} - \hat{n}_2) \\
    &= \int \dd^2 \hat{n} \sum_{\ell^{\prime} \ell^{\prime\prime}} u^{*}_{\ell^{\prime} 0} u_{\ell^{\prime\prime} 0} \sum_{m^{\prime}} \mathcal{D}^{\ell^{\prime} \, *}_{0 m^{\prime}}(-\hat{n}_1) \, Y^{*}_{\ell^{\prime} m^{\prime}}(\hat{n}) \sum_{m^{\prime\prime}} \mathcal{D}^{\ell^{\prime\prime}}_{0 m^{\prime\prime}}(-\hat{n}_2) \, Y_{\ell^{\prime\prime} m^{\prime\prime}}(\hat{n}) \\
    &= \sum_{\ell^{\prime} \ell^{\prime\prime}} u^{*}_{\ell^{\prime} 0} u_{\ell^{\prime\prime} 0} \sum_{m^{\prime} m^{\prime\prime}} \mathcal{D}^{\ell^{\prime} \, *}_{0 m^{\prime}}(-\hat{n}_1) \, \mathcal{D}^{\ell^{\prime\prime}}_{0 m^{\prime\prime}}(-\hat{n}_2) \, \delta_{\ell^{\prime} \ell^{\prime\prime}} \, \delta_{m^{\prime} m^{\prime\prime}} \\
    &= \sum_{\ell^{\prime} m^{\prime}} | u_{\ell^{\prime} 0} |^2 \, \mathcal{D}^{\ell^{\prime} \, *}_{0 m^{\prime}}(-\hat{n}_1) \, \mathcal{D}^{\ell^{\prime}}_{0 m^{\prime}}(-\hat{n}_2) \\
    &= \sum_{\ell} | u_{\ell 0} |^2 \mathcal{P}_{\ell}(\cos\left(\hat{n}_1-\hat{n}_2\right)).
\end{aligned}
\end{equation}
To get to the second equality we assumed that the halo profile $u(\hat{n}) \in \mathbb{R}$ has azimuthal symmetry implying that $\sum_m u_{\ell m} Y_{\ell m} = u_{\ell 0} Y_{\ell 0}$. Next we used the definition of the Wigner $D$-matrices to rotate the spherical harmonics. Then, we enforced the orthonormality condition $\int \dd^2 \hat{n} \, Y^{*}_{\ell m}(\hat{n}) Y_{\ell^{\prime} m^{\prime}}(\hat{n}) = \delta_{\ell \ell^{\prime}} \delta_{m m^{\prime}}$. In the last step we used the symmetries of the Wigner $D$-matrices and their relationship to spherical harmonics and Legendre polynomials $\mathcal{P}_{\ell}$ to simplify the equation. 

We perform a multipole expansion on the left hand side of Eq.~\eqref{eq:2pf1haloconfig}:
\begin{equation}
	\bar{\tau}^2 \xi^{1\halo} = \sum_{\ell=0}^{\infty} \sum_{m=-\ell}^{\ell} C_\ell^{1\halo} (\chi) \, Y^{*}_{\ell m}(\hat{n}_1) \, Y_{\ell m}(\hat{n}_2).
\end{equation}
We introduced the angular power spectrum of the $1$-halo term $C_\ell^{1\halo}$, here as a function of redshift. To simplify the right-hand side we use the property that $\frac{4\pi}{2l+1} \sum_m Y^{*}_{\ell m} Y_{\ell m} = \mathcal{P}_\ell$. Next we match the terms in $\ell$ from both expressions above:
\begin{equation}
	\bar{\tau}^{2} C_\ell^{1\halo}(\chi) = \frac{4\pi}{2l+1} \int \dd m \, n(\chi,m) \left[ P(\chi, m) \, u_{\ell 0}(\chi, m) \right]^2.
\end{equation}
To obtain the full sky-averaged power spectrum we need to integrate over the comoving radial coordinate $\chi$. Changing also the integration variable to redshift, \ie $\dd\chi = \dd z / H(z)$ where $H(z)$ is the Hubble constant, we obtain the final expression for the $1$-halo term:
\begin{equation}
\begin{gathered}
	\bar{\tau}^2 C_\ell^{1\halo} = \frac{4\pi}{2l+1} \int \dd z \, \frac{\chi(z)^2}{H(z)} \int \dd m \, n(z,m) \left[ P(z, m) \, u_{\ell 0}(z, m) \right]^2.
\end{gathered}
\end{equation}

\subsection{Two-halo term}\label{appx:2pf2halo}

For the second term we follow similar reasoning and steps as in the previous case, except now we need to consider the conversion probability inside two different halos. Labeling these by $i$ and $j$, located at arbitrary comoving distances $\chi_1$ and $\chi_2$ we find the average configuration space two-point function as:
\begin{equation}\label{eq:2pf2haloconfig}
\begin{aligned}
	\bar{\tau}^2 \xi^{2\halo} &= \ev{\sum_{i j} P^{i \, *}_{\gamma \to \darkph} (\chi_1, \hat{n}_1, m_i) P^j_{\gamma \to \darkph} (\chi_2, \hat{n}_2, m_j)} \\
		&= \left( \prod_{x=a,b} \int \dd m_x \int \dd^2 \hat{n}_x  \right) P(\chi_1, m_a) \, u(\hat{n}_1 - \hat{n}_a) P(\chi_2, m_b) \, u(\hat{n}_2 - \hat{n}_b) \\
        &\times \ev{\sum_{i j} \delta(m_a-m_i)\delta(m_b-m_j)\delta^2(\hat{n}_a-\hat{n}_i)\delta^2(\hat{n}_b-\hat{n}_j)}.
\end{aligned}
\end{equation}
The term in the brackets is related to the number density of each halo's characteristic $m$ and $z$, as well as to the correlation between positions of halos $\xi^{\rm hh}$ in the following way:
\begin{equation}
\begin{aligned}
     \ev{\sum_{i j} \delta(m_a-m_i)\delta(m_b-m_j)\delta^2(\hat{n}_a-\hat{n}_i)\delta^2(\hat{n}_b-\hat{n}_j)} = \\
     = n(\chi_1, m_1) \, n(\chi_2, m_2)& \, \xi^{\rm hh}(\chi_1, m_1, \chi_2, m_2).
\end{aligned}
\end{equation}
Known as the halo-halo auto-correlation function, $\xi^{hh}$ is proportional to the linear matter two-point function. To first order in linear theory, the following is true:
\begin{equation}\label{eq:twoptfuncthh}
	\xi^{hh}(\chi_1, m_a, \chi_2, m_b) \simeq b(\chi_1, m_a) b(\chi_2, m_b) \, \xi^\lin(\chi_1, \chi_2),
\end{equation}
where the bias function $b(z,m)$ is a deterministic function of the halo mass and redshift. We can further relate this to the linear matter power spectrum by doing a Fourier expansion over comoving wavenumbers:
\begin{equation}
\xi^\lin(\chi_1, \chi_2) = \int \frac{\dd^3k}{(2\pi)^3} \, e^{i k \cdot (\chi_1 - \chi_2)} \, P^\lin(k, \chi_1, \chi_2).
\end{equation}
On large scales, the linear matter power spectrum is well approximated by
\begin{equation}
    P^\lin(k, \chi_1, \chi_2) = \sqrt{P^\lin(k, \chi_1) P^\lin(k, \chi_2)}.
\end{equation}
Consider next the multipole expansion of a plane wave:
\begin{equation}
	e^{i k \cdot \chi} = 4\pi \sum_{\ell m} i^\ell j_\ell(k \chi) Y_{\ell m}^{*}(\hat{n}_k) Y_{\ell m}(\hat{n}),
\end{equation}
where $j_\ell(k \chi) \in \mathbb{R}$ is the spherical Bessel function. Performing the multipole expansion on $\xi^{hh}$ from~\eqref{eq:twoptfuncthh} and putting everything together, we obtain the halo-halo angular power spectrum to first order in linear theory:
\begin{equation}\label{eq:cellhh}
	C_\ell^{hh}(\chi_1, \chi_2, m_a, m_b) = \frac{2}{\pi} b(\chi_1, m_a) b(\chi_2, m_b) \int \dd k k^2 j_\ell(k \chi_1) j_\ell(k_2 \chi_2) P^\lin(k,\chi_1,\chi_2).
\end{equation}

We come back to the expression for the $2$-halo two-point function in Eq.~\ref{eq:2pf2haloconfig} and project the right hand side onto spherical harmonics. For this, we generalize a result we obtained in the previous section for the angle-dependent integrand:
\begin{equation}
	u(\hat{n}_1 - \hat{n}_a) \, u(\hat{n}_2 - \hat{n}_b) = \sum_{\ell^{\prime} m^{\prime}} u^{*}_{\ell^{\prime}}(\chi_1) \, \mathcal{D}^{\ell^{\prime}}_{0 m^{\prime}} (-\hat{n}_1) \, Y^{*}_{\ell^{\prime} m^{\prime}}(\hat{n}_a) \sum_{\ell^{\prime\prime} m^{\prime\prime}} u_{\ell^{\prime\prime}}(\chi_2) \, \mathcal{D}^{\ell^{\prime\prime} \, *}_{0 m^{\prime\prime}} (-\hat{n}_2) \, Y_{\ell^{\prime\prime} m^{\prime\prime}}(\hat{n}_b).
\end{equation}
Multiplying by the halo-halo auto-correlation and using the properties of $Y_{\ell m}$'s and $\mathcal{D}_{m m^{\prime}}^{\ell}$'s to simplify the expression we get
\begin{equation}
\begin{aligned}
	u(\hat{n}_1 - \hat{n}_a) \, &u(\hat{n}_2 - \hat{n}_b) \, \xi^{hh}(\chi_1, m_a, \chi_2, m_b) = \\
		&= \sum_{\ell m} u^{*}_{\ell}(\chi_1) \, \mathcal{D}^{\ell}_{0 m} (-\hat{n}_1) \, u_{\ell}(\chi_2) \, \mathcal{D}^{\ell \, *}_{0 m} (-\hat{n}_2) \, C_\ell^{hh} \\
		&= \sum_{\ell} u^{*}_{\ell}(\chi_1) u_{\ell}(\chi_2) \, \mathcal{P}_{\ell} (\hat{n}_1-\hat{n}_2) \, C_\ell^{hh}(\chi_1, \chi_2, m_a, m_b).
\end{aligned}
\end{equation}
The multipole expansion for the general $2$-halo two-point function is
\begin{equation}
	\bar{\tau}^2  \xi^{2\halo}(\chi_1, \chi_2) = \sum_{\ell}  C_\ell^{2\halo} (\chi_1, \chi_2) \, \mathcal{P}_{\ell} (\hat{n}_1-\hat{n}_2).
\end{equation}
Matching term by term we find
\begin{equation}
\begin{aligned}
	\bar{\tau}^2 C_\ell^{2\halo} (\chi_1, \chi_2) = \frac{4\pi}{2\ell+1} &\left( \int \dd m_a \, n(\chi_1, m_a) P(\chi_1, m_a) u_{\ell}(\chi_1, m_a) \right) \\
    &\left( \int \dd m_b \, n(\chi_2, m_b) P(\chi_2, m_b) u_{\ell}(\chi_2, m_b) \right)\, C_\ell^{hh}(\chi_1, m_a, \chi_2, m_b).
\end{aligned}
\end{equation}
Finally, we integrate over redshift to get the final expression for the $2$-halo term:
\begin{equation}
\begin{gathered}
    \bar{\tau}^2 C_\ell^{2\halo} = \frac{4\pi}{2\ell+1} \left[ \prod_{i=1,2}\int \dd z_i \, \frac{\chi(z_i)^2}{H(z_i)} \int \dd m_i \, n(z_i, m_i) b(z_i, m_i) P(z_i, m_i) u_{\ell}(z_i, m_i) \right] C_\ell^\lin(z_1,z_2), \\
	C_\ell^\lin(z_1,z_2) = \frac{2}{\pi} \int \dd k \, k^2 j_\ell(k \, \chi_1)\, j_\ell(k \, \chi_2)\, P^\lin(k,\chi_1,\chi_2).
\end{gathered}
\end{equation}

\section{Two-point correlators and quadratic estimators}\label{appx:qes}

In this appendix we enumerate the various two-point correlation functions among the temperature and polarization anisotropies. We then list all quadratic estimators for the un-screened CMB temperature and polarization anisotropies as well as the Thomson optical depth and photon to dark photon optical depth.

The temperature correlators are:
\begin{align}
    \ev{T^\Sc_{\ell_1 m_1} T^\Sc_{\ell_2 m_2}} =& (-1)^{m_1} C_{\ell_1}^{T^\Sc T^\Sc} \delta_{\ell_1 \ell_2} \delta_{m_1 m_2} \nonumber \\
    -& \sum_{\ell m} \tau_{\ell m}^\Th  (-1)^m     \begin{pmatrix}
				\ell_1 & \ell_2 & \ell \\
				m_1 & m_2 & -m 
	\end{pmatrix} \sqrt{2\ell+1} W^{0 0 0}_{\ell_1 \ell_2 \ell} \left[ C_{\ell_1}^{TT} + C_{\ell_2}^{TT} \right], \\
    \ev{T^\dSc_{\ell_1 m_1} T^\dSc_{\ell_2 m_2}} =& (-1)^{m_1} C_{\ell_1}^{T^\dSc T^\dSc} \delta_{\ell_1 \ell_2} \delta_{m_1 m_2} \nonumber \\
    +& \sum_{\ell m} T_{\ell m} (-1)^m     \begin{pmatrix}
				\ell_1 & \ell_2 & \ell \\
				m_1 & m_2 & -m 
	\end{pmatrix} \sqrt{2\ell+1} W^{0 0 0}_{\ell_1 \ell_2 \ell} \bar{T} \left[ C_{\ell_1}^{\tau\tau} + C_{\ell_2}^{\tau\tau} \right], \\ 
    \ev{T^\dSc_{\ell_1 m_1} T^\Sc_{\ell_2 m_2}} =& (-1)^{m_1} C_{\ell_1}^{T^\dSc T^\Sc} \delta_{\ell_1 \ell_2} \delta_{m_1 m_2} \nonumber \\
    +& \sum_{\ell m} T_{\ell m} (-1)^m     \begin{pmatrix}
				\ell_1 & \ell_2 & \ell \\
				m_1 & m_2 & -m 
	\end{pmatrix} \sqrt{2\ell+1} W^{0 0 0}_{\ell_1 \ell_2 \ell} \bar{T} C_{\ell_1}^{\tau\tau^\Th} \nonumber \\
    -& \sum_{\ell m} \tau_{\ell m}^{*} (-1)^m     \begin{pmatrix}
				\ell_1 & \ell_2 & \ell \\
				m_1 & m_2 & -m 
	\end{pmatrix} \sqrt{2\ell+1} W^{0 0 0}_{\ell_1 \ell_2 \ell} C_{\ell_2}^{TT},
\end{align}
where
\begin{align}
    C_{L}^{T^\Sc T^\Sc} =& \, C_{L}^{TT} + \sum_{\ell \ell^{\prime}} C_{\ell^{\prime}}^{\tau^\Th \tau^\Th} C_{\ell}^{TT} \left( W^{0 0 0}_{L \ell \ell^{\prime}} \right)^2, \\
    C_{L}^{T^\dSc T^\dSc} =& \, \bar{T}^2 \, C_{L}^{\tau\tau} + \sum_{\ell \ell^{\prime}} C_{\ell^{\prime}}^{\tau\tau} C_{\ell}^{TT} \left( W^{0 0 0}_{L \ell \ell^{\prime}} \right)^2, \\
    C_{L}^{T^\dSc T^\Sc} =& \sum_{\ell \ell^{\prime}} C_{\ell^{\prime}}^{\tau\tau^\Th} C_{\ell}^{TT} \left( W^{0 0 0}_{L \ell \ell^{\prime}} \right)^2.
\end{align}
The E-mode correlators are:
\begin{align}
    \ev{E^\Sc_{\ell_1 m_1} E^\Sc_{\ell_2 m_2}} =& (-1)^{m_1} C_{\ell_1}^{E^\Sc E^\Sc} \delta_{\ell_1 \ell_2} \delta_{m_1 m_2} \nonumber \\
    -& \sum_{\ell m} \tau_{\ell m}^\Th  (-1)^m     \begin{pmatrix}
				\ell_1 & \ell_2 & \ell \\
				m_1 & m_2 & -m 
	\end{pmatrix} \sqrt{2\ell+1} e_{\ell_1 \ell_2 \ell} W^{2 2 0}_{\ell_1 \ell_2 \ell} \left[ C_{\ell_1}^{EE} + C_{\ell_2}^{EE} \right], \\
    \ev{E^\dSc_{\ell_1 m_1} E^\dSc_{\ell_2 m_2}} =& (-1)^{m_1} C_{\ell_1}^{E^\dSc E^\dSc} \delta_{\ell_1 \ell_2} \delta_{m_1 m_2}, \\
    \ev{E^\dSc_{\ell_1 m_1} E^\Sc_{\ell_2 m_2}} =& (-1)^{m_1} C_{\ell_1}^{E^\dSc E^\Sc} \delta_{\ell_1 \ell_2} \delta_{m_1 m_2} \nonumber \\
    -& \sum_{\ell m} \tau_{\ell m}^{*} (-1)^m     \begin{pmatrix}
				\ell_1 & \ell_2 & \ell \\
				m_1 & m_2 & -m 
	\end{pmatrix} \sqrt{2\ell+1} e_{\ell_1 \ell_2 \ell} W^{2 2 0}_{\ell_1 \ell_2 \ell} C_{\ell_2}^{EE},
\end{align}
where
\begin{align}
    C_{L}^{E^\Sc E^\Sc} =& \, C_{L}^{EE} + \sum_{\ell \ell^{\prime}} C_{\ell^{\prime}}^{\tau^\Th \tau^\Th} C_{\ell}^{EE} e_{L \ell \ell^{\prime}} \left( W^{2 2 0}_{L \ell \ell^{\prime}} \right)^2, \\
    C_{L}^{E^\dSc E^\dSc} =& \sum_{\ell \ell^{\prime}} C_{\ell^{\prime}}^{\tau\tau} C_{\ell}^{EE} e_{L \ell \ell^{\prime}} \left( W^{2 2 0}_{L \ell \ell^{\prime}} \right)^2, \\
    C_{L}^{E^\dSc E^\Sc} =& \sum_{\ell \ell^{\prime}} C_{\ell^{\prime}}^{\tau\tau^\Th} C_{\ell}^{EE} e_{L \ell \ell^{\prime}} \left( W^{2 2 0}_{L \ell \ell^{\prime}} \right)^2
\end{align}
and
\begin{equation}
e_{\ell \ell^{\prime} \ell^{\prime\prime}} \equiv \frac{1}{2} \left[ 1 + (-1)^{\ell+\ell^{\prime}+\ell^{\prime\prime}} \right].
\end{equation}
The B-mode correlators are:
\begin{align}
    \ev{i B^\Sc_{\ell_1 m_1} i B^\Sc_{\ell_2 m_2}} =& (-1)^{m_1} C_{\ell_1}^{B^\Sc B^\Sc} \delta_{\ell_1 \ell_2} \delta_{m_1 m_2},  \\
    \ev{i B^\dSc_{\ell_1 m_1} i B^\dSc_{\ell_2 m_2}} =& (-1)^{m_1} C_{\ell_1}^{B^\dSc B^\dSc} \delta_{\ell_1 \ell_2} \delta_{m_1 m_2},  \\
    \ev{i B^\dSc_{\ell_1 m_1} i B^\Sc_{\ell_2 m_2}} =& (-1)^{m_1} C_{\ell_1}^{B^\dSc B^\Sc} \delta_{\ell_1 \ell_2} \delta_{m_1 m_2},
\end{align}
where
\begin{align}
    C_{L}^{B^\Sc B^\Sc} =& \sum_{\ell \ell^{\prime}} C_{\ell^{\prime}}^{\tau^\Th \tau^\Th} C_{\ell}^{EE} o_{L \ell \ell^{\prime}} \left( W^{2 2 0}_{L \ell \ell^{\prime}} \right)^2, \\
    C_{L}^{B^\dSc B^\dSc} =& \sum_{\ell \ell^{\prime}} C_{\ell^{\prime}}^{\tau\tau} C_{\ell}^{EE} o_{L \ell \ell^{\prime}} \left( W^{2 2 0}_{L \ell \ell^{\prime}} \right)^2, \\
    C_{L}^{B^\dSc B^\Sc} =& \sum_{\ell \ell^{\prime}} C_{\ell^{\prime}}^{\tau\tau^\Th} C_{\ell}^{EE} o_{L \ell \ell^{\prime}} \left( W^{2 2 0}_{L \ell \ell^{\prime}} \right)^2,
\end{align}
and
\begin{equation}
o_{\ell \ell^{\prime} \ell^{\prime\prime}} \equiv \frac{1}{2} \left[ 1 - (-1)^{\ell+\ell^{\prime}+\ell^{\prime\prime}} \right].
\end{equation}
The E-B correlators are:
\begin{align}
    \ev{E^\Sc_{\ell_1 m_1} i B^\Sc_{\ell_2 m_2}} =& - \sum_{\ell m} \tau_{\ell m}^\Th  (-1)^m \begin{pmatrix}
				\ell_1 & \ell_2 & \ell \\
				m_1 & m_2 & -m 
	\end{pmatrix} \sqrt{2\ell+1} o_{\ell_1 \ell_2 \ell} W^{2 2 0}_{\ell_2 \ell_1 \ell} C_{\ell_1}^{EE}, \\
    \ev{E^\dSc_{\ell_1 m_1} i B^\dSc_{\ell_2 m_2}} =& \ev{E^\dSc_{\ell_1 m_1} i B^\Sc_{\ell_2 m_2}} = 0, \\
    \ev{E^\Sc_{\ell_1 m_1} i B^\dSc_{\ell_2 m_2}} =& - \sum_{\ell m} \tau_{\ell m} (-1)^m \begin{pmatrix}
				\ell_1 & \ell_2 & \ell \\
				m_1 & m_2 & -m 
	\end{pmatrix} \sqrt{2\ell+1} o_{\ell_1 \ell_2 \ell} W^{2 2 0}_{\ell_2 \ell_1 \ell} C_{\ell_1}^{EE}.
\end{align}
The T-E correlators are:
\begin{align}
    \ev{T^\Sc_{\ell_1 m_1} E^\Sc_{\ell_2 m_2}} =& (-1)^{m_1} C_{\ell_1}^{T^\Sc E^\Sc} \delta_{\ell_1 \ell_2} \delta_{m_1 m_2} \nonumber \\
    -& \sum_{\ell m} \tau_{\ell m}^\Th  (-1)^m     \begin{pmatrix}
				\ell_1 & \ell_2 & \ell \\
				m_1 & m_2 & -m 
	\end{pmatrix} \sqrt{2\ell+1} \left[ W^{0 0 0}_{\ell_1 \ell_2 \ell} C_{\ell_2}^{TE} + W^{2 2 0}_{\ell_2 \ell_1 \ell} C_{\ell_1}^{TE} \right], \\
    \ev{T^\dSc_{\ell_1 m_1} E^\dSc_{\ell_2 m_2}} =& (-1)^{m_1} C_{\ell_1}^{T^\dSc E^\dSc} \delta_{\ell_1 \ell_2} \delta_{m_1 m_2} \nonumber \\
    +& \sum_{\ell m} E_{\ell m} (-1)^m     \begin{pmatrix}
				\ell_1 & \ell_2 & \ell \\
				m_1 & m_2 & -m 
	\end{pmatrix} \sqrt{2\ell+1} e_{\ell_1 \ell_2 \ell} W^{2 2 0}_{\ell_2 \ell \ell_1} \bar{T} C_{\ell_1}^{\tau\tau}, \\ 
    \ev{T^\Sc_{\ell_1 m_1} E^\dSc_{\ell_2 m_2}} =& (-1)^{m_1} C_{\ell_1}^{T^\Sc E^\dSc} \delta_{\ell_1 \ell_2} \delta_{m_1 m_2} \nonumber \\
    -& \sum_{\ell m} \tau_{\ell m} (-1)^m     \begin{pmatrix}
				\ell_1 & \ell_2 & \ell \\
				m_1 & m_2 & -m 
	\end{pmatrix} \sqrt{2\ell+1} e_{\ell_1 \ell_2 \ell} W^{2 2 0}_{\ell_2 \ell_1 \ell} C_{\ell_1}^{TE}, \\
    \ev{T^\dSc_{\ell_1 m_1} E^\Sc_{\ell_2 m_2}} =& (-1)^{m_1} C_{\ell_1}^{T^\dSc E^\Sc} \delta_{\ell_1 \ell_2} \delta_{m_1 m_2} \nonumber \\
    +& \sum_{\ell m} E_{\ell m} (-1)^m     \begin{pmatrix}
				\ell_1 & \ell_2 & \ell \\
				m_1 & m_2 & -m 
	\end{pmatrix} \sqrt{2\ell+1} e_{\ell_1 \ell_2 \ell} W^{2 2 0}_{\ell_2 \ell \ell_1} \bar{T} C_{\ell_1}^{\tau\tau^\Th} \nonumber \\
    -& \sum_{\ell m} \tau_{\ell m}^{*} (-1)^m     \begin{pmatrix}
				\ell_1 & \ell_2 & \ell \\
				m_1 & m_2 & -m 
	\end{pmatrix} \sqrt{2\ell+1} W^{0 0 0}_{\ell_1 \ell_2 \ell} C_{\ell_2}^{TE},
\end{align}
where
\begin{align}
    C_{L}^{T^\Sc E^\Sc} =& \, C_{L}^{TE} + \sum_{\ell \ell^{\prime}} C_{\ell^{\prime}}^{\tau^\Th \tau^\Th} C_{\ell}^{TE} W^{0 0 0}_{L \ell \ell^{\prime}} W^{2 2 0}_{L \ell \ell^{\prime}},  \\
    C_{L}^{T^\dSc E^\dSc} =& \sum_{\ell \ell^{\prime}} C_{\ell^{\prime}}^{\tau\tau} C_{\ell}^{TE} W^{0 0 0}_{L \ell \ell^{\prime}} W^{2 2 0}_{L \ell \ell^{\prime}}, \\
    C_{L}^{T^\dSc E^\Sc} =& \sum_{\ell \ell^{\prime}} C_{\ell^{\prime}}^{\tau\tau^\Th} C_{\ell}^{TE} W^{0 0 0}_{L \ell \ell^{\prime}} W^{2 2 0}_{L \ell \ell^{\prime}}.
\end{align}
The T-B correlators are:
\begin{align}
    \ev{T^\Sc_{\ell_1 m_1} i B^\Sc_{\ell_2 m_2}} =& \ev{T^\dSc_{\ell_1 m_1} i B^\dSc_{\ell_2 m_2}} = \ev{T^\Sc_{\ell_1 m_1} i B^\dSc_{\ell_2 m_2}} = 0, \\
    \ev{T^\dSc_{\ell_1 m_1} i B^\Sc_{\ell_2 m_2}} =& \sum_{\ell m} E_{\ell m} (-1)^m \begin{pmatrix}
				\ell_1 & \ell_2 & \ell \\
				m_1 & m_2 & -m 
	\end{pmatrix} \sqrt{2\ell+1} o_{\ell_1 \ell_2 \ell} W^{2 2 0}_{\ell_2 \ell \ell_1} \bar{T} C_{\ell_1}^{\tau \tau^\Th}.
\end{align}
Moving on to the quadratic estimators, there are two estimators for the un-screened temperature anisotropies
\begin{align}
    \hat{T}_{LM} =& N_L^{T; T^\dSc T^\dSc} \sum_{\ell m} \sum_{\ell^{\prime} m^{\prime}} (-1)^M 
    \begin{pmatrix}
				\ell & \ell^{\prime} & L \\
				m & m^{\prime} & -M 
	\end{pmatrix} \sqrt{2L+1} \,G_{\ell \ell^{\prime} L}^{T; T^\dSc T^\dSc} T^\dSc_{\ell m} T^\dSc_{\ell^{\prime} m^{\prime}}, \\
    \hat{T}_{LM} =& N_L^{T; T^\dSc T^\Sc} \sum_{\ell m} \sum_{\ell^{\prime} m^{\prime}} (-1)^M \begin{pmatrix}
				\ell & \ell^{\prime} & L \\
				m & m^{\prime} & -M 
	\end{pmatrix} \sqrt{2L+1} \,G_{\ell \ell^{\prime} L}^{T; T^\dSc T^\Sc} T^\dSc_{\ell m} T^\Sc_{\ell^{\prime} m^{\prime}}.
\end{align}
The weights and prefactors are given by
\begin{align}
    G_{\ell \ell^{\prime} L}^{T; T^\dSc T^\dSc} =& \frac{W^{0 0 0}_{\ell \ell^{\prime} L} \bar{T} \left[C_{\ell}^{\tau\tau} + C_{\ell^{\prime}}^{\tau\tau} \right]}{2 C^{T^\dSc T^\dSc}_{\ell} C^{T^\dSc T^\dSc}_{\ell^{\prime}}} , &  N_{L}^{T; T^\dSc T^\dSc} =& \left[ \sum_{\ell \ell^{\prime}} \frac{\bar{T}^2 \left| W^{0 0 0}_{\ell \ell^{\prime} L} \left[C_{\ell}^{\tau\tau} + C_{\ell^{\prime}}^{\tau\tau} \right] \right|^2}{2 C^{T^\dSc T^\dSc}_{\ell} C^{T^\dSc T^\dSc}_{\ell^{\prime}}} \right]^{-1}, \\
    G_{\ell \ell^{\prime} L}^{T; T^\dSc T^\Sc} =& \frac{W^{0 0 0}_{\ell \ell^{\prime} L} \bar{T} C_{\ell}^{\tau\tau^\Th}}{C^{T^\dSc T^\dSc}_{\ell} C^{T^\Sc T^\Sc}_{\ell^{\prime}}}, & N_{L}^{T; T^\dSc T^\Sc} =& \left[ \sum_{\ell \ell^{\prime}} \frac{\bar{T}^2 \left| W^{0 0 0}_{\ell \ell^{\prime} L} C_{\ell}^{\tau\tau^\Th} \right|^2}{C^{T^\dSc T^\dSc}_{\ell} C^{T^\Sc T^\Sc}_{\ell^{\prime}}} \right]^{-1}.
\end{align}
There are three estimators for the un-screened E-mode polarization anisotropies
\begin{align}
    \hat{E}_{LM} =& N_L^{E; T^\dSc E^\dSc} \sum_{\ell m} \sum_{\ell^{\prime} m^{\prime}} (-1)^M 
    \begin{pmatrix}
				\ell & \ell^{\prime} & L \\
				m & m^{\prime} & -M 
	\end{pmatrix} \sqrt{2L+1} \,G_{\ell \ell^{\prime} L}^{E; T^\dSc E^\dSc} T^\dSc_{\ell m} E^\dSc_{\ell^{\prime} m^{\prime}}, \\
    \hat{E}_{LM} =& N_L^{E; T^\dSc E^\Sc} \sum_{\ell m} \sum_{\ell^{\prime} m^{\prime}} (-1)^M \begin{pmatrix}
				\ell & \ell^{\prime} & L \\
				m & m^{\prime} & -M 
	\end{pmatrix} \sqrt{2L+1} \,G_{\ell \ell^{\prime} L}^{E; T^\dSc E^\Sc} T^\dSc_{\ell m} E^\Sc_{\ell^{\prime} m^{\prime}}, \\
    \hat{E}^{*}_{LM} =& N_L^{E; T^\dSc B^\Sc} \sum_{\ell m} \sum_{\ell^{\prime} m^{\prime}} (-1)^M \begin{pmatrix}
				\ell & \ell^{\prime} & L \\
				m & m^{\prime} & -M 
	\end{pmatrix} \sqrt{2L+1} \,G_{\ell \ell^{\prime} L}^{E; T^\dSc B^\Sc} T^\dSc_{\ell m} B^\Sc_{\ell^{\prime} m^{\prime}},
\end{align}
where
\begin{align}
    G_{\ell \ell^{\prime} L}^{E; T^\dSc E^\dSc} =& \frac{e_{\ell \ell^{\prime} L} W^{2 2 0}_{\ell^{\prime} L \ell} \bar{T} C_{\ell}^{\tau\tau}}{C^{T^\dSc T^\dSc}_{\ell} C^{E^\dSc E^\dSc}_{\ell^{\prime}}}, & N_{L}^{E; T^\dSc E^\dSc} =& \left[ \sum_{\ell \ell^{\prime}} \frac{e_{\ell \ell^{\prime} L} \bar{T}^2 \left| W^{2 2 0}_{\ell^{\prime} L \ell} C_{\ell}^{\tau\tau} \right|^2}{C^{T^\dSc T^\dSc}_{\ell} C^{E^\dSc E^\dSc}_{\ell^{\prime}}} \right]^{-1}, \\
    G_{\ell \ell^{\prime} L}^{E; T^\dSc E^\Sc} =& \frac{e_{\ell \ell^{\prime} L} W^{2 2 0}_{\ell^{\prime} L \ell} \bar{T} C_{\ell}^{\tau\tau^\Th}}{C^{T^\dSc T^\dSc}_{\ell} C^{E^\Sc E^\Sc}_{\ell^{\prime}}}, & N_{L}^{E; T^\dSc E^\Sc} =& \left[ \sum_{\ell \ell^{\prime}} \frac{e_{\ell \ell^{\prime} L} \bar{T}^2 \left| W^{2 2 0}_{\ell^{\prime} L \ell} C_{\ell}^{\tau\tau^\Th} \right|^2}{C^{T^\dSc T^\dSc}_{\ell} C^{E^\Sc E^\Sc}_{\ell^{\prime}}} \right]^{-1}, \\
    i G_{\ell \ell^{\prime} L}^{E; T^\dSc B^\Sc} =& \frac{o_{\ell \ell^{\prime} L} W^{2 2 0}_{\ell^{\prime} L \ell} \bar{T} C_{\ell}^{\tau\tau^\Th}}{C^{T^\dSc T^\dSc}_{\ell} C^{B^\Sc B^\Sc}_{\ell^{\prime}}}, & N_{L}^{E; T^\dSc B^\Sc} =& \left[ \sum_{\ell \ell^{\prime}} \frac{o_{\ell \ell^{\prime} L} \bar{T}^2 \left| W^{2 2 0}_{\ell^{\prime} L \ell} C_{\ell}^{\tau\tau^\Th} \right|^2}{C^{T^\dSc T^\dSc}_{\ell} C^{B^\Sc B^\Sc}_{\ell^{\prime}}} \right]^{-1}.
\end{align}
There are four estimators for the Thomson optical depth:
\begin{align}
    \hat{\tau}^\Th_{LM} =& - N_L^{\tau^\Th ; T^\Sc T^\Sc} \sum_{\ell m} \sum_{\ell^{\prime} m^{\prime}} (-1)^M 
    \begin{pmatrix}
				\ell & \ell^{\prime} & L \\
				m & m^{\prime} & -M 
	\end{pmatrix} \sqrt{2L+1} \,G_{\ell \ell^{\prime} L}^{\tau^\Th ; T^\Sc T^\Sc} T^\Sc_{\ell m} T^\Sc_{\ell^{\prime} m^{\prime}}, \\
    \hat{\tau}^\Th_{LM} =& - N_L^{\tau^\Th ; E^\Sc E^\Sc} \sum_{\ell m} \sum_{\ell^{\prime} m^{\prime}} (-1)^M \begin{pmatrix}
				\ell & \ell^{\prime} & L \\
				m & m^{\prime} & -M 
	\end{pmatrix} \sqrt{2L+1} \,G_{\ell \ell^{\prime} L}^{\tau^\Th ; E^\Sc E^\Sc} E^\Sc_{\ell m} E^\Sc_{\ell^{\prime} m^{\prime}}, \\
    \hat{\tau}^\Th_{LM} =& - N_L^{\tau^\Th ; E^\Sc B^\Sc} \sum_{\ell m} \sum_{\ell^{\prime} m^{\prime}} (-1)^M 
    \begin{pmatrix}
				\ell & \ell^{\prime} & L \\
				m & m^{\prime} & -M 
	\end{pmatrix} \sqrt{2L+1} \,G_{\ell \ell^{\prime} L}^{\tau^\Th ; E^\Sc B^\Sc} E^\Sc_{\ell m} B^\Sc_{\ell^{\prime} m^{\prime}}, \\
    \hat{\tau}^\Th_{LM} =& - N_L^{\tau^\Th ; T^\Sc E^\Sc} \sum_{\ell m} \sum_{\ell^{\prime} m^{\prime}} (-1)^M \begin{pmatrix}
				\ell & \ell^{\prime} & L \\
				m & m^{\prime} & -M 
	\end{pmatrix} \sqrt{2L+1} \,G_{\ell \ell^{\prime} L}^{\tau^\Th ; T^\Sc E^\Sc} T^\Sc_{\ell m} E^\Sc_{\ell^{\prime} m^{\prime}},
\end{align}
where
\begin{align}
    G_{\ell \ell^{\prime} L}^{\tau^\Th ; T^\Sc T^\Sc} =& \frac{W^{0 0 0}_{\ell \ell^{\prime} L} \left[C_{\ell}^{TT} + C_{\ell^{\prime}}^{TT} \right]}{2 C^{T^\Sc T^\Sc}_{\ell} C^{T^\Sc T^\Sc}_{\ell^{\prime}}}, & N_{L}^{\tau^\Th ; T^\Sc T^\Sc} =& \left[ \sum_{\ell \ell^{\prime}}  \frac{\left| W^{0 0 0}_{\ell \ell^{\prime} L} \left[C_{\ell}^{TT} + C_{\ell^{\prime}}^{TT} \right] \right|^2}{2 C^{T^\Sc T^\Sc}_{\ell} C^{T^\Sc T^\Sc}_{\ell^{\prime}}} \right]^{-1}, 
    \\
    G_{\ell \ell^{\prime} L}^{\tau^\Th ; E^\Sc E^\Sc} =& \frac{e_{\ell \ell^{\prime} L} W^{2 2 0}_{\ell \ell^{\prime} L} \left[C_{\ell}^{\tau\tau} + C_{\ell^{\prime}}^{\tau\tau} \right]}{2 C^{E^\Sc E^\Sc}_{\ell} C^{E^\Sc E^\Sc}_{\ell^{\prime}}}, & N_{L}^{\tau^\Th ; E^\Sc E^\Sc} =& \left[ \sum_{\ell \ell^{\prime}}  \frac{e_{\ell \ell^{\prime} L} \left| W^{2 2 0}_{\ell \ell^{\prime} L} \left[C_{\ell}^{\tau\tau} + C_{\ell^{\prime}}^{\tau\tau} \right]  \right|^2}{2 C^{E^\Sc E^\Sc}_{\ell} C^{E^\Sc E^\Sc}_{\ell^{\prime}}} \right]^{-1},
    \\
    i G_{\ell \ell^{\prime} L}^{\tau^\Th ; E^\Sc B^\Sc} =& \frac{o_{\ell \ell^{\prime} L} W^{2 2 0}_{\ell^{\prime} \ell L} C_{\ell}^{EE}}{C^{E^\Sc E^\Sc}_{\ell} C^{B^\Sc B^\Sc}_{\ell^{\prime}}}, & N_{L}^{\tau^\Th ; E^\Sc B^\Sc} =& \left[ \sum_{\ell \ell^{\prime}}  \frac{o_{\ell \ell^{\prime} L} \left| W^{2 2 0}_{\ell^{\prime} \ell L} C_{\ell}^{EE} \right|^2}{C^{E^\Sc E^\Sc}_{\ell} C^{B^\Sc B^\Sc}_{\ell^{\prime}}} \right]^{-1},
    \\
    G_{\ell \ell^{\prime} L}^{\tau^\Th ; T^\Sc E^\Sc} =& \frac{W^{0 0 0}_{\ell \ell^{\prime} L} C_{\ell^{\prime}}^{TE} + W^{2 2 0}_{\ell^{\prime} \ell L} C_{\ell}^{TE}}{C^{T^\Sc T^\Sc}_{\ell} C^{E^\Sc E^\Sc}_{\ell^{\prime}}}, & N_{L}^{\tau^\Th ; T^\Sc E^\Sc} =& \left[ \sum_{\ell \ell^{\prime}}  \frac{\left| W^{0 0 0}_{\ell \ell^{\prime} L} C_{\ell^{\prime}}^{TE} + W^{2 2 0}_{\ell^{\prime} \ell L} C_{\ell}^{TE} \right|^2}{C^{T^\Sc T^\Sc}_{\ell} C^{E^\Sc E^\Sc}_{\ell^{\prime}}} \right]^{-1}.
\end{align}
There are five estimators for the photon to dark photon optical depth:
\begin{align}
    \hat{\tau}^{*}_{LM} =& - N_L^{\tau; T^\dSc T^\Sc} \sum_{\ell m} \sum_{\ell^{\prime} m^{\prime}} (-1)^M 
    \begin{pmatrix}
				\ell & \ell^{\prime} & L \\
				m & m^{\prime} & -M 
	\end{pmatrix} \sqrt{2L+1} \,G_{\ell \ell^{\prime} L}^{\tau; T^\dSc T^\Sc} T^\dSc_{\ell m} T^\Sc_{\ell^{\prime} m^{\prime}}, \\
    \hat{\tau}^{*}_{LM} =& - N_L^{\tau; E^\dSc E^\Sc} \sum_{\ell m} \sum_{\ell^{\prime} m^{\prime}} (-1)^M \begin{pmatrix}
				\ell & \ell^{\prime} & L \\
				m & m^{\prime} & -M 
	\end{pmatrix} \sqrt{2L+1} \,G_{\ell \ell^{\prime} L}^{\tau; E^\dSc E^\Sc} E^\dSc_{\ell m} E^\Sc_{\ell^{\prime} m^{\prime}}, \\
    \hat{\tau}_{LM} =& - N_L^{\tau; E^\Sc B^\dSc} \sum_{\ell m} \sum_{\ell^{\prime} m^{\prime}} (-1)^M 
    \begin{pmatrix}
				\ell & \ell^{\prime} & L \\
				m & m^{\prime} & -M 
	\end{pmatrix} \sqrt{2L+1} \,G_{\ell \ell^{\prime} L}^{\tau; E^\Sc B^\dSc} E^\Sc_{\ell m} B^\dSc_{\ell^{\prime} m^{\prime}}, \\
    \hat{\tau}^{*}_{LM} =& - N_L^{\tau; T^\dSc E^\Sc} \sum_{\ell m} \sum_{\ell^{\prime} m^{\prime}} (-1)^M \begin{pmatrix}
				\ell & \ell^{\prime} & L \\
				m & m^{\prime} & -M 
	\end{pmatrix} \sqrt{2L+1} \,G_{\ell \ell^{\prime} L}^{\tau; T^\dSc E^\Sc} T^\dSc_{\ell m} E^\Sc_{\ell^{\prime} m^{\prime}}, \\
    \hat{\tau}_{LM} =& - N_L^{\tau; T^\Sc E^\dSc} \sum_{\ell m} \sum_{\ell^{\prime} m^{\prime}} (-1)^M \begin{pmatrix}
				\ell & \ell^{\prime} & L \\
				m & m^{\prime} & -M 
	\end{pmatrix} \sqrt{2L+1} \,G_{\ell \ell^{\prime} L}^{\tau; T^\Sc E^\dSc} T^\Sc_{\ell m} E^\dSc_{\ell^{\prime} m^{\prime}},
\end{align}
where
\begin{align}
    G_{\ell \ell^{\prime} L}^{\tau; 
    T^\dSc T^\Sc}
    =& \frac{W^{0 0 0}_{\ell \ell^{\prime} L} C_{\ell^{\prime}}^{TT}}{
    C^{T^\dSc T^\dSc}_{\ell} 
    C^{T^\Sc T^\Sc}_{\ell^{\prime}}}, & N_{L}^{\tau; 
    T^\dSc T^\Sc}
    =& \left[ \sum_{\ell \ell^{\prime}}  \frac{\left| W^{0 0 0}_{\ell \ell^{\prime} L} C_{\ell^{\prime}}^{TT} \right|^2}{
    C^{T^\dSc T^\dSc}_{\ell} 
    C^{T^\Sc T^\Sc}_{\ell^{\prime}}} \right]^{-1},
    \\
    G_{\ell \ell^{\prime} L}^{\tau; 
    E^\dSc E^\Sc} 
    =& \frac{W^{2 2 0}_{\ell \ell^{\prime} L} C_{\ell^{\prime}}^{EE}}{
    C^{E^\dSc E^\dSc}_{\ell} 
    C^{E^\Sc E^\Sc}_{\ell^{\prime}}}, & N_{L}^{\tau; 
    E^\dSc E^\Sc} 
    =& \left[ \sum_{\ell \ell^{\prime}}  \frac{\left| W^{2 2 0}_{\ell \ell^{\prime} L} C_{\ell^{\prime}}^{EE} \right|^2}{
    C^{E^\dSc E^\dSc}_{\ell} 
    C^{E^\Sc E^\Sc}_{\ell^{\prime}}} \right]^{-1},
    \\
    i G_{\ell \ell^{\prime} L}^{\tau; 
    E^\Sc B^\dSc} 
    =& \frac{o_{\ell \ell^{\prime} L} W^{2 2 0}_{\ell^{\prime} \ell L} C_{\ell}^{EE}}{
    C^{E^\Sc E^\Sc}_{\ell} 
    C^{B^\dSc B^\dSc}_{\ell^{\prime}}}, & N_{L}^{\tau; 
    E^\Sc B^\dSc} 
    =& \left[ \sum_{\ell \ell^{\prime}}  \frac{o_{\ell \ell^{\prime} L} \left| W^{2 2 0}_{\ell^{\prime} \ell L} C_{\ell}^{EE} \right|^2}{
    C^{E^\Sc E^\Sc}_{\ell} 
    C^{B^\dSc B^\dSc}_{\ell^{\prime}}} \right]^{-1},
    \\
    G_{\ell \ell^{\prime} L}^{\tau; 
    T^\dSc E^\Sc} 
    =& \frac{W^{0 0 0}_{\ell \ell^{\prime} L} C_{\ell^{\prime}}^{TE}}{
    C^{T^\dSc T^\dSc}_{\ell} 
    C^{E^\Sc E^\Sc}_{\ell^{\prime}}}, & N_{L}^{\tau; 
    T^\dSc E^\Sc} 
    =& \left[ \sum_{\ell \ell^{\prime}}  \frac{\left| W^{0 0 0}_{\ell \ell^{\prime} L} C_{\ell^{\prime}}^{TE} \right|^2}{
    C^{T^\dSc T^\dSc}_{\ell} 
    C^{E^\Sc E^\Sc}_{\ell^{\prime}}} \right]^{-1},
    \\
    G_{\ell \ell^{\prime} L}^{\tau; 
    T^\Sc E^\dSc} 
    =& \frac{e_{\ell \ell^{\prime} L} W^{2 2 0}_{\ell^{\prime} \ell L} C_{\ell}^{TE}}{
    C^{T^\Sc T^\Sc}_{\ell} 
    C^{E^\dSc E^\dSc}_{\ell^{\prime}}}, & N_{L}^{\tau; 
    T^\Sc E^\dSc} 
    =& \left[ \sum_{\ell \ell^{\prime}}  \frac{e_{\ell \ell^{\prime} L} \left| W^{2 2 0}_{\ell^{\prime} \ell L} C_{\ell}^{TE} \right|^2}{
    C^{T^\Sc T^\Sc}_{\ell} 
    C^{E^\dSc E^\dSc}_{\ell^{\prime}}} \right]^{-1}.
\end{align}

\bibliography{DarkPhoton.bib}
\bibliographystyle{JHEP}

\end{document}